\documentclass[12pt]{article}
\usepackage[top=1.2in, bottom=1.2in, left=0.75in, right=0.75in]{geometry}

\usepackage{natbib}
\setcitestyle{nosort}
\usepackage{BOONDOX-cal}
\usepackage{epsfig, color}
\usepackage{graphicx}
\usepackage{amsmath,amsthm,amsfonts,amssymb, bm,bbm}
\allowdisplaybreaks[4]
\usepackage{authblk}
\usepackage[figuresright]{rotating}

\usepackage{makeidx}
\usepackage{listings}
\usepackage[colorlinks,
linkcolor=black,
anchorcolor=black,
citecolor=black
]{hyperref}
\usepackage{enumerate}

\usepackage{rotating}
\usepackage{graphicx}
\usepackage{tikz}
\usepackage{multirow}
\usepackage{subfigure}
\usepackage{floatrow}
\usepackage{cases}

\newtheorem{theorem}{\hspace{6mm}Theorem}[section]
\newtheorem{proposition}{\hspace{6mm}Proposition}[section]
\newtheorem{lemma}{\hspace{6mm}Lemma}[section]

\newtheorem{definition}{\hspace{6mm}Definition}[section]
\newtheorem{assumption}{\hspace{6mm}Assumption}[section]

\def\be{\begin{equation}}
\def\ee{\end{equation}}

\allowdisplaybreaks[4]

\title{Constrained portfolio optimization in a life-cycle model: A deep pricing kernel approach} 

\author[$1$]{Wenyuan Li}
\author[$2$]{Pengyu Wei}

\affil[$1$]{Department of Statistics and Actuarial Science, The University of Hong Kong}
\affil[$2$]{Division of Banking and Finance, Nanyang Business School, Nanyang Technological University}

\date{\today}

\begin{document}

\maketitle

\begin{abstract}
This paper considers constrained portfolio optimization in a generalized life-cycle model. An individual with stochastic income manages a portfolio consisting of stocks, a bond, and life insurance to maximize their consumption, death benefit, and terminal wealth. Meanwhile, the individual faces a convex-set trading constraint, with the non-tradeable asset constraint, no short-selling constraint, and no borrowing constraint as special cases. We build the artificial markets to solve this problem by manipulating the compensated drift terms of the underlying assets to meet the trading constraints. By dual transform, we propose a deep pricing kernel approach to compute tight lower and upper bounds for the primal problem, which can be used when the value function lacks an explicit solution due to the pricing kernel's conditional expectation. Finally, we conclude that when considering the trading constraints, the individual will reduce their consumption, demand for life insurance and annuities, and wealth levels due to the restricted market.
\end{abstract}


\clearpage
\section{Introduction}
The constrained portfolio optimization problem is an extension of the classical portfolio allocation problem. It considers trading constraints, such as non-tradable assets (an incomplete market), no short-selling, no borrowing, etc., and thus adjusts the theoretical model to a more realistic scenario. Compared to the classical problem,  the constrained problem does not always have an explicit solution. The incompleteness arising from trading constraints undermines the uniqueness of the martingale measure and renders the traditional martingale approach inadequate.

Several seminal papers generalize the martingale approach via the convex duality method. \cite{karatzas1991martingale} propose a ``fictitious completion'' method to deal with the portfolio optimization problem in the incomplete market. They introduce additional stocks and build a ``fictitious'' complete market. By manipulating the drift term of these additional stocks, they can guarantee that the individual will not invest in them in the original complete market. \cite{cvitanic1992convex} study a general constrained portfolio problem in which the proportions invested in risky assets belong to a non-empty, closed, and convex set. Using a dual control method, they construct a group of artificial markets that can invest without trading constraints, which provides upper bounds for the primal problem. Finally, they prove that the optimal strategy under the smallest artificial market is optimal for the primal problem. Their framework contains an incomplete market, no short-selling, and no-borrowing constraints as special cases.  \cite{he1993labor} add labor income to the constrained portfolio optimization problem. They use a dual control approach and transform a no-borrowing problem into a variational inequality in the dual space. Several examples of deterministic labor income have been studied in their paper. \cite{cuoco1997optimal} extends \cite{cvitanic1992convex} to the case with stochastic income. He focuses on the optimal amount instead of the optimal proportion allocating among the assets and includes \cite{he1993labor}'s work (no-borrowing constraint) as special cases. For more recent work, we refer to \cite{tepla2000optimal,hu2005constrained,hainaut2006life,haugh2006evaluating,chabakauri2013dynamic,bick2013solving,aase2015life,aase2016life,gu2020constrained,hu2023constrained, wang2024life}.

In life-cycle investment, more and more researchers apply constrained portfolio optimization to address trading constraints in an individual's lifetime investment. \cite{zeng2016optimal} extend \cite{he1993labor}'s work to the actuarial field and studies the wealth-constraint effect on the life insurance purchase.  \cite{branger2019hedging} considers the no borrowing and short-selling constraint in a life-cycle model and study the unemployment risk by a truncated finite difference method. \cite{hambel2022solving} build a group of artificial insurance markets to solve a life-cycle model with unhedgeable biometric shocks. However, most existing literature only focuses on one or two trading constraints, and a general framework is lacking in the context of life-cycle investment.

This paper considers a constrained portfolio optimization problem within a generalized life-cycle model. The individual has a stochastic income and aims to find the optimal trading and insurance strategies to maximize his or her expected consumption utility, bequest utility, and terminal wealth utility. Inspired by the existing literature, we restrict the trading strategy to a non-empty, closed, and convex set that includes many trading constraints (non-tradeable asset constraint, no short-selling constraint, no borrowing constraint, portfolio mix constraint) as special cases. Following \cite{cuoco1997optimal}'s framework, we build a group of artificial markets by adding compensations to the drift terms of stocks and bonds. Due to the lack of uniqueness of martingale measures under trading constraints, we first derive a group of static budget constraints from the individual's wealth process. Next, a dual problem is obtained through the Lagrangian dual control method, which is an upper bound for the primal problem. A candidate strategy can be derived from the upper bound, which then yields a lower bound by substituting the candidate strategy into the wealth process. 

We introduce a neural network to study the best compensated returns for the underlying assets. We first derive a semi-explicit upper bound in the form of a conditional $p$-th moment of the pricing kernel. We simulate this upper bound using Monte Carlo methods and minimize it using stochastic gradient descent. From the upper bound, we induce a candidate strategy in the form of a conditional $p$-th moment of the pricing kernel. Finally, we use deep regression to estimate those conditional expectations and substitute the candidate strategy into the wealth process to obtain a tight lower bound. Our numerical examples show that our ``deep pricing kernel approach'' significantly reduces the duality gap compared to the ``simulation
of artificial markets strategies'' (SAMS) approach  \citep{bick2013solving}. Specifically, for the risk-averse individual $\gamma=1.5$, our approach lowers the relative duality gap from $1.97\%$ under the SAMS approach to only $0.06\%$. For the risk-seeking individual $\gamma=0.8$, it reduces the relative duality gap from $0.71\%$ to $0.03\%$. Moreover, due to the flexibility of the semi-explicit upper bound, the ``deep pricing kernel approach" can be applied to other problems when the value function lacks an explicit solution because of the pricing kernel's conditional $p$-th moment. Finally, both methods show that when considering trading constraints, the individual reduces their consumption, wealth levels, and demand for life insurance and annuities.

To the best of our knowledge, this is the first application of a neural network to compute the best trading and insurance strategies for a constrained portfolio optimization problem. We make four contributions to the existing literature: First, we study the constrained portfolio optimization problem in a life-cycle model with stochastic income and insurance provided. A general dual control framework is constructed that includes many trading constraints, such as no short-selling and no borrowing constraints, as special cases. 
Second, by constructing artificial markets, we change the original dynamic budget constraint into a group of static budget constraints. Subsequently, a Lagrangian upper bound and induced strategies for the lower bound are provided for the primal problem. Third, we propose a dual-control ``deep pricing kernel approach'' to compute the constrained life-cycle model. We derive a semi-explicit solution for the upper bound in terms of the pricing kernel's $p$-th moment, and then use deep regression to induce the corresponding strategy for the lower bound.  Compared to \cite{bick2013solving}, our approach studies high-dimensional, non-linear forms of compensated control $v_t$, significantly reduces the duality gap, and can solve more complex cases that lack explicit solutions. Fourth, we find that individuals reduce their consumption, wealth, and demand for life insurance and annuities when accounting for trading constraints. This implies that although trading constraints help stabilize the financial market, they partially reduce individuals' purchasing power and lifetime welfare.

The rest of the paper is organized as follows: Section \ref{sec_model_settings} introduces our model settings for the financial market, insurance market, wealth process, preference, and the trading constraint set. Section \ref{sec_ai_market} explains the construction of the artificial market and derives the static budget constraint for the wealth process. Section \ref{sec_primal_prob_and_dual_prob} describes the Lagrangian dual control approach and derives the corresponding dual problem. Section \ref{sec_numerical_analysis} conducts the numerical simulation and compares our algorithm with existing literature. Section \ref{sec_conclusion} concludes. All proofs are relegated to the supplementary materials.

\section{Model settings} \label{sec_model_settings}
We consider a constrained portfolio optimization problem in a generalized life-cycle model. The model contains three important dates: a random death time $T_x$ (defined later), a deterministic retirement time $T_R$, and a deterministic time horizon of the family $T$. During the decision period $[0, T \wedge T_x)$, where $T \wedge T_x = \min(T, T_x)$, the individual is allowed to purchase stocks, a bond, and life insurance to improve their consumption level, death benefit, and terminal wealth. 

\subsection{Financial market}
Let $(\Omega, \mathcal{F}, \mathbb{P})$ be a filtered complete probability space. The financial risk is described by a $n$-dimensional Brownian motion $Z_t$ adapted to the filtration $\mathbb{F}=\{\mathcal{F}_t\}_{t \in [0,T]}$.

In the financial market, there are $n+1$ assets. The first asset is the bond, which is locally risk-free and pays no dividends. Its price process is given by
\begin{equation*}
	B_t = \exp\left(\int_0^t r_s ds\right),
\end{equation*}
where $r_t$ is the interest rate process generated by $Z_t$.

\begin{assumption}\label{assumption_1}
	The interest rate process $r_t$ satisfies
	\begin{equation*}
		E\left[\exp\left(\int_0^T |r_t| dt\right)\right] <\infty,
	\end{equation*}
 where $|\cdot|$ means the absolute value.
\end{assumption}

The price process of the risky assets are $S=(S_1, ..., S_n)$ with a cumulative dividend process $D=(D_1,...,D_n)$ satisfying the Ito process 
\begin{eqnarray*}
	S_t + D_t = S_0 + \int_0^t I_{S,u}\mu_u du + \int_0^t I_{S,u}\sigma_u dZ_u, 
\end{eqnarray*}
where $I_{S,t}$ denotes the $n \times n$ diagonal matrix with element $S_t$ and 
\begin{equation*}
	\int_0^T |I_{S,t}\mu_t| dt + \int_0^T |I_{S,t}\sigma_t|^2 dt < \infty.
\end{equation*}

\begin{assumption}\label{assumption_2}
	The volatility matrix $\sigma_t$ satisfies the nondegeneracy condition 
	\begin{equation*}
		x^{\top}\sigma_t \sigma_t^{\top} x \geq \epsilon |x|^2, \text{P-a.s.}
	\end{equation*}
	for any $(x,t)\in \mathbb{R}^n \times [0,T]$ and $\epsilon > 0$. Moreover, denote the market price of the risk vector by
	\begin{equation*}
		\kappa_{0,t} = -\sigma^{-1}_t(\mu_t-r_t\bar{1}_n),
	\end{equation*}
	where $\bar{1}_n = (1,...,1)^{\top} \in \mathbb{R}^n$, we assume a Novikov condition
	\begin{eqnarray*}
		E\left[ \exp\left( \frac{1}{2} \int_0^T |\kappa_{0,t}|^2 dt \right)\right]<\infty.
	\end{eqnarray*}
   in order to ensure the existence of an equivalent martingale measure.
\end{assumption}

\subsection{Mortality}
Denote by $T_x$ the future lifetime of the individual aged $x$, which is a random variable independent of the filtration $\mathbb{F}$ in the financial market. Next, we can introduce the following actuarial notations
\begin{eqnarray*}
	&_{t}p_x = \mathbb{P}[T_x>t], ~_{t}q_x = \mathbb{P}[T_x\leq t] = 1- {_{t}p_x}, ~\lim \limits_{t\rightarrow \infty} {_{t}p_x} =0, ~\lim \limits_{t\rightarrow \infty} {_{t}q_x} =1,
\end{eqnarray*}
where $_{t}p_x$ is the probability that the individual alive at age $x$ survives to at least age $x+t$, $_{t}q_x$ is the probability that the individual aged $x$ dies before $x+t$. Following actuarial practice, we also define the force of mortality (hazard rate)
\begin{equation}\label{hazard_rate}
	\lambda_{x+t} = \frac{1}{_{t}p_x}\frac{d}{dt} {_{t}q_x} = - \frac{1}{_{t}p_x}\frac{d}{dt} {_{t}p_x}.
\end{equation}
Subsequently, the survival and death probabilities can be rewritten as
\begin{eqnarray*}
{ _{t}p_x} = \exp\left\{ - \int_0^t \lambda_{x+s} ds\right\},~{ _{t}q_x} = \int_0^t {_{s}p_x} \lambda_{x+s} ds.	
\end{eqnarray*}
The probability density function of $T_x$ satisfies
\begin{equation*}
	f_{T_x}(t) = {_{t}p_x}  \lambda_{x+t}, ~\text{for} ~t>0.
\end{equation*}

\subsection{Wealth process}
At time 0, the individual at age $x$ starts to manage a portfolio until the first time of the death time $T_x$ and the family's time horizon $T$. Let $T_R$ denote the retirement time. Before the death time $T_x$ and the retirement time $T_R<T$, the individual receives a stochastic non-negative income $Y_t$ generated by the  $n$-dimensional Brownian motion $Z_t$. 

Define the trading strategy $(\alpha, \theta)$ under the price coefficients $\mathcal{P}(r, \mu, \sigma)$, where $\alpha$ and $\theta_k$ represent the money amounts invested at time $t$ in the bond and $k-$th risky asset, respectively. A trading strategy is called admissible if 
\begin{equation}\label{admissible}
	\int_0^T |\alpha_t r_t| dt + \int_0^T |\theta^{\top}_t \mu_t| dt + \int_0^T |\theta^{\top}_t \sigma_t|^2 dt <\infty.
\end{equation} 
We use $\Theta$ to denote the admissible set of trading strategies. Before the individual's death or the family's time horizon, the wealth process satisfies
\begin{eqnarray}
	&&W_t = \alpha_t + \sum \limits_{k=1}^{n} \theta_{k,t}, ~0 \leq t < \min(T_x,T),\label{wealth}\\
	&&W_t = w_0 + \int_0^t (\alpha_s r_s + \theta^{\top}_s \mu_s) ds + \int_0^t \theta_s^{\top} \sigma_s dZ_s - \int_0^t (c_s + I_s - Y_s) ds - C_t, \label{dynamic_budget_constraint}\\
	&&W_t \geq -K, ~ K \in \mathbb{R}^+, \label{wealth_lower_bound}\\
	&&W_T \geq 0, \label{terminal_wealth_bound}
\end{eqnarray}
where $c_t$ is the consumption rate, $I_t$ is the life insurance premium, and $C_t$ is the free disposal of wealth. Free disposal of wealth is the amount of money the individual chooses not to reinvest up to time $t$. Equation \eqref{dynamic_budget_constraint} is usually called the ``dynamic budget constraint''. Equations \eqref{wealth_lower_bound} and \eqref{terminal_wealth_bound} show that the individual is allowed to borrow against the future income but needs to pay the debt at the terminal time. Lastly, equation \eqref{wealth_lower_bound} admits a uniform lower bound to eliminate the arbitrage opportunity, such as the doubling strategy in \cite{harrison1979martingales}. At the death time $T_x$, the individual's wealth has a jump from the insurance payment
\begin{equation*}
	W_{T_x} = W_{T_x-} + \frac{I_{T_x}}{\lambda_{x+T_x}},
\end{equation*} 
where $\lambda_t$ is the force of mortality defined in \eqref{hazard_rate}, and $I_{T_x}/\lambda_{x+T_x}$ is the life insurance payout.

\subsection{Preference and feasibility}
The individual's objective is to choose an investment and insurance strategy $(\alpha, \theta, I)$ to optimize the expected utility of consumption when the individual is alive, the wealth level at the death time, and the terminal wealth at the family's time horizon.
\begin{equation*}
	\sup \limits_{(\alpha, \theta) \in A, I} E\left[\int_0^T U_1(c_t,t)\mathbbm{1}_{\{t<T_x\}} dt +  U_2\left(W_{T_x},T_x\right)\mathbbm{1}_{\{T_x<T\}} + U_3(W_{T},T)\mathbbm{1}_{\{T_x\geq T\}}\right],	
\end{equation*}
where $A$ is the portfolio constraint set in $\mathbb{R}^{n+1}$, $U_1$ is the consumption utility, $U_2$ is the bequest utility, and $U_3$ is the terminal utility. We assume all the utilities satisfy the following properties.

\begin{definition}\label{def_utility_function}	
	Utility functions $U_i : (0,\infty) \times [0,T] \rightarrow \mathbb{R}, i=1,2,3$ are increasing, strictly concave, and continuously differentiable in its first variable and continuous in the second variable. Moreover, it satisfies the Inada condition
\begin{equation}\label{Inada_condition}
	U'_i(0+,t) = \infty, U'_i(\infty,t) = 0+, ~\text{for}~ \forall t \in [0,T],
\end{equation}
in which $U'_i$ is the first-order derivative with respect to the first variable. 
\end{definition}

Since the individual's time to death $T_x$ is independent of the filtration $\mathbb{F}$ in the financial market, we have the equivalent preference
\begin{eqnarray}
	&&\sup \limits_{(\alpha, \theta) \in A, I} E\left[\int_0^T U_1(c_t,t)\mathbbm{1}_{\{t<T_x\}} dt +  U_2\left(W_{T_x},T_x\right)\mathbbm{1}_{\{T_x<T\}} + U_3(W_{T},T)\mathbbm{1}_{\{T_x\geq T\}}\right] \notag\\
	&=&\sup \limits_{(\alpha, \theta) \in A, I} E\left[\int_0^T {_{t}p_x}U_1(c_t,t) dt + \int_0^T f_{T_x}(t) U_2\left(W_t+\frac{I_t}{\lambda_{x+t}},t \right)dt \right.\notag\\
	&&\left.+ \int_T^{\infty} f_{T_x}(t) U_3(W_T,T)dt
	\right]	\notag\\
	&=& \sup \limits_{(\alpha, \theta) \in A, I} E\left[\int_0^T {_{t}p_x}U_1(c_t,t) dt + \int_0^T {_{t}p_x}\lambda_{x+t} U_2\left(W_t+\frac{I_t}{\lambda_{x+t}},t \right)dt + {_{T}p_x} U_3(W_T,T)\right]\notag\\
	&:=& \sup \limits_{(\alpha, \theta) \in A, I} E\left[\int_0^T {_{t}p_x}U_1(c_t,t) dt + \int_0^T {_{t}p_x}\lambda_{x+t} U_2\left(M_t,t \right)dt + {_{T}p_x} U_3(W_T,T)\right],
\end{eqnarray} 
where $M_t = W_t+ I_t/\lambda_{x+t}$. 

Before moving to the feasibility of strategies, we first define the consumption and bequest set. Consider the set $G$
\begin{equation} \label{consumption_and_bequest_set}
	G:=\left\{ (c,M,W_T) : E^{Q_0}\left[\int_0^T \left|c_t\right| + \left|M_t\right| dt + \left|W_T\right|\right]<\infty, \text{P-a.s.} \right\},
\end{equation}
where $Q_0$ is the risk neutral measure such that $dZ_{0,t} = dZ_t - \kappa_{0,t}dt$ is a Brownian motion (see Assumption \ref{assumption_2}). Moreover, Let $G_+$ denote the orthant of $(c,M,W_T)$ that $c_t> 0$, $M_t > 0$, and $W_T > 0$.

Given price coefficients $\mathcal{P}=(r,\mu,\sigma)$, a consumption and bequest plan $(c,M,W_T)\in G_+$ is called ``feasible'' if there exists an admissible trading strategy $(\alpha,\theta)\in \Theta$ for $\forall t \in [0,T]$, and a non-negative increasing free disposal $C$ satisfying the dynamic budget constraint from \eqref{wealth} to \eqref{terminal_wealth_bound}. In addition,  the plan $(c,M,W_T)\in G_+$ is said to be ``$A$-feasible'' if it is feasible and $(\alpha,\theta)\in A$ for $\forall t \in [0,T]$. In both cases, the trading strategy is said to ``finance'' $(c,M,W_T)$. We use $\mathcal{B}(\mathcal{P},A)$ to denote the set of $A$-feasible consumption and bequest plan given the pricing coefficient $\mathcal{P}$.

\subsection{Portfolio constraint set}\label{port_constraint_set}

We assume that the agent's portfolio $(\alpha, \theta)$ is constrained to take values in a portfolio constraint set $A$, which is a non-empty, closed, and convex subset of $\mathbb{R}^{n+1}$. It can describe various trading constraints such as short-sale prohibitions, non-tradeable assets, or minimal capital requirements. For $v=(v_0, v_{-}) \in \mathbb{R} \times \mathbb{R}^n$, define
\begin{equation}\label{supporting_function}
	\delta(v) = \sup \limits_{(\alpha,\theta)\in A} -(\alpha v_0 + \theta^{\top}v_{-}),
\end{equation}
which is the support function of $-A$. This function can easily reach $+\infty$, and hence it is important to define its effective domain as
\begin{equation*}
	\widetilde{A} = \left\{ v \in \mathbb{R}^{n+1} : \delta(v)<\infty \right\}.
\end{equation*}
In the convex analysis, it is well-known that $\delta$ is a positively homogeneous, lower semi-continuous, and proper convex function on $\mathbb{R}^{n+1}$ and $\widetilde{A}$ is a closed convex cone. We assume the support function satisfies the following constraint
\begin{assumption}\label{assumption_3}
	The function $\delta$ is upper semi-continuous and bounded above on $\widetilde{A}$. Moreover, $v_0 \geq 0$ for all $v \in \widetilde{A}$.
\end{assumption}
$v_0 \geq 0$ for all $v \in \widetilde{A}$ is immediately obtained if $(\alpha,0) \in A$ for any $\alpha$ large enough, i.e., as long as lending and investing nothing in the risky assets is admissible. Moreover, since $\delta$ is positively homogeneous and $\widetilde{A}$ is a cone, the function $\delta$ bounded above on $\widetilde{A}$ is equivalent to $\delta$ being non-positive on $\widetilde{A}$. Specifically, if $A$ is a cone, then $\delta \equiv 0$ on $\widetilde{A}$. Below, we provide some examples of constraint sets $A$ satisfying Assumption \ref{assumption_3}, together with the associated support functions and dual sets (see Section 3 in \cite{cuoco1997optimal} for details).

\begin{enumerate}[(a)]
	\item No constraints:
	\begin{eqnarray*}
		&&A = \mathbb{R}^{n+1},\\
		&&\widetilde{A} = \{0\},\\
		&&\delta(v) = 0 ~\text{for}~ \forall v \in \widetilde{A}.
	\end{eqnarray*}
	
	\item Nontradeable assets (incomplete market):
	\begin{eqnarray*}
		&&A = \{(\alpha,\theta)\in \mathbb{R}^{n+1}: \theta_k=0, ~k= m+1,..., n\},\\
		&&\widetilde{A} = \{v\in \mathbb{R}^{n+1} : v_k = 0, ~k=0,...,m\},\\
		&&\delta(v) = 0 ~\text{for}~ v \in \widetilde{A}.
	\end{eqnarray*}	
	
	\item Short-sale constraint
	\begin{eqnarray*}
		&&A = \{ (\alpha,\theta) \in \mathbb{R}^{n+1} : \theta_k \geq 0, ~k=m+1,...,n \},\\
		&&\widetilde{A} = \{ v \in \mathbb{R}^{n+1} : v_k=0, ~k = 1,...,m; v_k \geq 0, ~k=m+1,...,n \},\\
		&&\delta(v) = 0 ~\text{for}~ v \in \widetilde{A}.
	\end{eqnarray*}
	
	\item Buying constraints
	\begin{eqnarray*}
		&&A = \{(\alpha,\theta)\in \mathbb{R}^{n+1}: \theta_k \leq 0, ~k=m+1,...,n\},\\
		&&\widetilde{A} =\{ v\in \mathbb{R}^{n+1}: v_k=0, ~k=1,...,m; v_k\leq 0, ~k=m+1,...,n\},\\
		&&\delta(v)=0 ~\text{for}~ v\in \widetilde{A}. 
	\end{eqnarray*}
	
	\item Portfolio-mix constraint
	\begin{equation*}
		A = \left\{ (\alpha,\theta) \in \mathbb{R}^{n+1}: \alpha + \sum\limits_{k=1}^n \theta_k \geq 0,~\theta \in D 
		\left(\alpha + \sum\limits_{k=1}^n \theta_k\right)\right\},
	\end{equation*}
	where $D$ is any nonempty, closed, convex subset of $\mathbb{R}^n$ containing the origin,
	\begin{eqnarray*}
		&&\widetilde{A} = \{ v\in \mathbb{R}^{n+1}: v^{\top}(\alpha,\theta) \geq 0,~\forall (\alpha,\theta)\in A \},\\
		&&\delta(v) = 0 ~\text{for}~ v\in \widetilde{A}. 
	\end{eqnarray*}
	
	\item Minimum capital requirement
	\begin{equation*}
		A = \left\{ (\alpha,\theta) \in \mathbb{R}^{n+1}: \alpha + \sum \limits_{k=1}^n \theta_k \geq K\right\},
	\end{equation*} 
	where $K \geq 0$,
	\begin{eqnarray*}
		&&\widetilde{A} = \{k\bar{1}_{n}: k\geq 0\},\\
		&&\delta(v) = -K v_0 ~\text{for}~ v\in \widetilde{A}.
	\end{eqnarray*}
	
	\item Collateral constraints
	\begin{equation*}
		A = \left\{(\alpha,\theta) \in \mathbb{R}^{n+1}: \Psi_0 \alpha + \sum \limits_{k=1}^n \Psi_k \theta_k \geq \gamma (\Psi_0 \alpha^+ + \sum \limits_{k=1}^n \Psi_k \theta^+_k)\right\},
	\end{equation*}
	where $\Psi_k \in [0,1]$ for $k=0,1,...,n$ denotes the fraction of the amount of asset $k$ can be borrowed using the asset as collateral and $\gamma \in [0,1]$,
	\begin{eqnarray*}
		&&\widetilde{A} = \{ v\in \mathbb{R}^{n+1}: v^{\top}(\alpha,\theta) \geq 0, \forall (\alpha,\theta)\in A\},\\
		&& \delta(v)=0 ~\text{for}~ v\in \widetilde{A}.
	\end{eqnarray*}
	
	\item Any combination of above constraints.
\end{enumerate}

\section{Artificial market and static budget constraint}\label{sec_ai_market}

Inspired by \cite{cuoco1997optimal}, we define the artificial market to solve the constrained portfolio optimization. Given a constraint set $A$, let $\mathcal{N}$ denote the $\widetilde{A}$ valued process satisfying
\begin{equation*}
	E\left[\int_0^T |v_t|^2 dt\right]<\infty.
\end{equation*}
For each $v\in\mathcal{N}$, the processes
\begin{eqnarray}
	&&\beta_{v,t} = \exp\left(-\int_0^t r_s + v_{0,s}ds\right), \notag\\
	&&\kappa_{v,t} = - \sigma^{-1}_t(\mu_t + v_{-,t} - (r_t + v_{0,t})\bar{1}_n), \notag\\
	&&\xi_{v,t} = \exp\left(\int_0^t \kappa^{\top}_{v,s}dZ_s - \frac{1}{2}\int_0^t |\kappa_{v,s}|^2 ds\right), \notag\\
	&&\pi_{v,t} = \beta_{v,t} \xi_{v,t},\notag\\
	&&dZ_{v,t} = dZ_t - \kappa_{v,t}dt, \label{change of measure}
\end{eqnarray}
define an artificial market $\mathcal{M}_v$, where $\xi_v$ is a strictly positive local martingale. We further use $\mathcal{N}^*$ to denote the subset of elements $v$ in $\mathcal{N}$ for which $\xi_v$ is exactly a martingale. Note that $\mathcal{N}^*$ is nonempty given the Novikov condition and the fact that $\widetilde{A}$ is a cone ensuring that $0 \in \mathcal{N^*}$. Then, each $\pi_{v,t}, v\in \mathcal{N}^*$ can be interpreted as the unique pricing kernel/state-price density in a fictitious unconstrained market $\mathcal{M}_v$ with price coefficients $\mathcal{P}=(r+v_0,\mu+v_-,\sigma)$. With the adjustment of drift term by $v=(v_0,v_-)$, the stocks can become more attractive or less attractive compared to the bond. Subsequently, ``$A$-feasible'' trading strategies can be built by the change of an individual's preference between stocks and the bond. More generally, each $\pi_{v,t}$ with $v \in \mathcal{N}^*$ constitutes an arbitrage-free state-price density in the original economy when the portfolio policies are constrained to be in $A$, and the fulfillment of a budget constraint with respect to all of these state-price densities is sufficient to guarantee the $A$-feasibility.

To satisfy the lower boundedness property \eqref{wealth_lower_bound} of wealth process $W_t$, we add the following assumption to the income process $Y_t$
\begin{assumption}\label{assumption_4}
	\begin{equation}\label{income_constraint}
		\sup \limits_{ v \in \mathcal{N}^*} E^{Q_v}\left[\int_0^T {_tp_x}e^{-\int_0^t r_s ds}Y_t dt\right] \leq K_y,	
	\end{equation}
	for some positive constant $K_y>0$.
\end{assumption}

Assumption \ref{assumption_4} includes the uniformly bounded income case studied in \cite{cuoco1997optimal}. Next, we show the equivalent static budget constraint of the $A$-feasible dynamic constraint, which means the individual's benefit (terminal wealth plus death benefit plus consumption) is less and equal to the endowment (initial wealth plus income plus supporting function).
\begin{theorem}\label{theorem_1}
	A consumption and bequest plan $(c,M,W_T)\in G^*_+$ is $A$-feasible if and only if 
	\begin{eqnarray}
		&&E^{Q_v}\left[ {_Tp_x}\beta_{v,T}  W_T + \int_0^T {_tp_x}\lambda_{x+t} \beta_{v,t}  M_t dt + \int_0^T {_tp_x} \beta_{v,t} (c_t-Y_t)dt \right] \notag\\
		&&\leq w_0 + E^{Q_v} \left[\int_0^T {_tp_x}\beta_{v,t}\delta(v_t)dt\right] ~\text{for}~ \forall v\in \mathcal{N}^*. \label{static_budget_constraint}
	\end{eqnarray}	
\end{theorem}

\section{Primal problem and dual problem} \label{sec_primal_prob_and_dual_prob}
From \eqref{static_budget_constraint}, we can formulate the primal problem with the dynamic budget constraint \eqref{dynamic_budget_constraint} as a problem with a static budget constraint.
\begin{align}
	& \hspace{4cm} \sup \limits_{(c,M,W_T) \in G_+} J(c,M,W_T) \notag\\
	&\text{s.t.} ~E^{Q_v}\left[ {_{T}p_x}\beta_{v,T} W_T + \int_0^T {_{t}p_x}\lambda_{x+t}\beta_{v,t}M_t dt + \int_0^T {_{t}p_x}\beta_{v,t} (c_t-Y_t) dt\right]	\tag{P} \label{primal_problem}\\
	&  \leq w_0 + E^{Q_v}\left[\int_0^T {_{t}p_x}\beta_{v,t} \delta(v_t)dt\right],\notag
\end{align}
for $\forall v\in \mathcal{N}^*$, where
\begin{eqnarray*}
	J(c,M,W_T) &=& E\left[ \int_0^T {_{t}p_x} U_1(c_t,t) dt + \int_0^T {_{t}p_x}\lambda_{x+t} U_2(M_t,t) dt + {_{T}p_x}U_3(W_T,T)\right].
\end{eqnarray*}
Since $0\in \mathcal{N}^*$, problem \eqref{primal_problem} can be considered as a convex optimization problem on a closed, norm bounded subset of $L^1(\bar{\lambda} \times Q_0)$, where $\bar{\lambda}$ is the Lebesgue measure on $[0,T]$. However, $L^1$ spaces are not reflexive so lack compactness. The existing literature circumvents this difficulty using the Lagrangian dual control method. Because the set $\{\pi_v: v\in \mathcal{N}^*\}$ is convex, this suggests the existence of pricing kernel $\pi_{v^*}$, a Lagrangian multiplier $\psi^*>0$ such that $(c^*,M^*, W^*_T,\psi^*,v^*)$ is a saddle point of the Lagrangian
\begin{eqnarray*}
	&&\mathcal{L}(c,M,W_T,\psi,v)= E\left[ \int_0^T {_{t}p_x} U_1(c_t,t) dt + \int_0^T {_{t}p_x} \lambda_{x+t}U_2(M_t,t) dt + {_{T}p_x} U_3(W_T,T)\right]\\
	&&+\psi \left\{ w_0 - E\left[ \int_0^T {_{t}p_x} \pi_{v,t} [c_t + \lambda_t M_t -Y_t -\delta(v_t)]dt  + {_{T}p_x}\pi_{v,T} W_T\right] \right\}.
\end{eqnarray*}
It is easy to see that maximizing the Lagrangian $\mathcal{L}$ with $(c, M, W_T)$ and minimizing $(\psi, v)$ provides an upper bound for the primal problem. We show this in the general case, starting at time 0.
\begin{eqnarray}
	&&J(c,M,W_T) := E\left[ \int_0^T {_tp_x} U_1(c_t,t) dt + \int_0^T {_tp_x}\lambda_{x+t}  U_2(M_t,t) dt + {_Tp_x} U_3(W_T,T) \right] \notag\\
	&&= E\left[ \int_0^T {_tp_x} [U_1(c_t,t)-c_t\psi \pi_{v,t}] dt + \int_0^T  {_tp_x} \lambda_{x+t}  [U_2(M_t,t)-M_t \psi \pi_{v,t}] dt \right.\notag\\
	&&\left.  + {_Tp_x}  [U_3(W_T,T)-W_T\psi \pi_{v,T}] \right] + \psi E\left[\int_0^T {_tp_x} \pi_{v,t} c_t dt + \int_0^T {_tp_x}\lambda_{x+t} \pi_{v,t} M_t dt + {_Tp_x}\pi_{v,T} W_T\right]\notag\\
	&&\leq E\left[ \int_0^T {_tp_x} \sup \limits_{c_t > 0}[U_1(c_t,t)-c_t\psi \pi_{v,t}] dt + \int_0^T {_tp_x}\lambda_{x+t} \sup \limits_{M_t > 0} [U_2(M_t,t)-M_t \psi \pi_{v,t}] dt \right.\notag\\
	&&\left.  + {_Tp_x} \sup \limits_{W_T > 0} [U_3(W_T,T)-W_T\psi \pi_{v,T}] \right]+\psi\left\{w_0 + E\left[\int_0^T {_tp_x}\pi_{v,t}[Y_t+\delta(v_t)]dt\right]\right\}\notag\\
 	&&:=\sup \limits_{c > 0, M > 0, W_T> 0}\mathcal{L}(c, M, W_T, \psi, v), \label{lag_inequality}
\end{eqnarray}
where the first inequality is by taking the supremum with $(c, M, W_T)$ and the static budget constraint \eqref{static_budget_constraint}. Finally, taking the supremum on the left and infimum on the right of \eqref{lag_inequality}, we arrive at
\begin{equation}
   \sup \limits_{c > 0, M > 0, W_T> 0} J(c,M,W_T) \leq \inf \limits_{(\psi,v)\in (0,\infty)\times \mathcal{N}^*} \sup \limits_{c > 0, M > 0, W_T> 0}\mathcal{L}(c, M, W_T, \psi, v).\label{lag_upperbound}
\end{equation}

Let $\widetilde{J}(\psi,v)$ denote the Lagrangian $\mathcal{L}$ maximized over $(c,M,W_T)$, and then we define the dual problem, which is an upper bound for the primal problem \eqref{primal_problem}
\begin{align}
	&\hspace{4cm} \inf \limits_{(\psi,v)\in (0,\infty)\times \mathcal{N}^*}\widetilde{J}(\psi,v)\notag\\
	&=\inf \limits_{(\psi,v)\in (0,\infty)\times \mathcal{N}^*}E\left[ \int_0^T {_{t}p_x} \widetilde{U}_1(\psi \pi_{v,t},t) dt +\int_0^T {_{t}p_x}\lambda_{x+t} \widetilde{U}_2(\psi \pi_{v,t},t) dt \right. \tag{D} \label{dual_problem}\\
	&\left.+{_{T}p_x}\widetilde{U}_3(\psi \pi_{v,T},T) + \psi \left\{  w_0 + \int_0^T {_{t}p_x}\pi_{v,t} [Y_t + \delta(v_t)] dt\right\}\right], \notag
\end{align}
where dual utilities are given by
\begin{eqnarray*}
	&&\widetilde{U}_i(z,t) = \sup \limits_{x > 0}\left\{ U_i(x,t) - zx\right\},
\end{eqnarray*}
for $z>0$ and each $U_i, i=1,2,3$, and satisfies the Inada condition \eqref{Inada_condition}.

For $\widetilde{U}_1(z,t),~z>0$, by the concavity of $U_1$, we have a $c^*$ such that
\begin{equation}\label{tildeU1_derivation}
	\widetilde{U}_1(z,t) = U_1(c^*,t) - zc^*,
\end{equation}
where $U'_1(c^*,t)-z=0$, i.e. $c^* = U'^{-1}_1(z,t)$, and $U'^{-1}_1(z,t)$ is the inverse of $U'(c,t)$ with respect to the first variable. Next, take the first order derivative with $z$ on both sides of \eqref{tildeU1_derivation} and by  $U'_1(c^*,t)-z=0$, we have
\begin{equation*}
	\frac{\partial \widetilde{U}_1(z,t)}{\partial z} = U'_1(c^*,t)\frac{\partial c^*}{\partial z} - c^* - z\frac{\partial c^*}{\partial z} = -c^*,
\end{equation*}
i.e.
\begin{equation*}
	c^* = U'^{-1}_1(z,t) = -\frac{\partial \widetilde{U}_1(z,t)}{\partial z}.
\end{equation*}
Define the function $f_i(z,t) =  U'^{-1}_i(z,t) = -\frac{\partial}{\partial z}\widetilde{U}_i(z,t), i=1,2,3$, similarly to the argument above, we have
\begin{equation}\label{opt_f}
	c^* = f_1(z,t), M^* = f_2(z,t), W^* = f_3(z,T).
\end{equation} 
Then, by Definition \ref{def_utility_function}, we can derive the following properties for dual utility.
\begin{lemma}\label{dual_utility_property}
	The dual utilities $\widetilde{U}_i(\cdot,t):(0,\infty)\rightarrow \mathbbm{R}, i=1,2,3$ are strictly decreasing and strictly convex with respect to the first variable. They have the explicit representations
	\begin{equation}\label{tildeU1_derivation2}
		\widetilde{U}_i(z,t) = U_i(f_i(z,t),t) - zf_i(z,t), ~\text{where}~ i=1,2,3.
	\end{equation} 
	and derivatives $\frac{\partial }{\partial z}\widetilde{U}_i(z,t)=-f_i(z,t)=-U'^{-1}_i(z,t)$. Furthermore,
	\begin{equation*}
		\widetilde{U}_i(0+,t) = U_i(\infty,t), ~~\widetilde{U}_i(\infty,t) = U_i(0+,t).
	\end{equation*}
\end{lemma}

Furthermore, under each artificial market $\mathcal{M}_v$, we can derive the following proposition for the dual problem (\ref{dual_problem}).
\begin{proposition}\label{find_opt_psi_prop}
	For an arbitrary $v \in \mathcal{N}^*$, there exists a unique optimal $\psi_v$ minimizing $\widetilde{J}(\psi,v)$ such that
	\begin{equation*}
		\frac{\partial \widetilde{J}(\psi_v,v)}{\partial \psi} = 0.
	\end{equation*}
	In addition, the optimal wealth under $(\psi_v,v)$ is given by
	\begin{eqnarray}
		&& W_{v,t} = E^{Q_{v}} \left[  \int_t^T e^{-\int_t^s r_u + v_{0,u} + \lambda_{x+u} du}[f_1(\psi_v\pi_{v,s},s) -Y_s + \lambda_{x+s}f_2(\psi_v\pi_{v,s},s) - \delta(v_s)] ds \right. \notag\\
		&&\left.+ e^{-\int_t^T r_s + v_{0,s} + \lambda_{x+s} ds}f_3(\psi_v\pi_{v,T},T)|\mathcal{F}_{t} \right],\label{optimal_wealth_v} 
	\end{eqnarray}
	and the optimal free disposal $C^*_{v,t} \equiv 0$ under $(\psi_v,v)$.
\end{proposition}

In the next section, we will show how to design an algorithm to compute the upper and lower bounds for the primal problem.

\section{Algorithm design and numerical analysis}\label{sec_numerical_analysis}

Following the parameter settings in \cite{huang2008portfolio}, we assume that an individual is 35 years old at the initial time, retires at the age of 65, and the family stops making investment decisions at the individual's age of 95, so $T_R = 30$ and $T = 60$. The individual's force of mortality follows the Gompertz law
\begin{equation*}
	\lambda_{x+t} = \frac{1}{9.5} \exp{\left(\frac{x+t-86.3}{9.5}\right)}, ~x=35.
\end{equation*}

Before the first time of the family decision horizon $T$ and death time $T_x$, the individual is allowed to invest in a bond and a stock
\begin{eqnarray*}
	B_t &=& \exp\left(\int_0^t r(u) du \right),\\
	S_t + D_t &=& S_0 + \int_0^t \mu(u) S_u du + \int_0^t \sigma(u) S_u dZ_u,
\end{eqnarray*}
where $r(t), \mu(t), \sigma(t)$ are continuous functions of $t$ , $\sigma(t)>0$ for $t\in [0,T]$, and $Z_t$ is a one-dimensional Brownian motion. Moreover, the individual's income process has no idiosyncratic risk (only has Brownian motion from the financial market)
\begin{eqnarray}
	\left\{
	\begin{array}{ll}
		Y_t&= Y_0 + \mu_Y \int_0^t Y_u du + \sigma_Y \int_0^t Y_u dZ_u, ~~ 0\leq t < \min(T_x,T_R), \label{numerical_Yt_process}\\
		Y_t&= 0, ~~\min(T_x,T_R)\leq t\leq T,
	\end{array}
	\right.
\end{eqnarray}
where $\mu_Y$ and $\sigma_Y$ are two constants. 
We consider the portfolio-mix constraint (Section \ref{port_constraint_set} (e)) with $D = [0,1]$, then the portfolio constraint set $A$ and its effective domain $\widetilde{A}$ are given by 
\begin{eqnarray}
	A &=& \left\{ (\alpha, \theta) \in \mathbb{R}^2: \alpha + \theta \geq 0, \theta \in [0, \alpha+\theta]\right\} \label{numerical_A} \\
	&=& \left\{ (\alpha, \theta) \in \mathbb{R}^2: \alpha \geq 0, \theta \geq 0\right\},\notag\\
	\widetilde{A} &=& \left\{ (v_0, v_-): (\alpha, \theta)(v_0, v_-)^{\top} \geq 0, \forall (\alpha, \theta) \in A\right\} \notag\\
	&=& \left\{ (v_0, v_-): v_0 \geq 0, v_- \geq 0 \right\}.\notag
\end{eqnarray}
Moreover, $\delta(v_t) = 0$. As a result, the individual's wealth process \eqref{wealth} has the following equivalent form 
\begin{eqnarray}\label{numerical_wealth}
	W_t &=& W_0 + \int_0^t [ (r(s)+\lambda_{x+s}) W_s + (\mu(s)-r(s)) \theta_s] ds + \int_0^t \sigma(s) \theta_s dZ_s \notag\\
	&&  - \int_0^t (c_s + \lambda_{x+s}M_s - Y_s) ds - C_t,
\end{eqnarray}
where $ 0 \leq t \leq \min(T_x,T)$ and $M_t = W_t + \frac{I_t}{\lambda_{x+t}}$. 

We restrict utilities to the power utility
\begin{eqnarray*}
	\left\{
	\begin{array}{l}
		U_1(c_t,t) = e^{-\widetilde{\delta} t}\frac{c^{1-\gamma}_t}{1-\gamma},\\
		U_2(M_t,t) = e^{-\widetilde{\delta} t}\varphi V_B(t,M_t),\\
		U_3(W_T,T) = e^{-\widetilde{\delta} T}\frac{W^{1-\gamma}_T}{1-\gamma},
	\end{array}
	\right.
\end{eqnarray*}
where $V_B(t, M_t)$ is the value function of family investment after the individual dies, and the subscript ``B'' is short for bequest. Moreover, $\varphi$ is the bequest coefficient that measures the significance of the bequest demand. The same setting for bequest utility can be found in \cite{zeng2016optimal}, \cite{branger2019hedging}, and \cite{boyle2022annuity}.

We assume there is no trading constraint after the individual dies, so we can make fair comparisons between the cases with and without constraint when the individual is alive. Thus, the wealth process after the individual dies at time $t\in[0, T]$ is 
\begin{eqnarray}
	dW_s &=& [r(s)W_s + (\mu(s)-r(s))\theta_s] ds + \sigma(s)\theta_s dZ_s - c_s ds, s \in [t,T],\label{wealth_process_after_death}\\
	W_t &=& M_t.\notag
\end{eqnarray}
Furthermore, the value function of family investment after the individual dies follows
\begin{eqnarray}
	V_B(t,W_t) = \sup \limits_{\theta,c} E_t\left[ \int_t^T e^{-\widetilde{\delta}(s-t)}\frac{c^{1-\gamma}_t}{1-\gamma} ds + e^{-\widetilde{\delta}(T-t)}\frac{W^{1-\gamma}_T}{1-\gamma} \right],\label{VB_definiton}
\end{eqnarray}
where $E_t[\cdot]$ means the conditional expectation on the filtration $\mathcal{F}_t$. Subsequently, under the dynamic programming principle, we can derive the following lemma
\begin{lemma}\label{explicit_solution_0}
	The explicit solution of $V_B(t,M_t)$ is given by
	\begin{equation}
		V_B(t,W_t) = \frac{1}{1-\gamma} W_t^{1-\gamma}g(t)^{\gamma},\label{VB_explicit}
	\end{equation}
	where
	\begin{eqnarray}
		&&g(t) = \int_t^T e^{-\frac{\widetilde{\delta}}{\gamma}(s-t)}F_B(s-t,s)ds + e^{-\frac{\widetilde{\delta}}{\gamma}(T-t)}F_B(T-t,T),\label{gt}\\
		&&F_B(\tau,s) = \exp{\left\{-\int_0^{\tau} \frac{\gamma-1}{\gamma} r(s-u) du - \frac{1}{2}\frac{\gamma-1}{\gamma^2}\int_0^{\tau}\kappa^2_{0,s-u}du\right\}}. \notag
	\end{eqnarray}
\end{lemma}

\subsection{Algorithms}
We propose the following methods to make comparisons.
\begin{itemize}
	\item \textbf{Method 1: SAMS approach}\\
	Benchmark from \cite{bick2013solving}, assume $v_t=v(t)$ is affine in $t$, minimize the upper bound to obtain $v^*_t$, and then compute the lower bound under the $v^*_t$.
	\item \textbf{Method 2: Deterministic neural network approach}\\
	Restrict $v_t = v(t)$ as a neural network of time $t$, minimize the upper bound to obtain $v^*_t$, and then compute the lower bound under the $v^*_t$. 
	\item \textbf{Method 3: Deep pricing kernel approach}\\
	Restrict $v_t = v(t, Y_t)$ as a neural network of $(t, Y_t)$, minimize the upper bound to obtain $v^*_t$, deep regress on the conditional $p$-th moment of the pricing kernel, and then compute the lower bound under the $v^*_t$. 
\end{itemize}

Denote $(\alpha_v, \theta_v, c_v, I_v)$ as the general strategy and $((\alpha_v)^*, (\theta_v)^*, (c_v)^*, (I_v)^*)$ as the optimal strategy under the artificial market $\mathcal{M}_v$, then we derive the lower and upper bounds in each method.

\noindent
\subsubsection{Detailed explanations for Method 1 and Method 2}

When $v_t=v(t)$, i.e., $v_t$ is a function of $t$, we can derive the explicit solution of the upper bound for the primal problem \eqref{primal_problem}. 

\begin{proposition}\label{explicit_solution_1}
	Suppose that $v_t = v(t)$ and $t \in [T_R, T]$, then the upper bound of the primal problem \eqref{primal_problem} is given by
	\begin{equation}
		\widetilde{J}_R(t,W_{v,t};v) = \frac{1}{1-\gamma}\widetilde{F}_1(t,W_{v,t})^{1-\gamma}\widetilde{F}_2(t)^{\gamma},\label{upper_bound_JR}
	\end{equation}
	where the subscript ``R'' is short for retirement and 
	\begin{eqnarray*}
		&&\widetilde{F}_1(t,W_{v,t}) =  W_{v,t}, \, \widetilde{F}_2(t) = \int_t^T {_{s-t}p_{x+t}}e^{ - \frac{\widetilde{\delta}}{\gamma}(s-t)}(1+\lambda_{x+s}\varphi^{\frac{1}{\gamma}}g(s))F_3(s-t,s) ds \\
		&&+ {_{T-t}p_{x+t}}e^{-\frac{\widetilde{\delta}}{\gamma}(T-t)}F_3(T-t,T), \, F_2(\tau,s) = \exp{\left\{- \int_0^{\tau} r(s-u) + v_{0}(s-u) du\right\}},\\
		&&F_3(\tau,s) = \exp{\left\{-\int_0^{\tau} \frac{\gamma-1}{\gamma} (r(s-u)+v_{0}(s-u)) du - \frac{1}{2}\frac{\gamma-1}{\gamma^2}\int_0^{\tau} \kappa^2_{v,s-u} du\right\}},
	\end{eqnarray*}
	${_{s-t}p_{x+t}} ={_{s}p_{x}}/{_{t}p_{x}}$ and $g(s)$ follows \eqref{gt}. Moreover, the optimal strategies are
	\begin{eqnarray}
		&&(\theta_{v,t})^* = \text{Proj}_{[0,W_{v,t}]}\left\{-\frac{ \kappa_{v,t}}{\gamma \sigma(t)}\widetilde{F}_1(t,W_{v,t})\right\},\label{JR_strategies1}\\
		&&(c_{v,t})^* = \widetilde{F}_1(t,W_{v,t})/\widetilde{F}_2(t), ~~(M_{v,t})^* =[\widetilde{F}_1(t,W_{v,t})\varphi^{\frac{1}{\gamma}}g(t)]/\widetilde{F}_2(t),\label{JR_strategies2} 
	\end{eqnarray}
    where $\text{Proj}_{[0,W_{v,t}]}\{\cdot\}$ is a truncation projection to the interval $[0,W_{v,t}]$.
\end{proposition}

\begin{proposition}\label{explicit_solution_2}
	Suppose that $v_t = v(t)$ and $t \in [0, T_R]$, then the upper bound of the primal problem \eqref{primal_problem} is given by
	\begin{equation}
		\widetilde{J}(t,W_{v,t},Y_t;v) = \frac{1}{1-\gamma}\widetilde{F}_3(t,W_{v,t},Y_t)^{1-\gamma}\widetilde{F}_2(t)^{\gamma},\label{upper_bound_J}
	\end{equation}
	where
	\begin{eqnarray*}
		&&\widetilde{F}_3(t,W_{v,t},Y_t) =  W_{v,t} + Y_t\int_t^{T_R} {_{s-t}p_{x+t}}F_1(s-t,s) ds, \\
		&&\widetilde{F}_2(t) = \int_t^T {_{s-t}p_{x+t}}e^{-\frac{\widetilde{\delta}}{\gamma}(s-t)}(1+\lambda_{x+s}\varphi^{\frac{1}{\gamma}}g(s))F_3(s-t,s) ds + {_{T-t}p_{x+t}}e^{-\frac{\widetilde{\delta}}{\gamma}(T-t)}F_3(T-t,T),\\
		&&F_1(\tau,s) = \exp{\left\{\mu_Y\tau + \int_0^{\tau}-[r(s-u)+v_{0}(s-u)] + \kappa_{v,s-u} \sigma_Y du\right\}}, \\
        &&F_2(\tau,s) = \exp{\left\{ - \int_0^{\tau} r(s-u) + v_{0}(s-u) du\right\}},\\
		&&F_3(\tau,s) = \exp{\left\{-\int_0^{\tau} \frac{\gamma-1}{\gamma} (r(s-u)+v_{0}(s-u)) du - \frac{1}{2}\frac{\gamma-1}{\gamma^2}\int_0^{\tau} \kappa^2_{v,s-u} du\right\}},
	\end{eqnarray*}
	and $g(s)$ follows \eqref{gt}. Moreover, the optimal strategies are
	\begin{eqnarray}
		(\theta_{v,t})^* &=& \text{Proj}_{[0,W_{v,t}]}\left\{-\frac{\kappa_{v,t}}{\gamma \sigma(t)}\widetilde{F}_3(t,W_{v,t},Y_t) - \frac{\sigma_Y}{\sigma(t)}Y_t\int_t^{T_R} {_{s-t}p_{x+t}}F_1(s-t,s)ds\right\}\label{J_strategy_theta}\\
		(c_{v,t})^* &=&  \widetilde{F}_3(t,W_{v,t},Y_t)/\widetilde{F}_2(t),~~(M_{v,t})^* = [\widetilde{F}_3(t,W_{v,t},Y_t)\varphi^{\frac{1}{\gamma}}g(t)]/\widetilde{F}_2(t),\label{J_strategy_c_M}
	\end{eqnarray}
    where $\text{Proj}_{[0,W_{v,t}]}\{\cdot\}$ is a truncation projection to the interval $[0,W_{v,t}]$.
\end{proposition}

For Method 1, follow \cite{bick2013solving}, we separate $v_t$ at the retirement time $T_R$, i.e.
\begin{eqnarray*}
	v_t = v(t) =\left\{
	\begin{array}{ll}
		v^w(t) = (v^w_{0}(t),v^w_{-}(t)) = ((a_1 + a_2 t)_+, (a_3 + a_4 t)_+), ~~ 0\leq t < T_R,\\
		v^R(t) = (v^R_{0}(t),v^R_{-}(t)) = ((a_5 + a_6 t)_+, (a_7 + a_8 t)_+), ~~T_R\leq t\leq T,
	\end{array}
	\right.
\end{eqnarray*}
where superscript $w$ is short for ``working'', superscript $R$ is short for ``retirement'', and $(\cdot)_+$ is the positive part of a function.

For Method 2, we use one neural network ($v_{0}$, $v_{-}$) with state variable time $t$ to describe $v_t$. We let the neural network $\mathcal{NN}^{\phi_1}_1$ learn the retirement time $T_R$ by itself and therefore do not separate $v_t$ at $T_R$. 
\begin{eqnarray*}
	v_t = v(t) = (v_{0}(t), v_{-}(t))=\mathcal{NN}^{\phi_1}_1(t), ~~ 0\leq t \leq T,
\end{eqnarray*}
where $\phi_1$ are the weights and biases for the neural network. The detailed structure of $\mathcal{NN}^{\phi_1}_1$ are discussed in Section \ref{section_comparsion_three}.

After minimizing the explicit upper bound $\widetilde{J}(0,W_{v,0},Y_0;v)$ in Proposition \ref{explicit_solution_2}, we obtain the optimal $v^*_t$ for $0\leq t \leq T$. Then, we can define the candidate value function $\overline{J}(t,\overline{W}_{v^*,t},Y_t;v^*)$ as
\begin{eqnarray*}
	&&\overline{J}(t,\overline{W}_{v^*,t},Y_t;v^*) = E_t\left[ \int_t^T {_{s-t}p_{x+t}}e^{-\widetilde{\delta} (s-t)}\frac{((c_{v^*,s})^*)^{1-\gamma}}{1-\gamma} ds\right.\\
	&& \left.  + \int_t^T {_{s-t}p_{x+t}}\lambda_{x+s} e^{ -\widetilde{\delta} (s-t)}\frac{((M_{v^*,s})^*)^{1-\gamma}}{1-\gamma}\varphi [g(s)]^{\gamma} ds + {_{T-t}p_{x+t}} e^{ -\widetilde{\delta} (T-t)}\frac{((\overline{W}_{v^*,T})^*)^{1-\gamma}}{1-\gamma} \right], 
\end{eqnarray*}
where the candidate wealth process $\overline{W}_{v^*,t}$ is driven by the optimal strategies \eqref{JR_strategies1}, \eqref{JR_strategies2}, \eqref{J_strategy_theta}, and \eqref{J_strategy_c_M} 
\begin{eqnarray}
	d\overline{W}_{v^*,t} &=& \{ [r(t)+\lambda_{x+t}]\overline{W}_{v^*,t} + (\theta_{v^*,t})^* [\mu(t)-r(t)] \}dt + (\theta_{v^*,t})^* \sigma(t) dZ_t\notag\\
	&& - [(c_{v^*,t})^* + \lambda_{x+t} (M_{v^*,t})^* - Y_t] dt, \label{lb_wealth}\\
	\overline{W}_{v^*,0} &=& \overline{w}_0.\notag
\end{eqnarray}
The candidate value function $\overline{J}(t,\overline{W}_{v^*,t},Y_t;v^*)$ provides a lower bound for the primal Problem \eqref{primal_problem} because $\theta_{v^*,t}$ satisfies the portfolio constraint set \eqref{numerical_A} and $C_t \equiv 0$ is a sub-strategy for free disposal in \eqref{numerical_wealth}. 
From all things above, we obtain the tight lower and upper bounds for the primal Problem \eqref{primal_problem} 
\begin{equation*}
	\overline{J}(0,\overline{W}_{v^*,0},Y_0;v^*) \leq J(c,M,W_T) \leq \widetilde{J}(0,W_{v^*,0},Y_0;v^*).
\end{equation*} 

\subsubsection{Method 3: Deep pricing kernel approach}
When $v_t=v(t, Y_t)$, we can only obtain the semi-explicit solution for upper bounds.

\begin{proposition}\label{D_explicit_solution_1}
	Suppose that $v_t = v(t, Y_t)$ and $t \in [T_R, T]$, then the upper bound of the primal problem \eqref{primal_problem} is given by
	\begin{equation}
		\widetilde{J}^D_R(t,W_{v,t};v) = \frac{1}{1-\gamma}\widetilde{F}^D_1(t,W_{v,t})^{1-\gamma}\widetilde{F}^D_2(t)^{\gamma}, \label{Method3_value_function}
	\end{equation}
	where the superscript $D$ is short for ``Deep pricing kernel approach'', the subscript ``R'' is short for retirement, and
	\begin{eqnarray*}
		&&\widetilde{F}^D_1(t,W_{v,t}) =  W_{v,t}, \\
		&&\widetilde{F}^D_2(t) = \int_t^{T} {_{s-t}p_{x+t}} e^{- \frac{\widetilde{\delta}}{\gamma}(s-t)}[1+\lambda_{x+s}\varphi^{\frac{1}{\gamma}}g(s)]E_t\left[\left(\frac{\pi_{v,s}}{\pi_{v,t}}\right)^{-\frac{1-\gamma}{\gamma}}\right] ds \\
		&&+ {_{T-t}p_{x+t}} e^{-\frac{\widetilde{\delta}}{\gamma}(T-t)}E_t\left[\left(\frac{\pi_{v,T}}{\pi_{v,t}}\right)^{-\frac{1-\gamma}{\gamma}}\right],
	\end{eqnarray*}
	and $g(s)$ follows \eqref{gt}. Moreover, the optimal strategies are
	\begin{eqnarray}
		(\theta^D_{v,t})^* &=& \text{Proj}_{[0,W_{v,t}]}\left\{-\frac{\kappa_{v,t}}{\gamma \sigma(t)}\widetilde{F}^D_1(t,W_{v,t}) \right\},\label{D_J_strategy_theta}\\
		(c^D_{v,t})^* &=&  \widetilde{F}^D_1(t,W_{v,t})/\widetilde{F}^D_2(t),~~(M^D_{v,t})^* = [\widetilde{F}^D_1(t,W_{v,t})\varphi^{\frac{1}{\gamma}}g(t)]/\widetilde{F}^D_2(t),\label{D_J_strategy_c_M}
	\end{eqnarray}
    where $\text{Proj}_{[0,W_{v,t}]}\{\cdot\}$ is a truncation projection to the interval $[0,W_{v,t}]$.
\end{proposition}

\begin{proposition}\label{D_explicit_solution_2}
	Suppose that $v_t = v(t, Y_t)$ and $t \in [0, T_R]$, then the upper bound of the primal problem \eqref{primal_problem} is given by
	\begin{equation*}
		\widetilde{J}^D(t,W_{v,t},Y_t;v) = \frac{1}{1-\gamma}\widetilde{F}^D_3(t,W_{v,t},Y_t)^{1-\gamma}\widetilde{F}^D_4(t, Y_t)^{\gamma},
	\end{equation*}
	where the superscript $D$ is short for ``Deep pricing kernel approach'' and
	\begin{eqnarray*}
		&&\widetilde{F}^D_3(t,W_{v,t},Y_t) =  W_{v,t} + \int_t^T {_{s-t}p_{x+t}} Y_tE_t\left[\frac{\pi_{v^*,s}Y_s}{\pi_{v^*,t}Y_t}\right] ds, \\
		&&\widetilde{F}^D_4(t,Y_t) = \int_t^{T} {_{s-t}p_{x+t}} e^{- \frac{\widetilde{\delta}}{\gamma}(s-t)}[1+\lambda_{x+s}\varphi^{\frac{1}{\gamma}}g(s)]E_t\left[\left(\frac{\pi_{v,s}}{\pi_{v,t}}\right)^{-\frac{1-\gamma}{\gamma}}\right] ds \\
		&&+ {_{T-t}p_{x+t}} e^{-\frac{\widetilde{\delta}}{\gamma}(T-t)}E_t\left[\left(\frac{\pi_{v,T}}{\pi_{v,t}}\right)^{-\frac{1-\gamma}{\gamma}}\right],
	\end{eqnarray*}
	and $g(s)$ follows \eqref{gt}. 
    Moreover, the optimal strategies are
	\begin{eqnarray}
		(\theta^D_{v,t})^* &=& \text{Proj}_{[0,W_{v,t}]}\left\{-\frac{\kappa_{v,t}}{\gamma \sigma(t)}\widetilde{F}^D_3(t,W_{v,t},Y_t) \right. \notag\\
		&&\left. - \frac{\sigma_Y}{\sigma(t)}Y_t\left[ \frac{\partial \widetilde{F}^D_3}{\partial Y}(t,W_{v,t},Y_t)-\frac{\widetilde{F}^D_3(t,W_{v,t},Y_t)}{\widetilde{F}^D_4(t,Y_t)} \frac{\partial \widetilde{F}^D_4}{\partial Y}(t,Y_t)\right]\right\},\label{D_J_strategy_theta2}\\
		(c^D_{v,t})^* &=&  \widetilde{F}^D_3(t,W_{v,t},Y_t)/\widetilde{F}^D_4(t,Y_t),~~(M^D_{v,t})^* = [\widetilde{F}_3(t,W_{v,t},Y_t)\varphi^{\frac{1}{\gamma}}g(t)]/\widetilde{F}^D_4(t),\label{D_J_strategy_c_M2}
	\end{eqnarray}
    where $\text{Proj}_{[0,W_{v,t}]}\{\cdot\}$ is a truncation projection to the interval $[0,W_{v,t}]$.
\end{proposition}

We let neural network $\mathcal{NN}^{\phi_2}_2(t,Y_t)$ to learn the optimal $v_t$ minimizing the explicit upper bounds $\widetilde{J}^D(0,W_{v,0},Y_0;v)$ in Proposition \ref{D_explicit_solution_2}. In the optimal case, $v^*_t = v^*(t, Y_t)=\mathcal{NN}^{\phi^*_2}_2(t, Y_t)$ for $0\leq t \leq T$ where $\phi^*_2$ are the optimal weights and biases to minimize the upper bound. Next, we can use deep regression to approximate two unknown conditional expectations under $v^*_t$
    \begin{eqnarray}
        \text{Expection 1:} && E_t\left[\frac{\pi_{v^*,s}Y_s}{\pi_{v^*,t}Y_t}\right] \approx \exp(\mathcal{NN}^{\phi^*_3}_3(Y_t, s-t)), ~0\leq t<s<T, \label{conditional_expectation_1}\\
        \text{Expection 2:} && E_t\left[\left(\frac{\pi_{v^*,s}}{\pi_{v^*,t}}\right)^{-\frac{1-\gamma}{\gamma}}\right] \approx \exp(\mathcal{NN}^{\phi^*_4}_4(Y_t, s-t)), ~0\leq t<s \leq T. \label{conditional_expectation_2}
    \end{eqnarray}
In the expressions above, $\phi^*_3$ is the optimal weight and bias of $\mathcal{NN}^{\phi_3}_3$ minimizing the regression loss
\begin{equation*}
    \min \limits_{\phi_1} \sum \limits_{i=0}^{N}\sum \limits_{j=i}^{N} \left[\frac{\pi_{v^*,t_j}Y_{t_j}}{\pi_{v^*,t_i}Y_{t_i}}- \exp(\mathcal{NN}^{\phi_3}_3(Y_{t_i}, t_j-t_i))\right]^2,
\end{equation*}
where $t_i = i\Delta t, i=0, 1, ..., N$ and $\Delta t=T/N$. Moreover, $\phi^*_4$ is the optimal weight and bias of $\mathcal{NN}^{\phi_4}_4$ minimizing the regression loss
\begin{equation*}
    \min \limits_{\phi_2} \sum \limits_{i=0}^{N}\sum \limits_{j=i}^{N} \left[\left(\frac{\pi_{v^*,t_j}}{\pi_{v^*,t_i}}\right)^{-\frac{1-\gamma}{\gamma}} - \exp(\mathcal{NN}^{\phi_4}_4(Y_{t_i}, t_j-t_i))\right]^2.
\end{equation*}
For the derivatives $\partial \widetilde{F}^D_3/\partial Y$ and $\partial \widetilde{F}^D_4/\partial Y$ in the candidate strategy \eqref{D_J_strategy_theta2}, one can compute them via the finite difference of the neural networks $\exp(\mathcal{NN}^{\phi_3}_3)$ and $\exp(\mathcal{NN}^{\phi_4}_4)$.

Finally, similar to Methods 1 and 2, we plug the candidate strategy into the Monte Carlo simulation of the lower-bound wealth process \eqref{lb_wealth}, and the primal value function is bounded between the lower and upper bounds.
\begin{equation*}
	\overline{J}^D(0,\overline{W}_{v^*,0},Y_0;v^*) \leq J(c,M,W_T) \leq \widetilde{J}^D(0,W_{v^*,0},Y_0;v^*),
\end{equation*} 
where the superscript $D$ is short for ``Deep pricing kernel approach''. 

\subsection{Comparisons among three approaches}	\label{section_comparsion_three}
    Inspired by \cite{huang2008portfolio}, we set the base model parameters as 
\begin{eqnarray*}
	&&\widetilde{\delta} = 0.02,~~ \mu_Y = 0.01,~~ \sigma_Y = 0.05,~~ W_0 = 200.00,~~ Y_0 = 50.00, \notag\\
    &&\mu(t) = 0.07,~~ r(t) = 0.02,~~ \sigma(t) = 0.2,\label{base_parameters}
\end{eqnarray*} 
     where the units for $W_0$ and $Y_0$ are thousand US dollars.
     
    Table \ref{example1_table} shows the lower and upper bounds for each method. Methods 1 and 2 share a similar explicit upper bound. We use the trapezoidal rule to compute the double integral in this explicit upper bound, and the number of time intervals is set to 200. To make fair comparisons, we use gradient descent to find $v^*$ for the upper bounds across all three methods. The learning rates are initially set as $10^{-3}$, decay by a factor of 0.96 every 100 iterations, for a total of 500 iterations. For the lower bounds, we set all three methods to use the same number of paths, 100,000, and the same number of time intervals, 200. 
	
	For Method 2, we set the structure of the neural network $v_t=v(t)$ as ``1-30-30-30-2'', which means one node (time $t$) in the input layer, 30 nodes each hidden layer for three layers, and two nodes ($v_0$ and $v_-$) in the output layer (see Figure \ref{nn_method2_v}).

\begin{figure}[htbp]
	\centering
	\begin{tikzpicture}[x=1.5cm, y=1.3cm, >=stealth]
		
		\tikzset{%
			every neuron/.style={
				circle,
				draw,
				minimum size=1.0cm
			},
			neuron missing/.style={
				draw=none, 
				scale=3.5,
				text height=0.222cm,
				execute at begin node=\color{black}$\vdots$
			},
		}
		\foreach \m [count=\y] in {1}
		\node [every neuron/.try, neuron \m/.try, fill = green!95!black!10] (input-\m) at (0,4.2-2.4*\y){};
		
		\foreach \m [count=\y] in {1,missing,missing,2}
		\node [every neuron/.try, neuron \m/.try] (hidden1-\m) at (2,5.0-\y*1.3) {};
		
		\node [every neuron/.try, neuron 1/.try, fill = blue!95!black!10] (hidden1-1) at (2,5.0-1*1.3) {};
		\node [every neuron/.try, neuron 2/.try, fill = blue!95!black!10] (hidden1-2) at (2,5.0-4*1.3) {};
		
		\foreach \m [count=\y] in {1,missing,missing,2}
		\node [every neuron/.try, neuron \m/.try ] (hidden2-\m) at (4,5.0-\y*1.3) {};
		
		\node [every neuron/.try, neuron 1/.try, fill = blue!95!black!10] (hidden2-1) at (4,5.0-1*1.3) {};
		\node [every neuron/.try, neuron 2/.try, fill = blue!95!black!10] (hidden2-2) at (4,5.0-4*1.3) {};
		
		\foreach \m [count=\y] in {1,missing,missing,2}
		\node [every neuron/.try, neuron \m/.try ] (hidden3-\m) at (6,5.0-\y*1.3) {};
		
		\node [every neuron/.try, neuron 1/.try, fill = blue!95!black!10] (hidden3-1) at (6,5.0-1*1.3) {};
		\node [every neuron/.try, neuron 2/.try, fill = blue!95!black!10] (hidden3-2) at (6,5.0-4*1.3) {};
		
		\foreach \m [count=\y] in {1,2}
		\node [every neuron/.try, neuron \m/.try, fill = red!95!black!10] (output-\m) at (8,5.0-\y*2.2) {};
		
		\node at (input-1) {$t$};
		
		\foreach \l [count=\i] in {1}
		\node at (hidden1-\i) {$H^{(1)}_{\l}$};
		\node at (hidden1-2) {$H^{(1)}_{30}$};
		
		\foreach \l [count=\i] in {1}
		\node at (hidden2-\i) {$H^{(2)}_{\l}$};
		\node at (hidden2-2) {$H^{(2)}_{30}$};
		
		\foreach \l [count=\i] in {1}
		\node at (hidden3-\i) {$H^{(3)}_{\l}$};
		\node at (hidden3-2) {$H^{(3)}_{30}$};
		
		\node at (output-1) {$v_0$};
		\node at (output-2) {$v_-$};
		
		\foreach \i in {1}
		\foreach \j in {1,...,2}
		\draw [->] (input-\i) -- (hidden1-\j);
		
		\foreach \i in {1,...,2}
		\foreach \j in {1,...,2}
		\draw [->] (hidden1-\i) -- (hidden2-\j);
		
		\foreach \i in {1,...,2}
		\foreach \j in {1,...,2}
		\draw [->] (hidden2-\i) -- (hidden3-\j);

		\foreach \i in {1,...,2}
		\foreach \j in {1,2}
		\draw [->] (hidden3-\i) -- (output-\j);
		
		\foreach \l [count=\x from 0] in {Input, Hidden 1, Hidden 2, Hidden 3, Ouput}
		\node [align=center, above] at (\x*2,4.2) {\l};
		
	\end{tikzpicture}
	\caption{Neural network for $v_t=v(t)$ with a ``$1-30-30-30-2$'' structure for Method 2. 
	}\label{nn_method2_v}
\end{figure}
    For Method 3, we have three neural networks: $\mathcal{NN}^{\phi_2}_2(t,Y_t)$ for $v_t=v(t,Y_t)$, $\mathcal{NN}^{\phi_3}_3(Y_t,s-t)$ for the first conditional expectation \eqref{conditional_expectation_1}, and $\mathcal{NN}^{\phi_4}_4(Y_t,s-t)$ for the second conditional expectation \eqref{conditional_expectation_2}. We set their structure as ``2-30-30-30-2'', ``2-30-1'', ``2-30-30-1'', respectively. The existing path of the pricing kernel $\pi_{v^*,t}$ from the upper-bound simulation can be reused to perform deep regression to estimate the conditional expectations  \eqref{conditional_expectation_1} and \eqref{conditional_expectation_2} for the lower bound.

	In Table \ref{example1_table}, we design three quantities to compare the three methods. The first is the ``duality gap''. It is defined as the absolute difference between the lower and upper bounds. The second is the ``relative gap''. It is defined as the absolute ratio of the ``duality gap'' over the lower bound. The third is ``welfare loss''. Following \cite{bick2013solving}, we define the ``welfare loss'' as the upper bound of the fraction of wealth that an individual would like to through away to get access to an optimal strategy. More specifically, under the market $\mathcal{M}_{v^*}$, it is the proportion $L$ such that the following equation holds for the lower and upper bounds of the value function.  
	\begin{equation*}
		\overline{J}(0,\overline{W}_{v^*,0},Y_0;v^*) = \widetilde{J}(0,W_{v^*,0}[1-L],Y_0[1-L];v^*).
	\end{equation*} 
	From Proposition \ref{explicit_solution_2}, Proposition \ref{D_explicit_solution_2}, and \eqref{conditional_expectation_1}, we have 
	\begin{equation*}
		\widetilde{J}(0,W_{v^*,0}[1-L],Y_0[1-L];v^*) = (1-L)^{1-\gamma}\widetilde{J}(0,W_{v^*,0},Y_0;v^*).
	\end{equation*}
	Therefore, the upper bound of welfare loss is
	\begin{equation*}\label{welfare_loss}
		L = 1 - \left(\frac{\overline{J}(0,\overline{W}_{v^*,0},Y_0;v^*)}{\widetilde{J}(0,W_{v^*,0},Y_0;v^*)}\right)^{\frac{1}{1-\gamma}}.
	\end{equation*}
	
	From Table \ref{example1_table}, we see ``Deep pricing kernel method'' (Method 3) consistently outperforms Method 1 (SAMS approach) and Method 2 (deterministic NN approach)  in every aspect: smaller upper bound, bigger lower bound, smaller duality gap, smaller relative gap, and smaller welfare loss. In particular, for key error metrics such as the duality gap, relative gap, and welfare loss, Method 3 improves Method 1's results at least one order of magnitude. Specifically, for a risk-averse individual with $\gamma=1.5$, Method 3 reduces the duality gap from 0.182199 to 0.005304, the relative gap from 1.974759\% to 0.058574\%, and the welfare loss from 0.039105 to 0.001171. For a more risk-seeking individual with $\gamma=0.8$, Method 2 lowers the duality gap from 2.755830 to 1.148930, and Method 3 further decreases it to 0.136330. For the relative gap, Method 2 narrows it from 0.708001\% to 0.295164\%, and Method 3 further shrinks it to 0.034944\%. Lastly, for welfare loss, Method 2 cuts it from 0.034660 to 0.014628, and Method 3 further improves it to 0.001745.

	\begin{table}[htbp]
		\caption{\textbf{Lower and upper bounds for different methods}} 
		\smallskip
		\centering 
        \setlength{\tabcolsep}{5mm}
		\begin{tabular}{l r r r}\label{example1_table} \vspace{-0.3in}\\
                \multicolumn{4}{c}{$\gamma=1.5$}\\
                \hline
			           &  Method 1    & Method 2  & Method 3 \\
			$v_t$'s form  &  $a+b*t$    &$\mathcal{NN}(t)$      & $\mathcal{NN}(t, Y_t)$   \\
			\hline
			Upper bound  &  -9.044193	&-9.044817	&-9.049943 \\
			Lower bound  &  -9.226392	&-9.225890	&-9.055247\\
			Duality gap  &  0.182199	&0.181073	&0.005304\\
			Relative gap (\%) &  1.974759	&1.962662	&0.058574\\
			Welfare loss &    0.039105	&0.038868	&0.001171\\
			Time elapsed (secs)&  187.91	&205.50	&1583.13\\
			\hline
                \multicolumn{4}{c}{$\gamma=0.8$}\\
                \hline
			           &  Method 1    & Method 2  & Method 3 \\
			$v_t$'s form  &  $a+b*t$    &$\mathcal{NN}(t)$      & $\mathcal{NN}(t, Y_t)$   \\
			\hline
			Upper bound  &  391.997070	&390.400760	&390.269870\\
			Lower bound  &  389.241240	&389.251830	&390.133540\\
			Duality gap  &  2.755830	&1.148930	&0.136330\\
			Relative gap (\%) & 0.708001 &0.295164	&0.034944 \\
			Welfare loss & 0.034660	 &0.014628	&0.001745 \\
			Time elapsed (secs)& 191.51	& 204.64 &1565.19 \\
                \hline
		\end{tabular}
	\end{table}

\begin{figure}[htbp]
    \centering
    \subfigure[Upper bounds under $\gamma=1.5$]
    {
        \includegraphics[width=3.0in,height=2.5in]{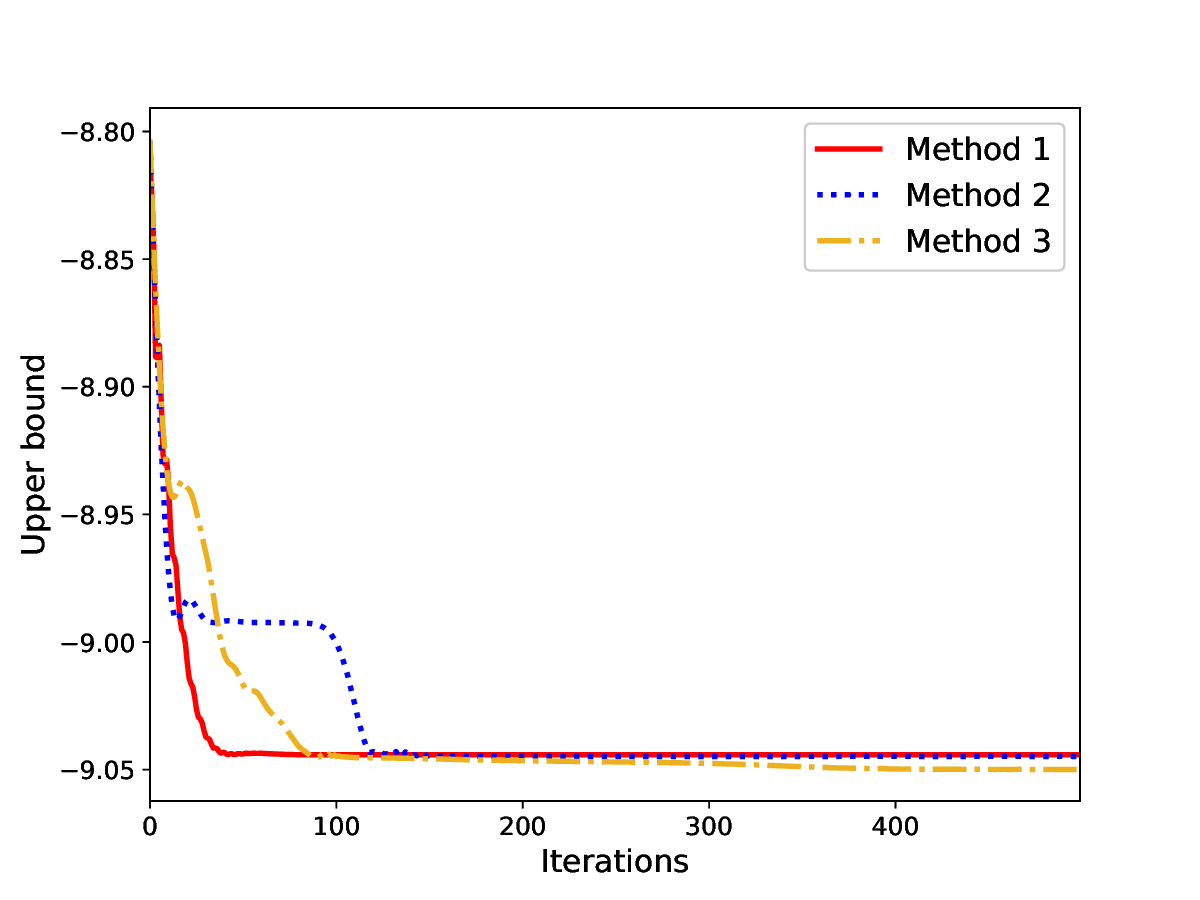}
        \label{fig_loss_curves:figure1}
    }
    \hspace{-0.45in}
    \subfigure[Regression under $\gamma=1.5$]
    {
	\includegraphics[width=3.0in,height=2.5in]{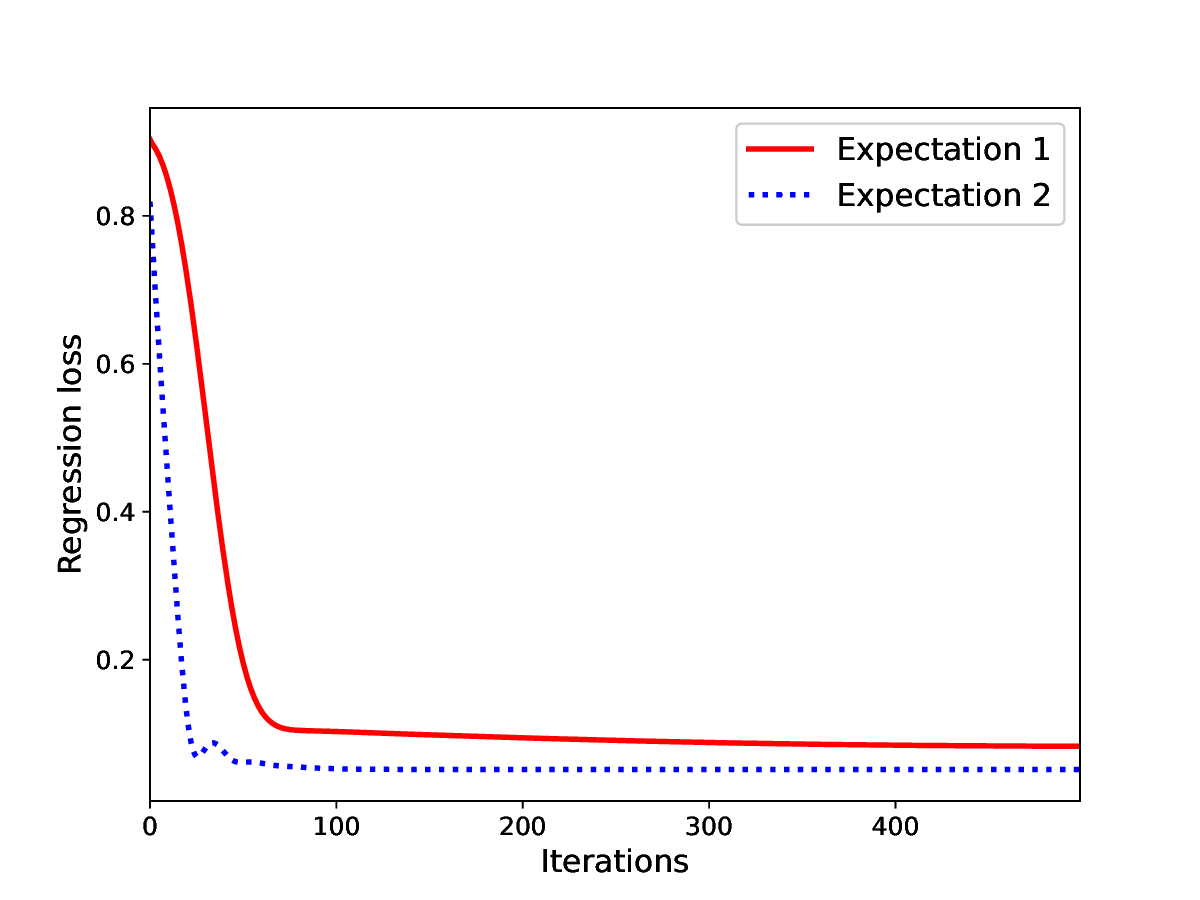}
        \label{fig_loss_curves:figure2}
    }
    \subfigure[Upper bounds under $\gamma=0.8$]
    {
        \includegraphics[width=3.0in,height=2.5in]{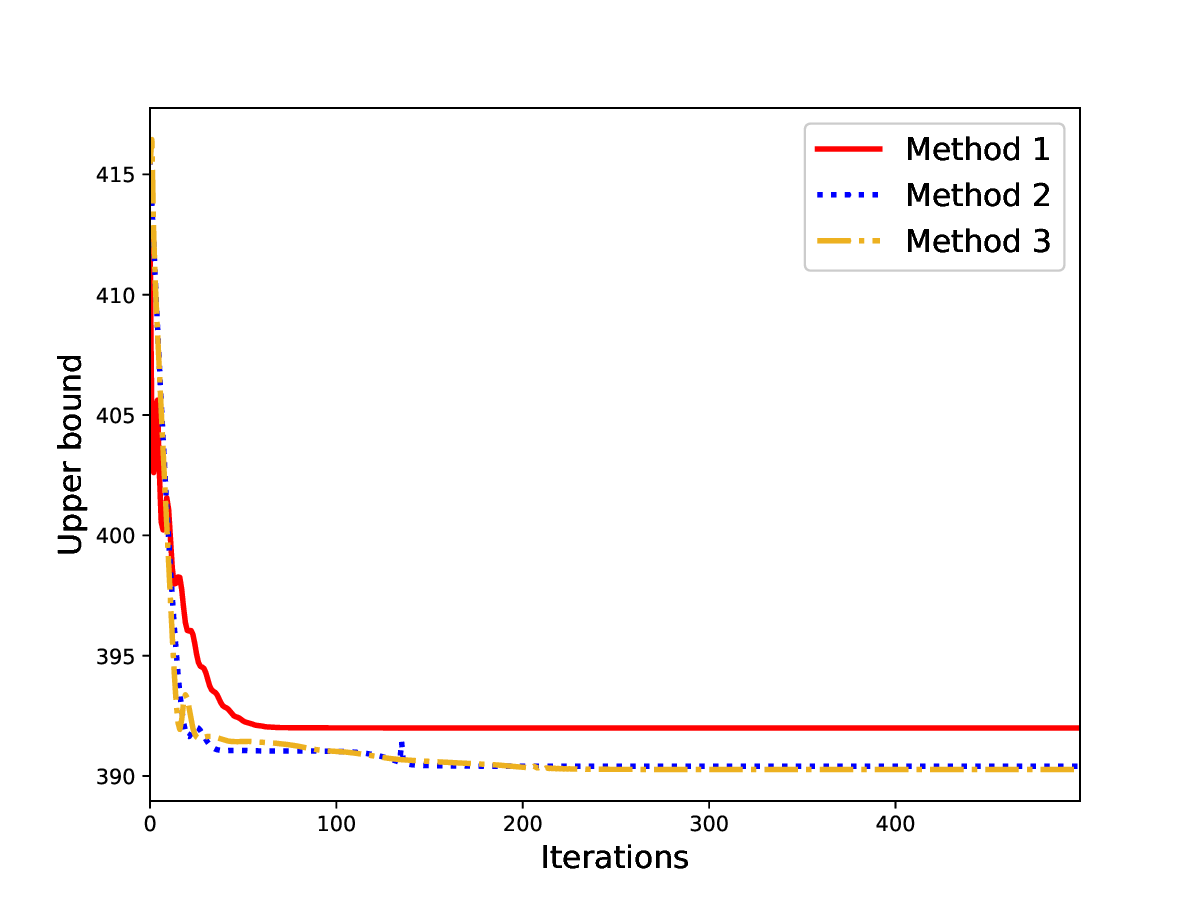}
        \label{fig_loss_curves:figure3}
    }
    \hspace{-0.45in}
    \subfigure[Regression under $\gamma=0.8$]
    {
	\includegraphics[width=3.0in,height=2.5in]{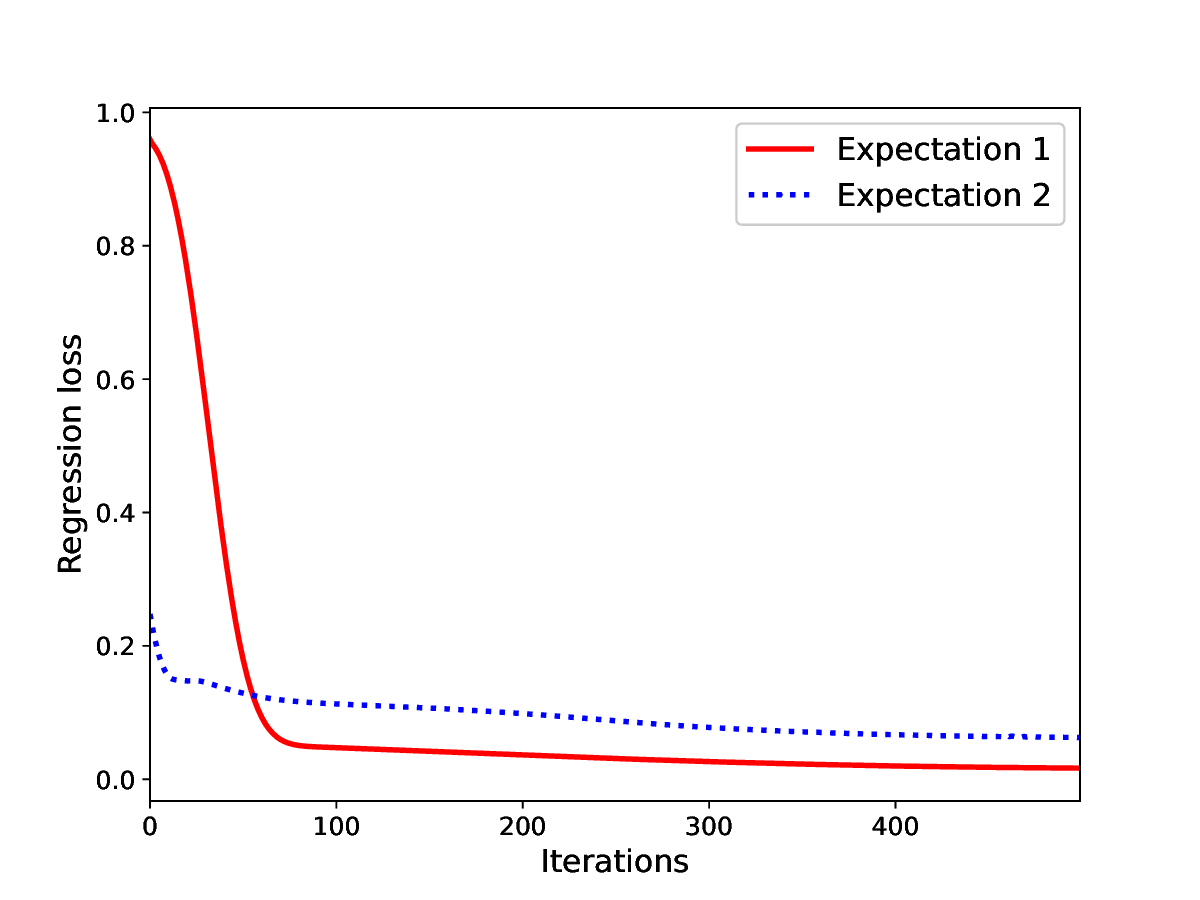}
        \label{fig_loss_curves:figure4}
    }
    \caption{Loss curves for training the upper bounds and deep regression. }
    \label{fig_loss_curves:figures}
\end{figure}

    Figure \ref{fig_loss_curves:figures} shows the loss curves for training the upper bounds and deep regression. In Figure \ref{fig_loss_curves:figure1} with $\gamma=1.5$, we see that the upper bound of Method 1 decreases faster than Methods 2 and 3 at the beginning. However, Methods 2 and 3 have smaller upper bounds in the end, which points out the limitation of Method 1. For a more risk-seeking case in \ref{fig_loss_curves:figure3} with $\gamma=0.8$, we observe that Methods 2 and 3 both obtain significantly smaller upper bounds, and Method 3 always has the smallest upper bound. Furthermore, Figures \ref{fig_loss_curves:figure2} and \ref{fig_loss_curves:figure4} both illustrate the stability of using the deep regression to estimate the conditional expectations of the pricing kernel.

\begin{figure}[htbp]
    \centering
    \subfigure[Bond compensated return]
    {
        \includegraphics[width=3.0in,height=2.5in]{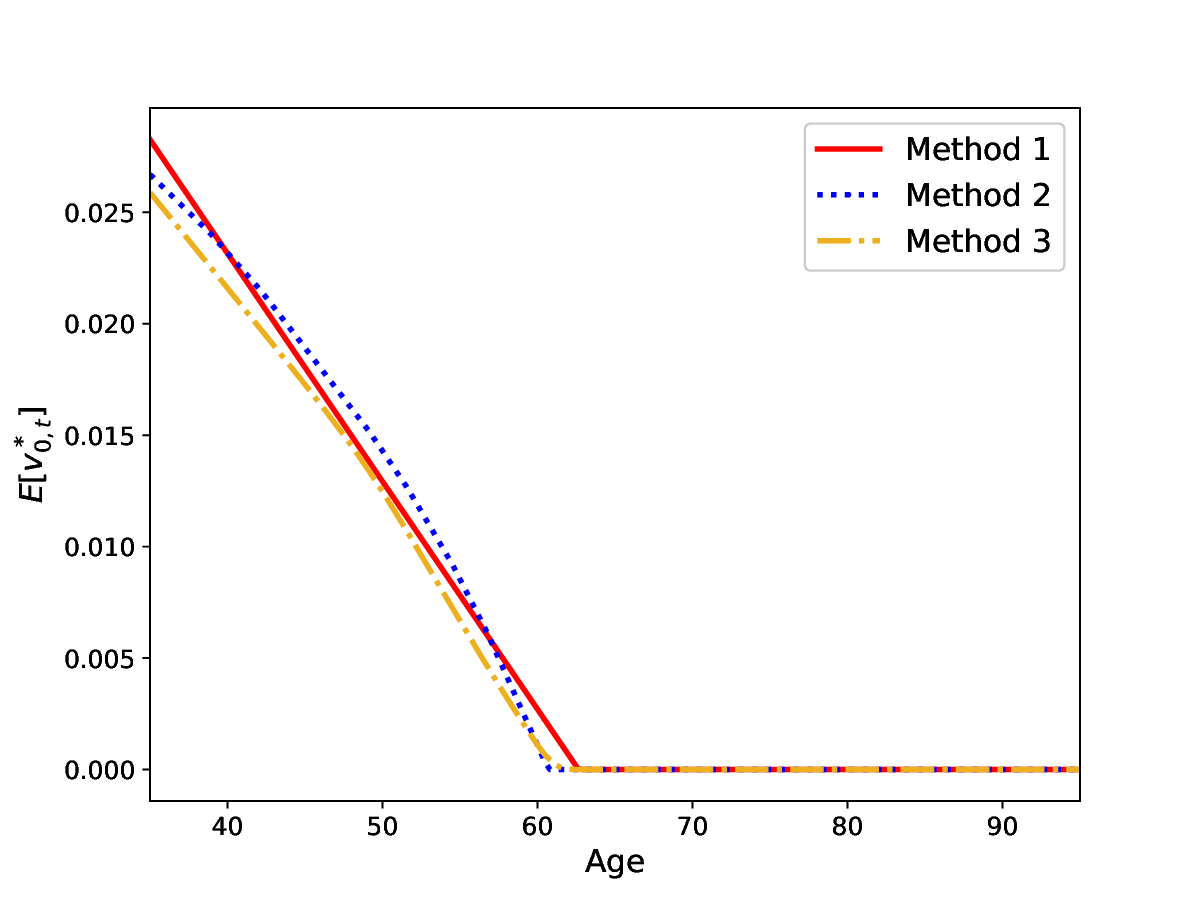}
        \label{fig_investment_gamma1point5:figure1}
    }
    \hspace{-0.45in}
    \subfigure[Stock compensated return]
    {
	\includegraphics[width=3.0in,height=2.5in]{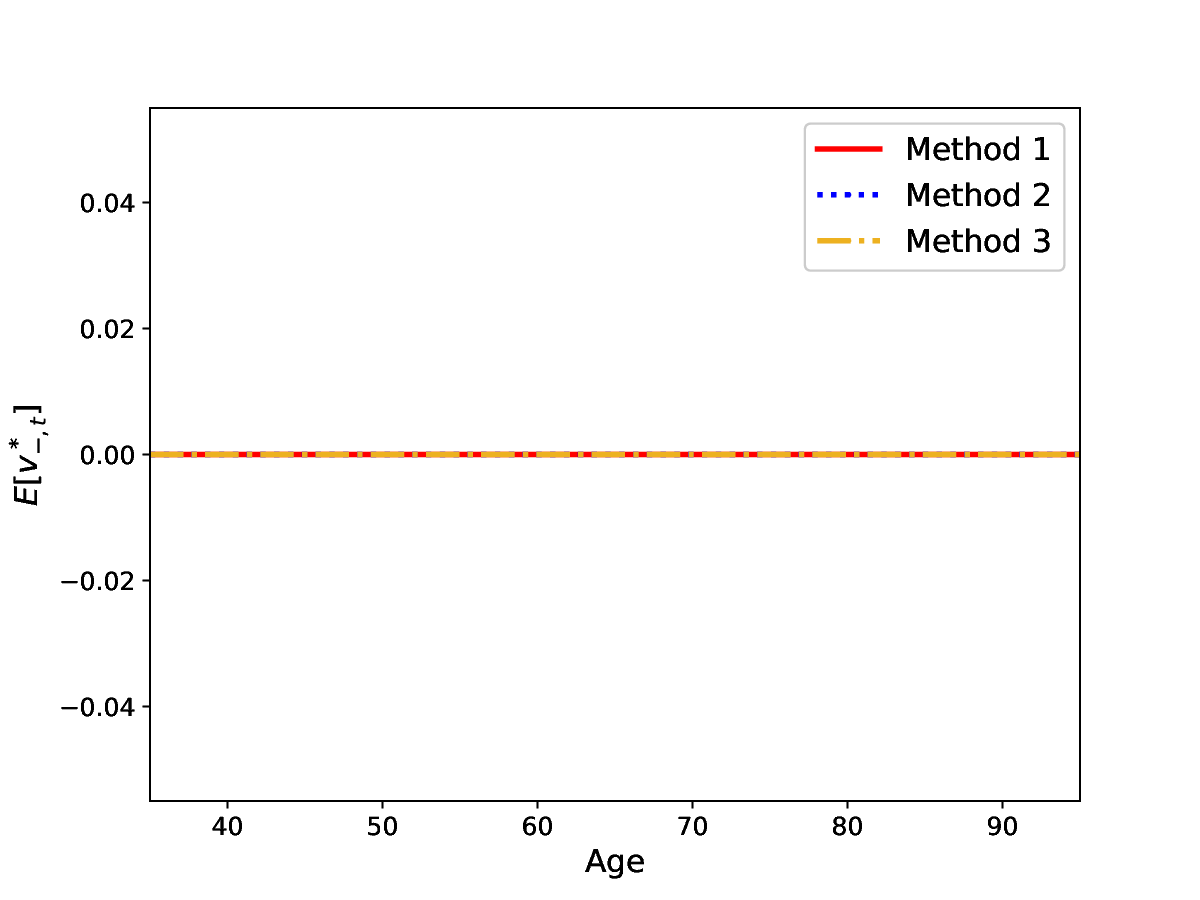}
        \label{fig_investment_gamma1point5:figure2}
    }
    \\
    \subfigure[Bond allocation amount]
    {
        \includegraphics[width=3.0in,height=2.5in]{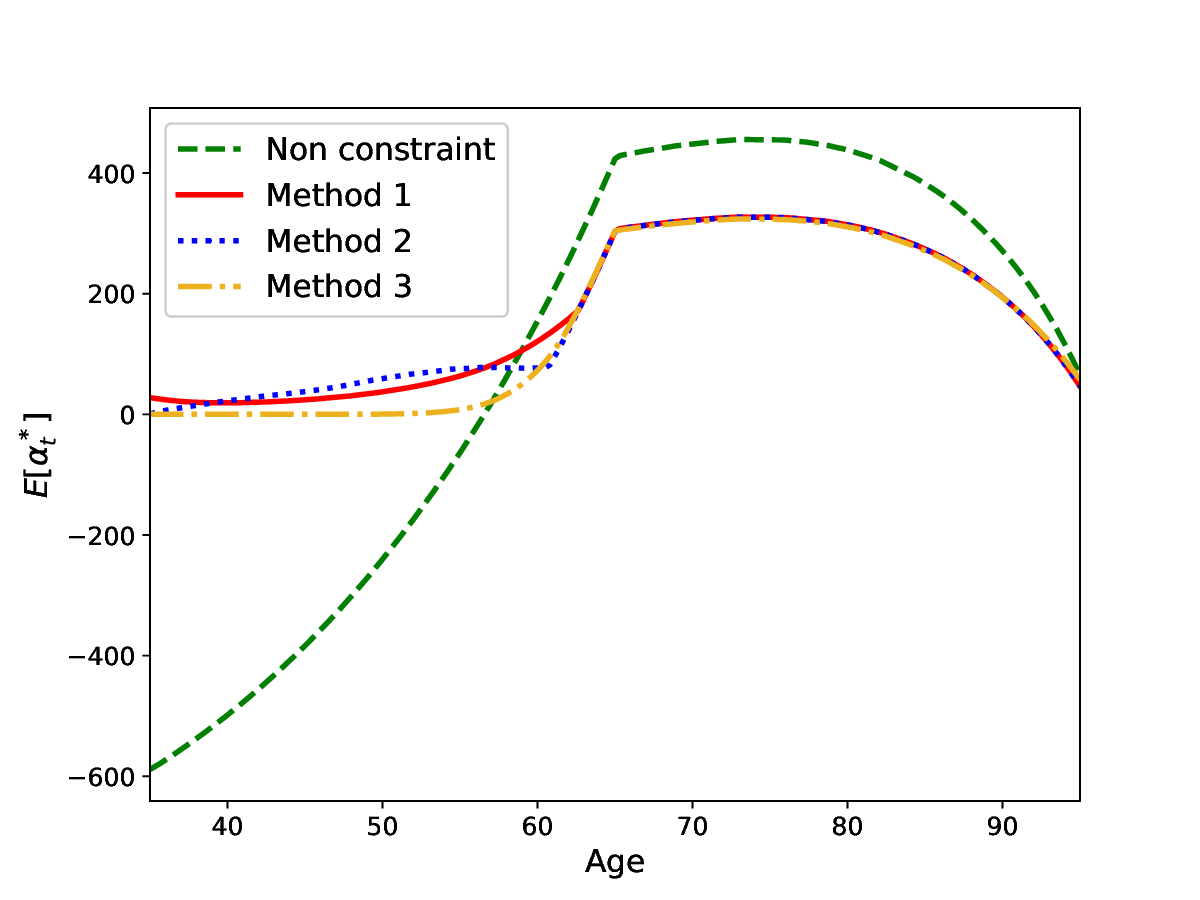}
        \label{fig_investment_gamma1point5:figure3}
    }
    \hspace{-0.45in}
    \subfigure[Stock allocation amount]
    {
	\includegraphics[width=3.0in,height=2.5in]{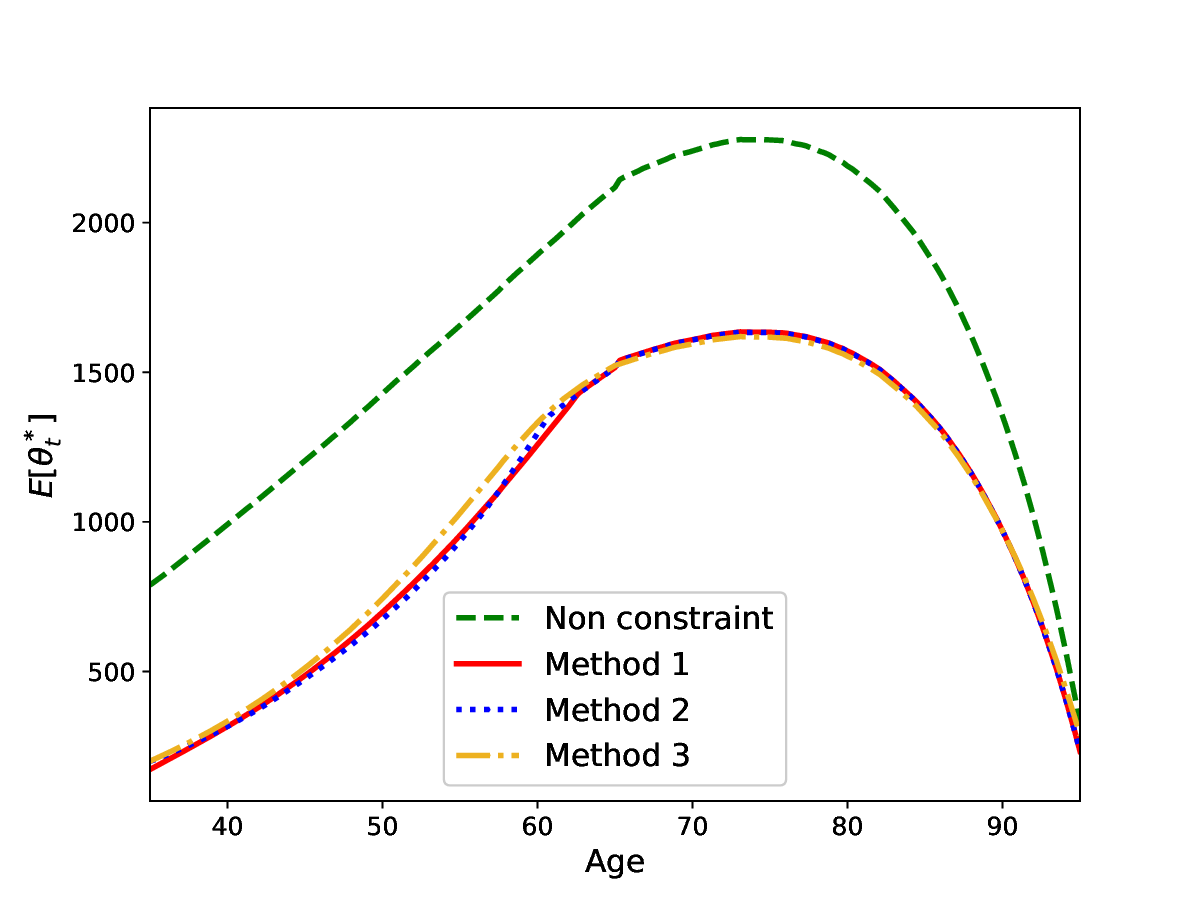}
        \label{fig_investment_gamma1point5:figure4}
    }
    \caption{Expected optimal investment strategies under $\gamma=1.5$}
    \label{fig_investment_gamma1point5:figures}
\end{figure}

    Figure \ref{fig_investment_gamma1point5:figures} plots the expected optimal investment strategies under $\gamma=1.5$. In the control $v_t=(v_{0, t}, v_{-, t})$, $v_{0, t}$ represents the compensated return for the interest rate $r$ and $v_{-, t}$ is the compensated return for stock appreciate rate $\mu$. For the compensated control $v_t$, we have three observations: (i) We observe $v_{0, t}\geq 0$ and $v_{-,t} \equiv 0$. The reason is that the artificial market increases the adjusted interest rate $r+v_{0,t}$, making the bond more attractive and discouraging shorting the bond to purchase the stock at the adjusted return $\mu+v_{-,t}$. (ii) For the monotonicity, we observe that $v_0$ generally decreases with time. This means the individual has a strong incentive to short the bond and buy the stock in their early years. To guarantee no short-selling constraint, the artificial market adds a greater $v_{0,t}$ to avoid shorting at the beginning. Moreover, $v_{-,t}$ decreases to 0 near the retirement under the risk-averse case $\gamma=1.5$, which means the risk-averse individual doesn't have the demand to short the bond and purchase the stock in old age. (iii) We observe that the $v_t$'s shapes are different under the three methods. Method 1 generates a linear $v_t$ in both cases and misses the short demand after retirement for case $\gamma=0.8$. Methods 2 and 3 both obtain nonlinear controls due to the neural network's generality and flexibility.

    For the investment strategy $(\alpha_t, \theta_t)$, we see that an individual without trading constraints tends to short the bond and long the stock in their early age. They reduce the short position in the bond and finally switch to a long position near retirement. When considering trading constraints, the individual will long the bond at an early age, which is consistent with the positive compensated control $v_0$ before retirement, as shown in Figure \ref{fig_investment_gamma1point5:figure1}. Finally, both allocation amounts follow a hump shape, increasing before retirement and decreasing after retirement.

\begin{figure}[htbp]
    \centering
    \subfigure[Bond compensated return]
    {
        \includegraphics[width=3.0in,height=2.5in]{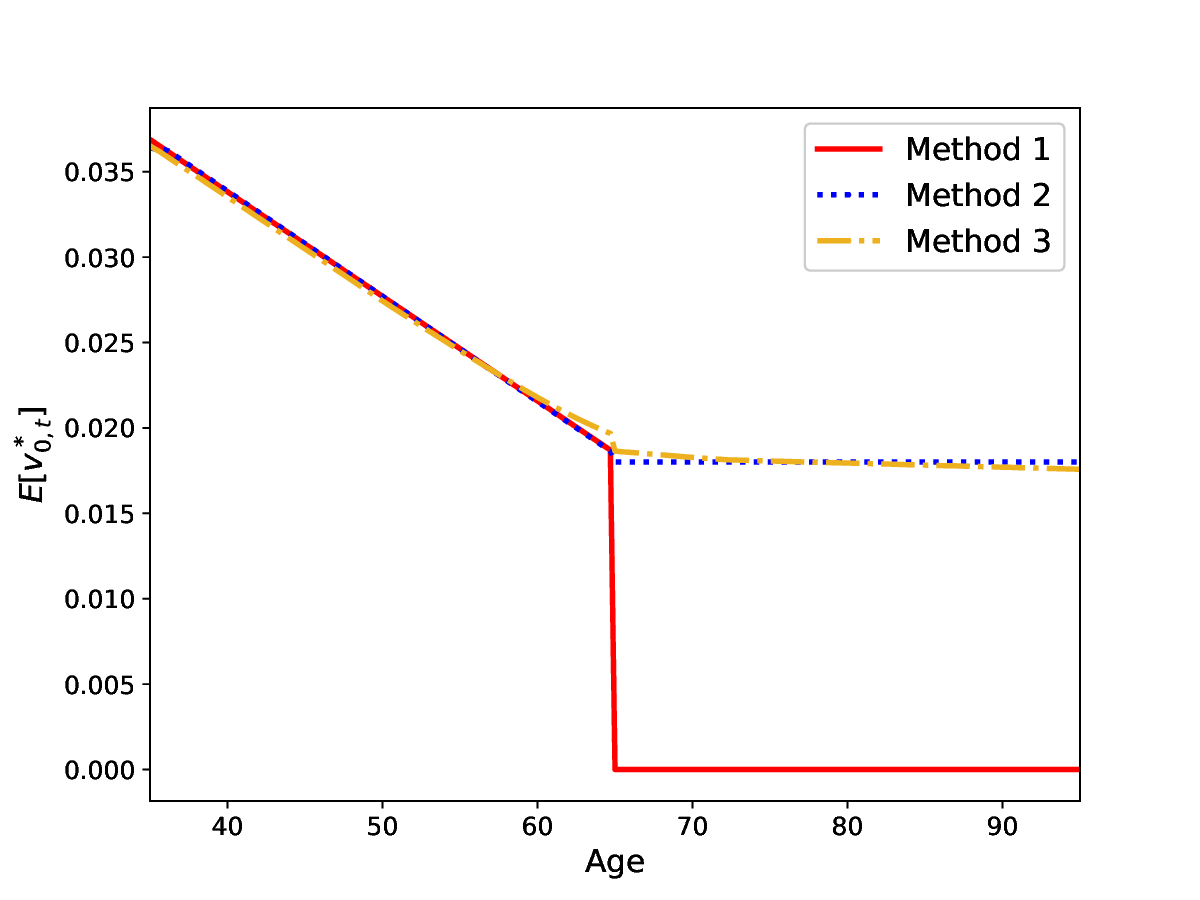}
        \label{fig_investment_gamma0point8:figure1}
    }
    \hspace{-0.45in}
    \subfigure[Stock compensated return]
    {
	\includegraphics[width=3.0in,height=2.5in]{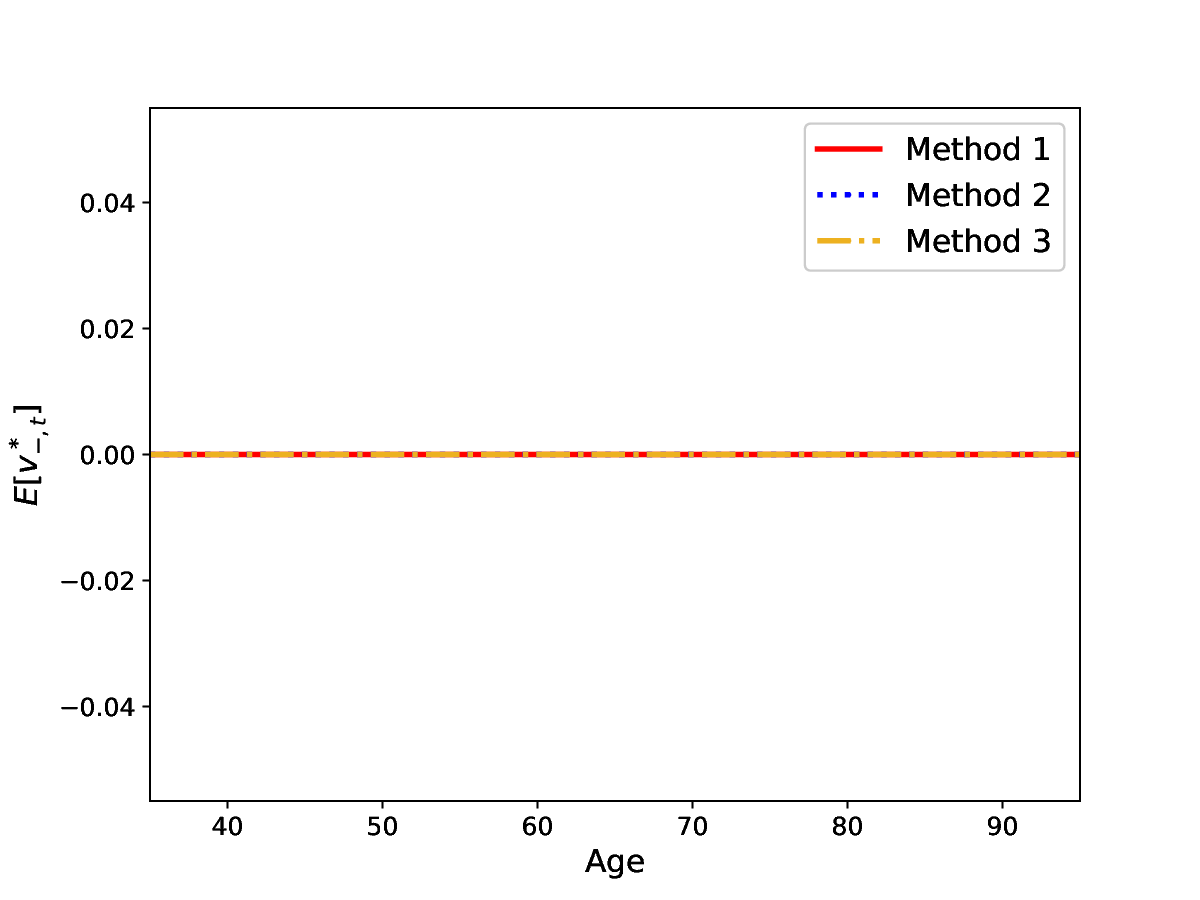}
        \label{fig_investment_gamma0point8:figure2}
    }
    \\
    \subfigure[Bond allocation amount]
    {
        \includegraphics[width=3.0in,height=2.5in]{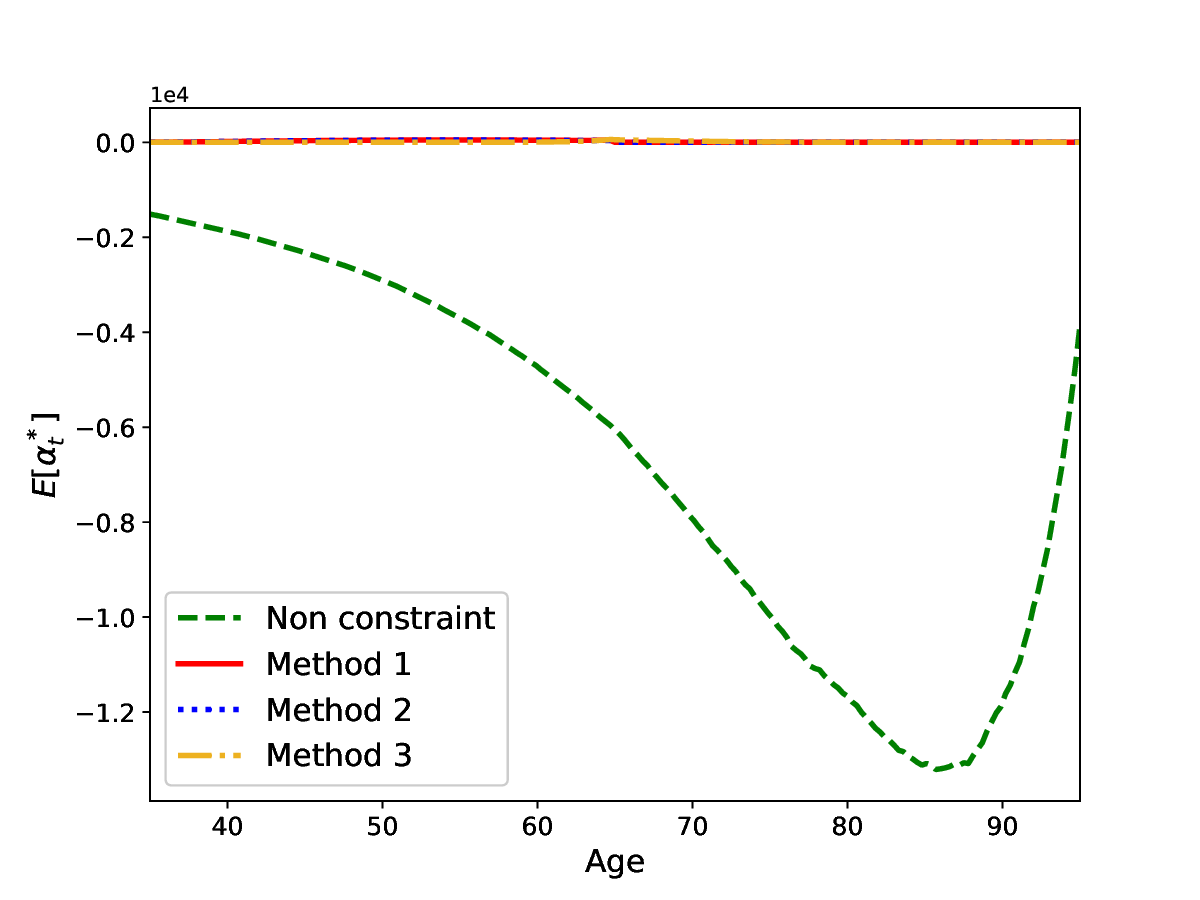}
        \label{fig_investment_gamma0point8:figure3}
    }
    \hspace{-0.45in}
    \subfigure[Stock allocation amount]
    {
	\includegraphics[width=3.0in,height=2.5in]{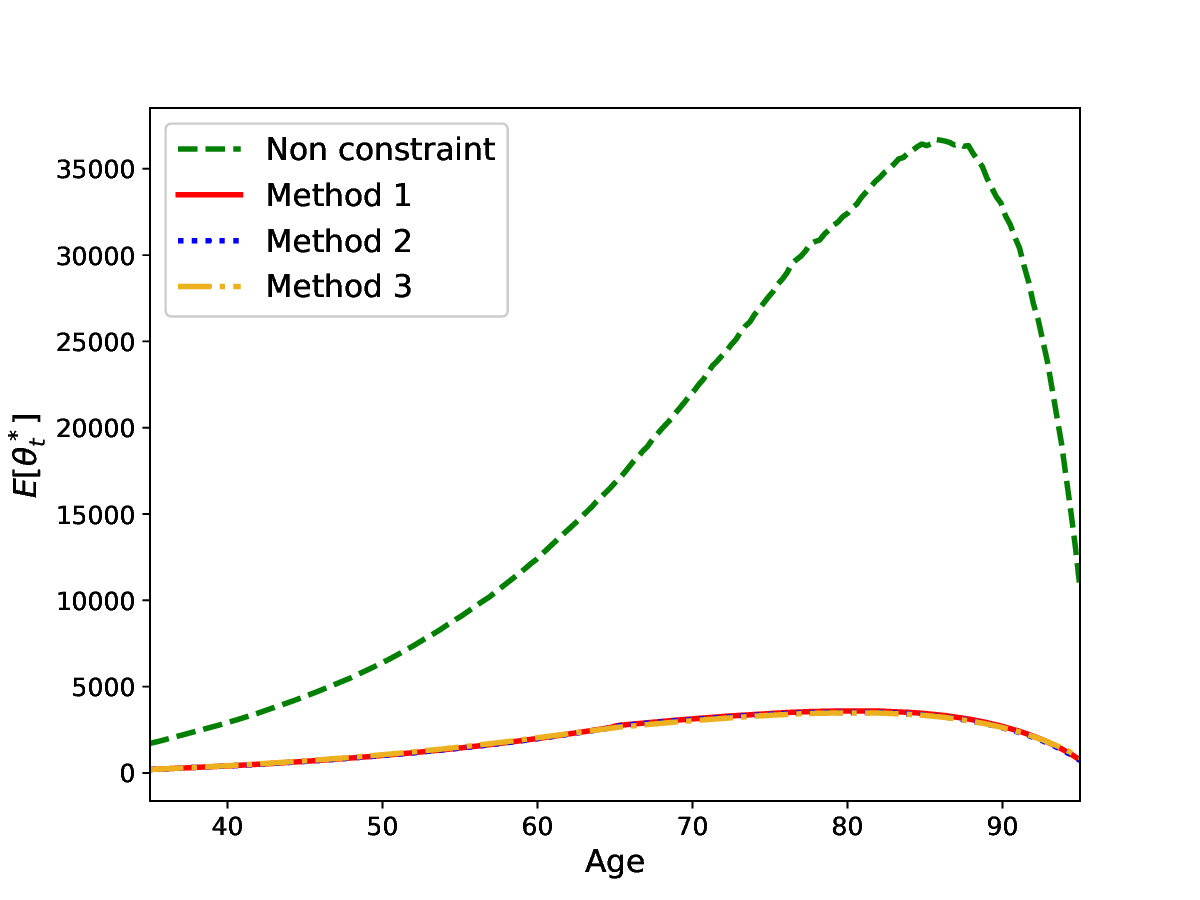}
        \label{fig_investment_gamma0point8:figure4}
    }
    \caption{Expected optimal investment strategies under $\gamma=0.8$}
    \label{fig_investment_gamma0point8:figures}
\end{figure}

    Figure \ref{fig_investment_gamma0point8:figures} illustrates expected optimal investment strategies under $\gamma=0.8$. We see that a risk-seeking individual tends to short the bond and long the stock throughout their life. Both Methods 2 and 3 identify this trend and yield a positive compensated return $v_0$ over time. However, Method 1 fails to identify this change and obtain a zero value for $v_0$ after retirement. Lastly, all three methods learn a zero $v_{-}$ over an individual's lifetime. The reason is that the optimal investment strategy is to short the bond and go long the stock, so the stock's compensation is redundant in the artificial market. 

\begin{figure}[htbp]
    \centering
    \vspace{-0.10in}
    \subfigure[Insurance premium]
    {
        \includegraphics[width=3.0in,height=2.5in]{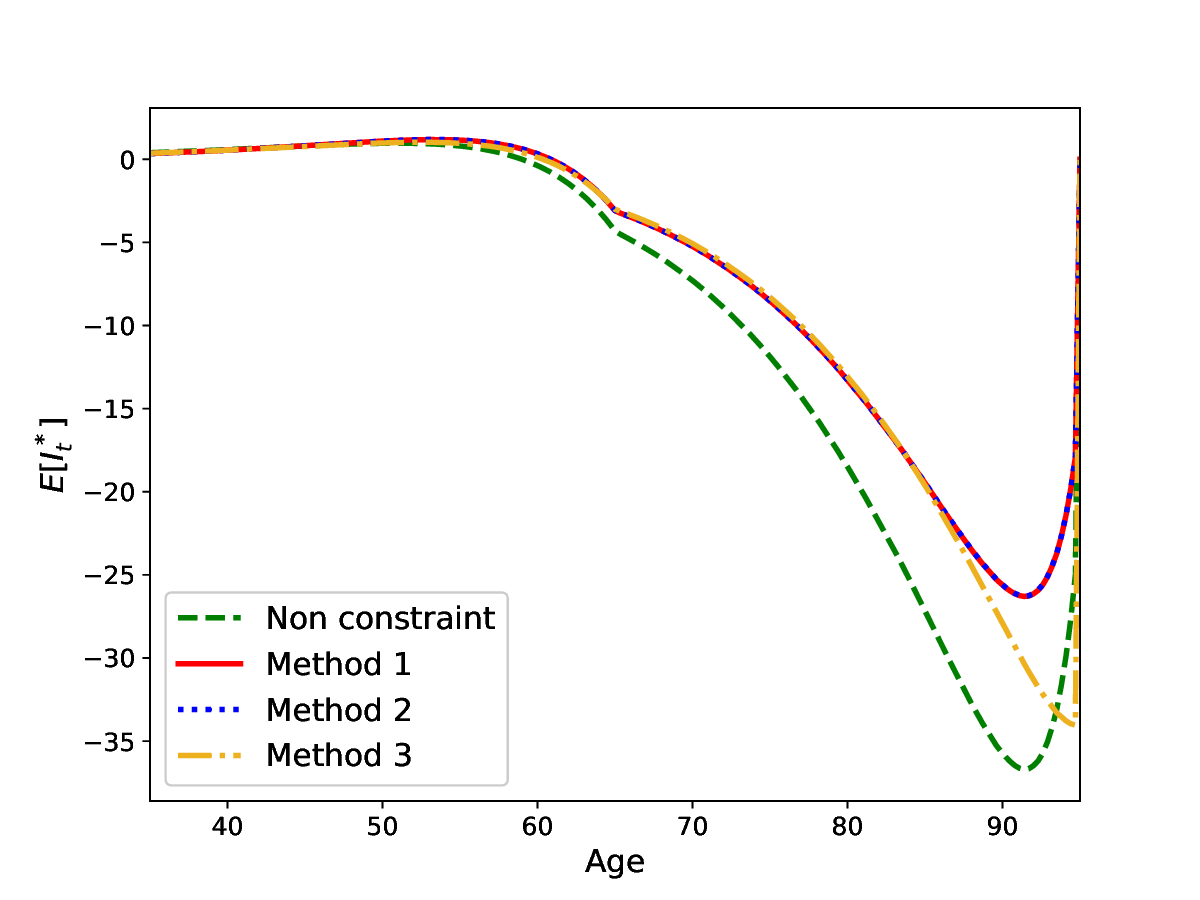}
        \label{fig_strategy_gamma1point5:figure1}
    }
    \hspace{-0.45in}
    \subfigure[Insurance payout]
    {
        \includegraphics[width=3.0in,height=2.5in]{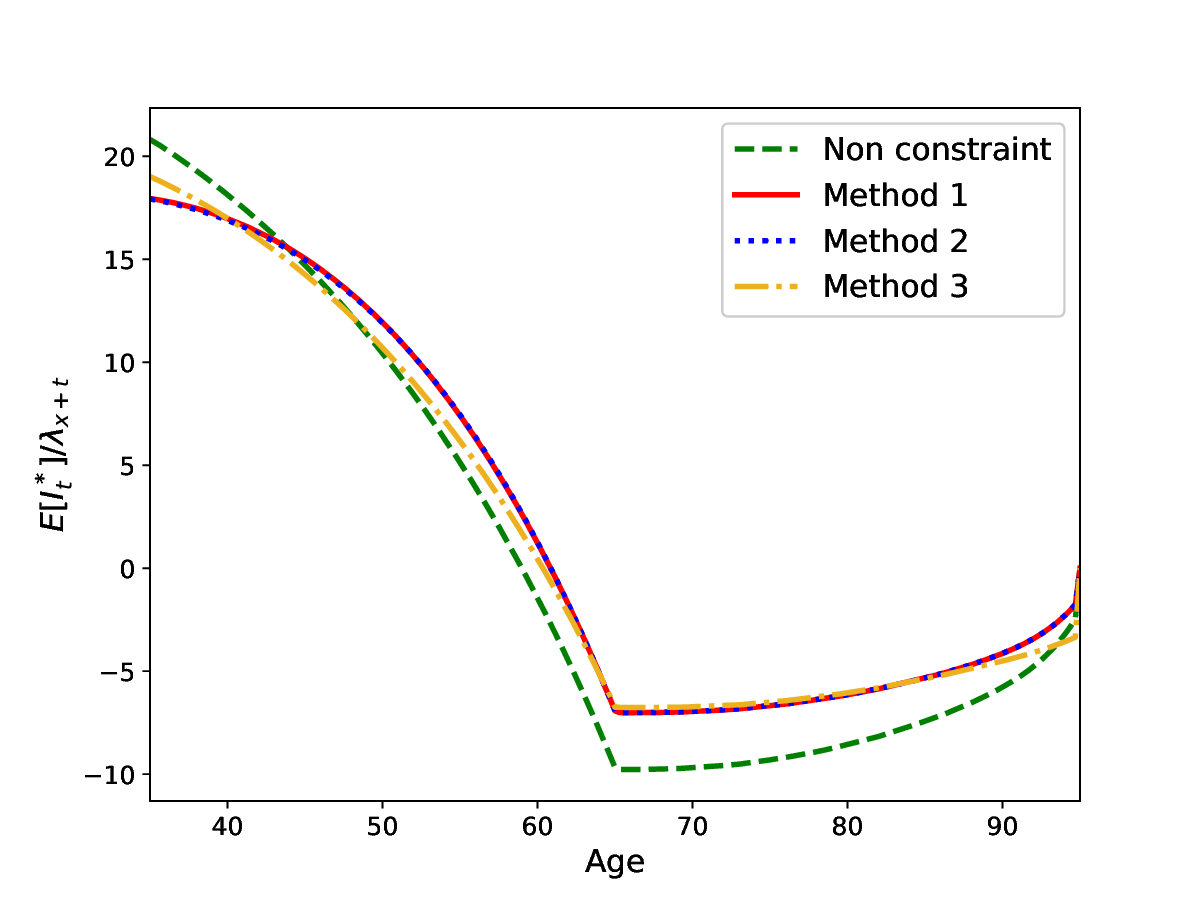}
        \label{fig_strategy_gamma1point5:figure2}
    }
    \\
    \vspace{-0.10in}
    \subfigure[Consumption]
    {
        \includegraphics[width=3.0in,height=2.5in]{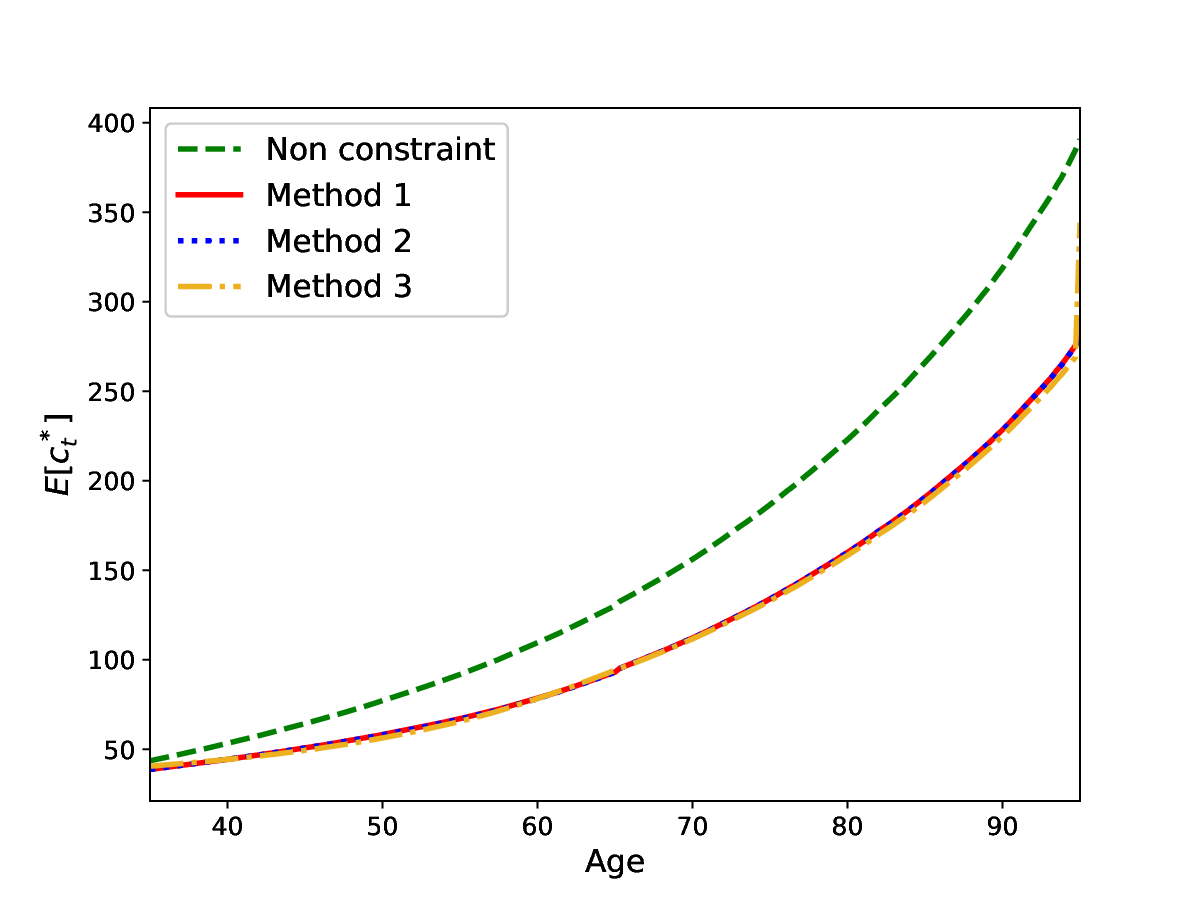}
        \label{fig_strategy_gamma1point5:figure3}
    }
    \hspace{-0.45in}
    \subfigure[Wealth]
    {
        \includegraphics[width=3.0in,height=2.5in]{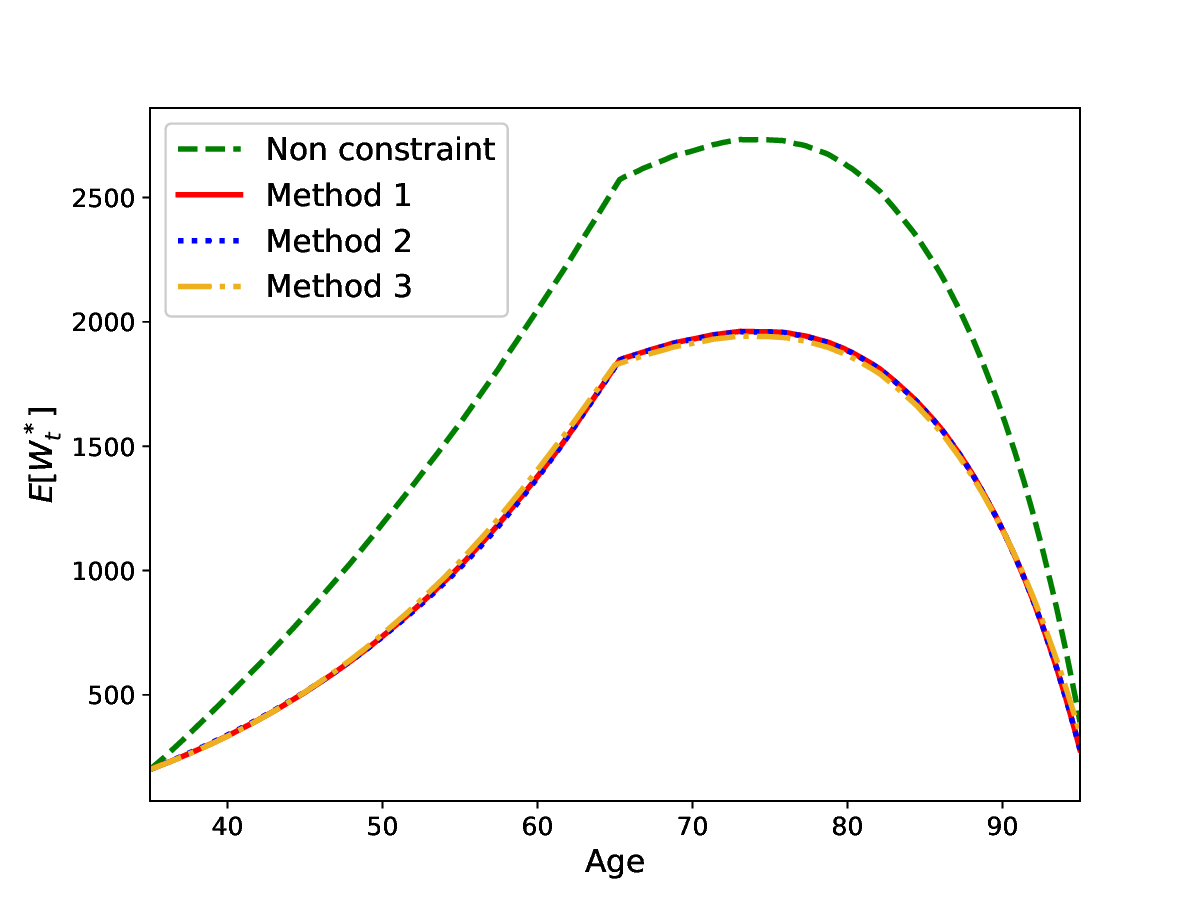}
        \label{fig_strategy_gamma1point5:figure4}
    }
    \caption{Expected optimal insurance and consumption strategies under $\gamma=1.5$. }
    \label{fig_strategy_gamma1point5:figures}
\end{figure}

    Figure \ref{fig_strategy_gamma1point5:figures} displays the expected optimal insurance and consumption strategies under $\gamma=1.5$. For the insurance strategies, we have three observations: (i) The expected life insurance premium $E[I^*_t]$ switches from positive to negative with age. This phenomenon stems from the contrasting roles of life insurance and annuities. Life insurance protects an individual's future income by charging premiums ($I_t>0$), while an annuity works as ``direct income'' ($I_t<0$) after an individual's retirement. Consequently, the individual purchases life insurance before retirement to protect income and switches to an annuity near retirement (see similar interpretations in \cite{fischer1973life} and \cite{li2025optimal}). (ii) The positive range of $E[I^*_t]$ is significantly smaller than its negative range. The evidence is that the maximum positive value in Figure \ref{fig_strategy_gamma1point5:figure1} is 0.96k USD and the maximum negative value is 36.71k USD for the non-constrained case. Same pattern for the constrained case. These values are consistent with the real market, where life insurance prices are much lower than annuity prices, as life insurance serves as income protection, whereas an annuity provides income directly. However, when looking at the insurance payout ($I_t/\lambda_{x+t}$) in Figure \ref{fig_strategy_gamma1point5:figure2}, we see that the payouts of these two products have the same range. In general, life insurance premiums/prices are much lower than annuities, but their payout are at the same level; both work as additional income at the payout stage. (iii) When considering the trading constraint, the individual tends to reduce their life insurance demand at an early age and lower their annuity demand in old age. This is because trading constraints narrow the admissible set and reduce the potential investment returns. As a result, demand for life insurance and annuities shrinks due to low purchasing power in a constrained portfolio. Furthermore, we find that the trading constraints lower the optimal consumption and wealth process. These are direct results of declining profit and purchasing power within the constrained portfolio. Lastly, among the three methods, Method 3 yields the highest initial life insurance purchase, the highest annuity purchase, a similar consumption level, and the highest terminal wealth, outperforming the other two methods. 
    
\begin{figure}[htbp]
    \centering
    \vspace{-0.10in}
    \subfigure[Insurance premium]
    {
        \includegraphics[width=3.0in,height=2.5in]{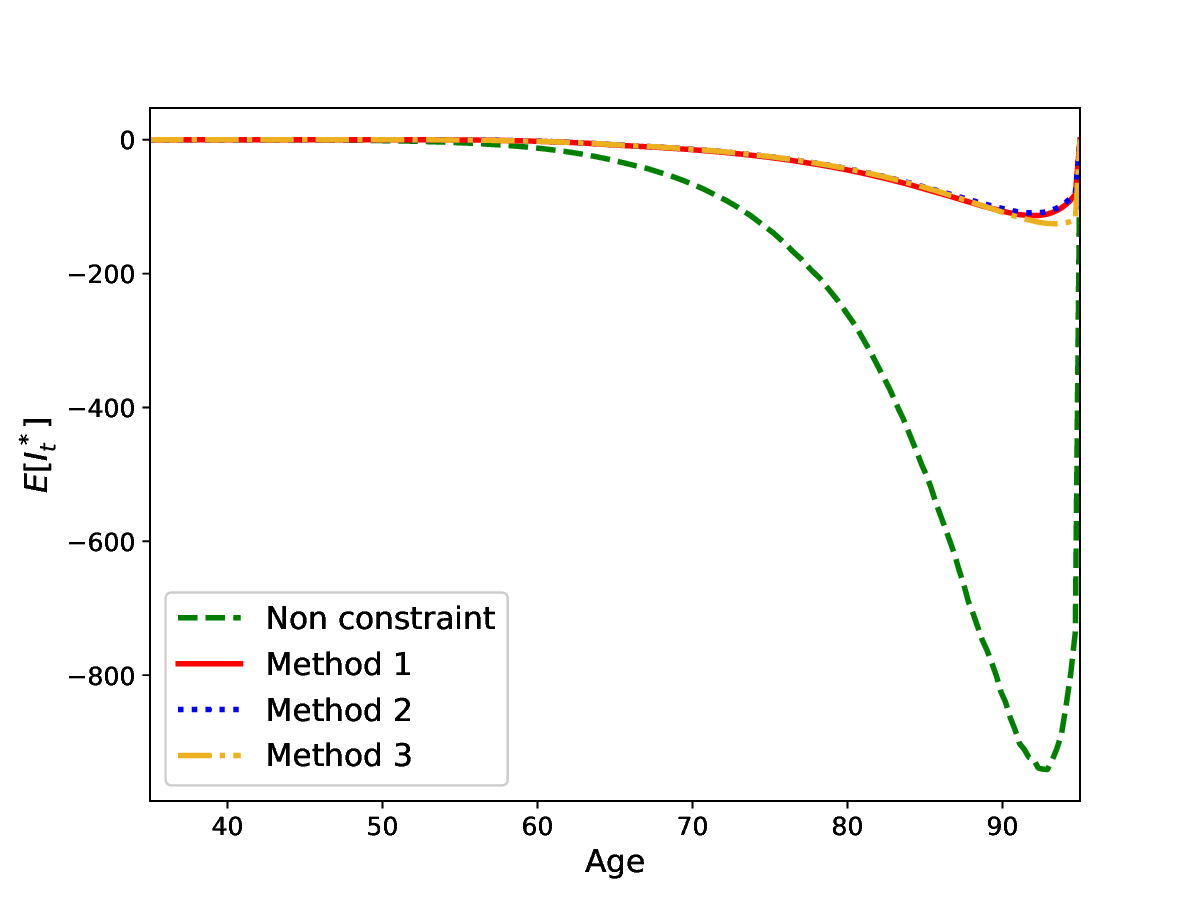}
        \label{fig_strategy_gamma0point8:figure1}
    }
    \hspace{-0.45in}
    \subfigure[Insurance payout]
    {
        \includegraphics[width=3.0in,height=2.5in]{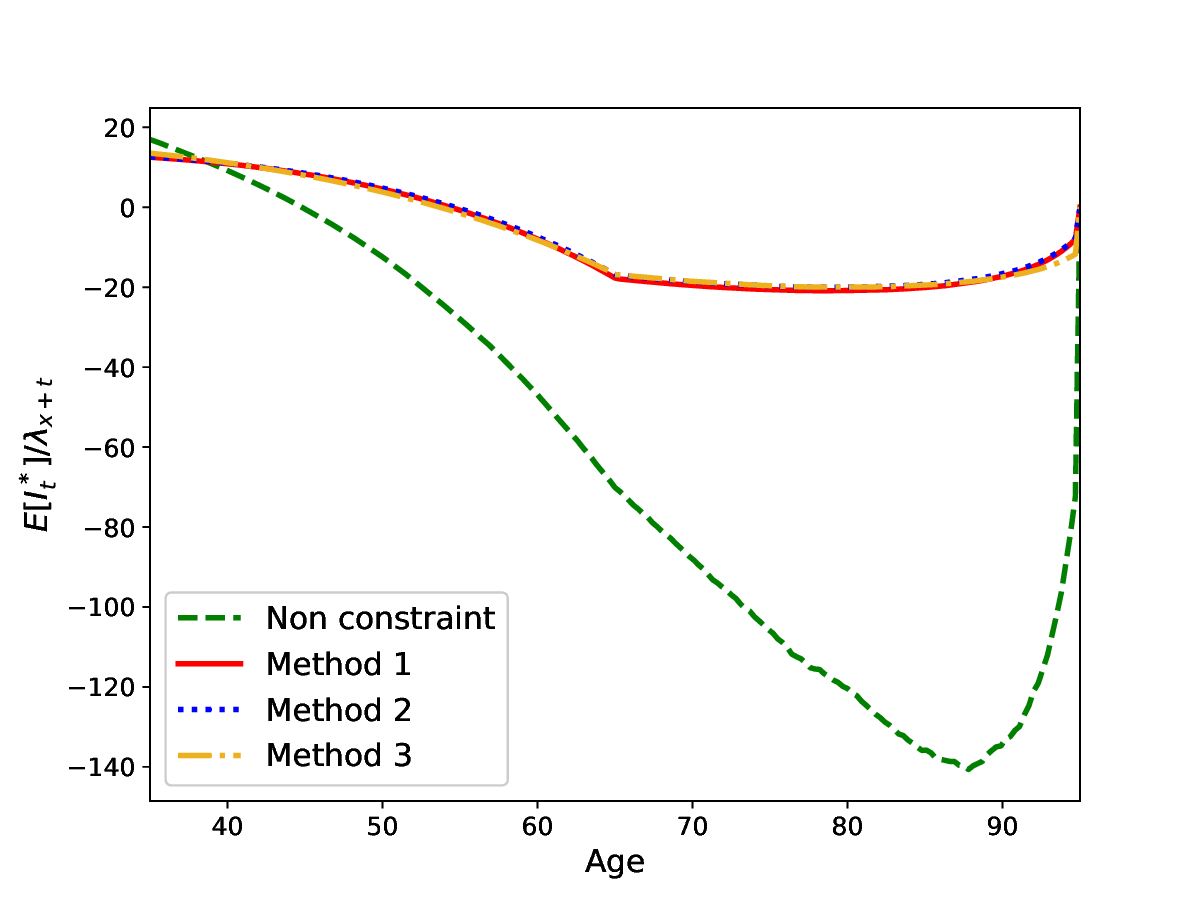}
        \label{fig_strategy_gamma0point8:figure2}
    }
    \\
    \vspace{-0.10in}
    \subfigure[Consumption]
    {
        \includegraphics[width=3.0in,height=2.5in]{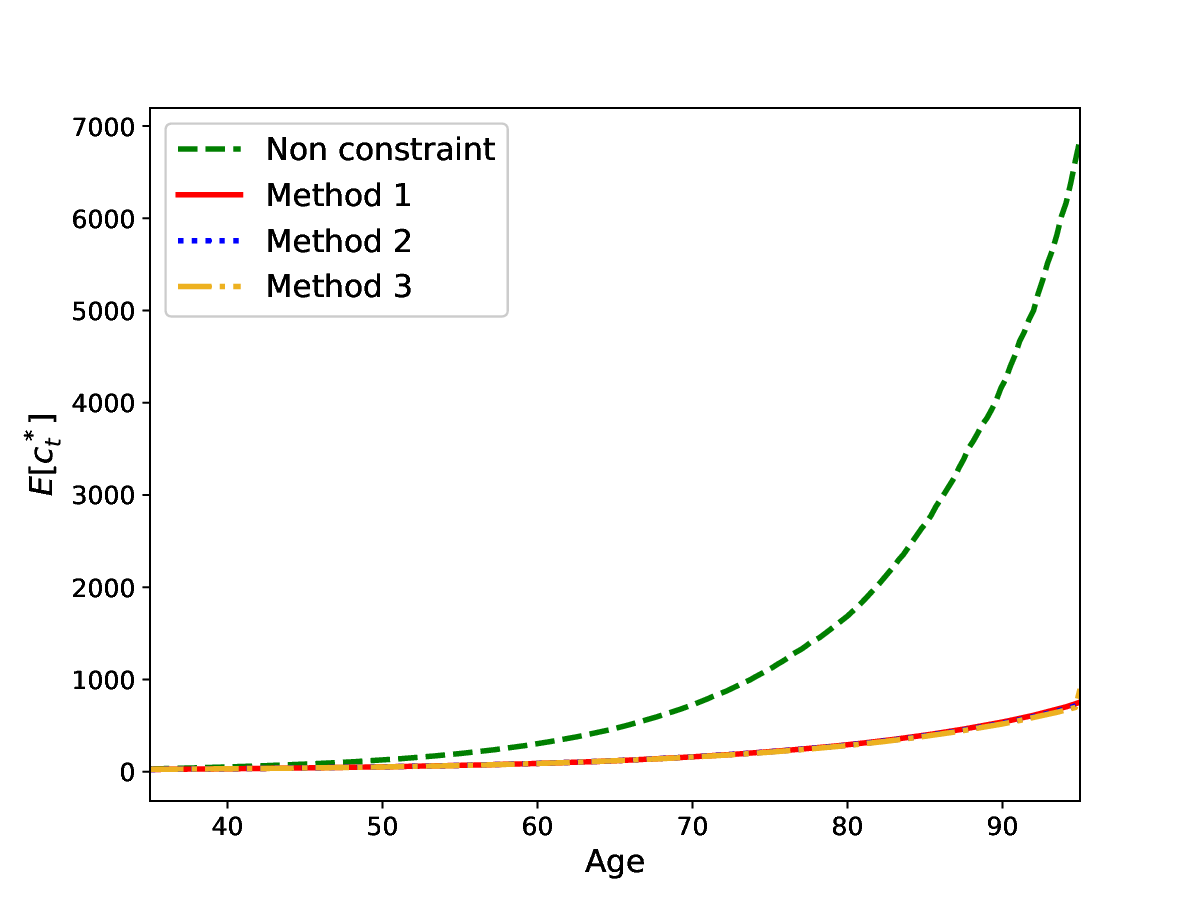}
        \label{fig_strategy_gamma0point8:figure3}
    }
    \hspace{-0.45in}
    \subfigure[Wealth]
    {
        \includegraphics[width=3.0in,height=2.5in]{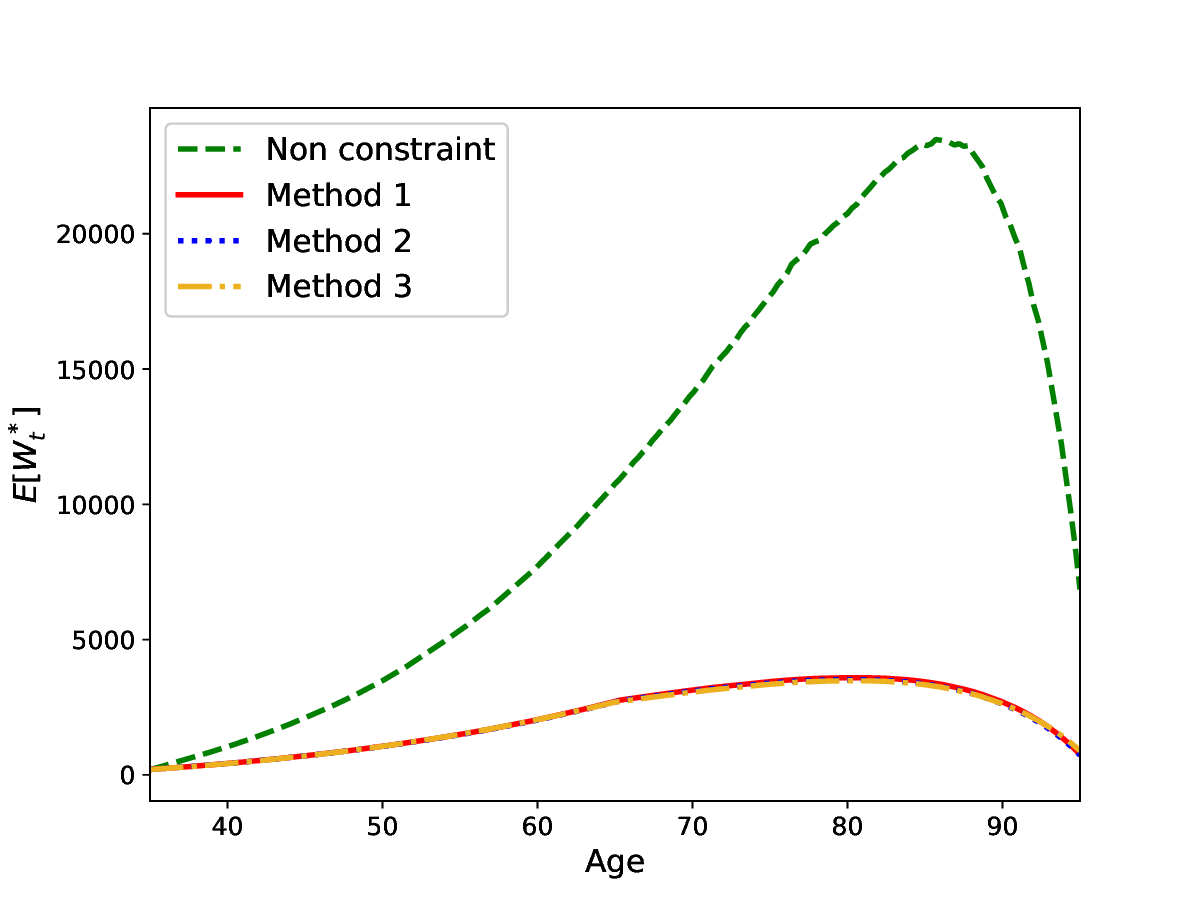}
        \label{fig_strategy_gamma0point8:figure4}
    }
    \caption{Expected optimal insurance and consumption strategies under $\gamma=0.8$. }
    \label{fig_strategy_gamma0point8:figures}
\end{figure}

    Figure \ref{fig_strategy_gamma0point8:figures} shows the expected optimal insurance and consumption strategies under $\gamma=0.8$. We see that a risk-seeking individual enjoys higher insurance and annuity payouts, a higher consumption level, and a higher wealth level than a risk-averse individual. For the non-constraint case, the risk-seeking individual behaves more ``crazy'' than the risk-averse individual.

\section{Conclusion}\label{sec_conclusion}
This paper studies the constrained portfolio optimization problem within a generalized life-cycle model. The individual has stochastic income and allocates their wealth among stocks, a bond, and life insurance to optimize consumption, death benefits, and terminal wealth. In addition, the individual's trading strategy is restricted to a non-empty, closed, convex set that includes non-tradeable assets, no short-selling, and no borrowing constraints as special cases. We define the artificial markets and transform the dynamic budget constraint in the primal problem into a group of static budget constraints. Next, using the Lagrangian dual control approach, we transform the primal problem into the dual problem and establish an upper bound for the primal problem. We propose a ``Deep pricing kernel method'' to induce a candidate strategy from the upper bound and provide tight lower and upper bounds for the primal problem. Our numerical simulations show that, overall, trading constraints can reduce an individual's purchasing power and benefits. Specifically, the individual will reduce their stock allocation, consume less, purchase less life insurance at an early age, reduce annuity demand after retirement, and accumulate less wealth over their lifetime. 

To the best of our knowledge, this is the first application of neural networks to the constrained portfolio optimization problem in the life-cycle model. Our deep learning approach significantly reduces the duality gap and can inspire future work on portfolio optimization problems without explicit solutions.

\subsection*{Acknowledgments}
This paper is in memory of Professor Pengyu Wei, a true scholar in Actuarial Science and Mathematical Finance. This research did not receive any specific grant from funding agencies.

\clearpage
\bibliographystyle{apalike}
\bibliography{ref}

@article{hainaut2006life,
  title={Life annuitization: Why and how much?},
  author={Hainaut, Donatien and Devolder, Pierre},
  journal={ASTIN Bulletin: the Journal of the IAA},
  volume={36},
  number={2},
  pages={629--654},
  year={2006},
  publisher={Cambridge University Press}
}

@article{aase2015life,
  title={Life insurance and pension contracts {I}: The time additive life cycle model},
  author={Aase, Knut K},
  journal={ASTIN Bulletin: The Journal of the IAA},
  volume={45},
  number={1},
  pages={1--47},
  year={2015},
  publisher={Cambridge University Press}
}

@article{aase2016life,
  title={Life insurance and pension contracts {II}: The life cycle model with recursive utility},
  author={Aase, Knut K},
  journal={ASTIN Bulletin: The Journal of the IAA},
  volume={46},
  number={1},
  pages={71--102},
  year={2016},
  publisher={Cambridge University Press}
}

@article{wang2024life,
  title={Life-cycle planning with {CEV} model and time-inconsistent preferences},
  author={Wang, Hao and Hu, Shujie and Siu, Tak Kuen and Wang, Rongming and Wang, Ning},
  journal={International Review of Economics \& Finance},
  volume={96},
  pages={103517},
  year={2024},
  publisher={Elsevier}
}

@article{branger2019hedging,
  title={Hedging recessions},
  author={Branger, Nicole and Larsen, Linda Sandris and Munk, Claus},
  journal={Journal of Economic Dynamics and Control},
  volume={107},
  pages={103715},
  year={2019},
  publisher={Elsevier}
}

@article{tepla2000optimal,
  title={Optimal portfolio policies with borrowing and shortsale constraints},
  author={Tepla, Lucie},
  journal={Journal of Economic Dynamics and Control},
  volume={24},
  number={11-12},
  pages={1623--1639},
  year={2000},
  publisher={Elsevier}
}

@article{hu2023constrained,
  title={Constrained monotone mean-variance problem with random coefficients},
  author={Hu, Ying and Shi, Xiaomin and Xu, Zuo Quan},
  journal={SIAM Journal on Financial Mathematics},
  volume={14},
  number={3},
  pages={838--854},
  year={2023},
  publisher={SIAM}
}

@article{gu2020constrained,
  title={Constrained utility deviation-risk optimization and time-consistent {HJB} equation},
  author={Gu, Jia-Wen and Si, Shijing and Zheng, Harry},
  journal={SIAM Journal on Control and Optimization},
  volume={58},
  number={2},
  pages={866--894},
  year={2020},
  publisher={SIAM}
}

@article{hu2005constrained,
  title={Constrained stochastic {LQ} control with random coefficients, and application to portfolio selection},
  author={Hu, Ying and Zhou, Xun Yu},
  journal={SIAM Journal on Control and Optimization},
  volume={44},
  number={2},
  pages={444--466},
  year={2005},
  publisher={SIAM}
}

@article{fischer1973life,
  title={A life cycle model of life insurance purchases},
  author={Fischer, Stanley},
  journal={International Economic Review},
  pages={132--152},
  year={1973},
  publisher={JSTOR}
}

@article{li2025optimal,
  title={Optimal life insurance and annuity decisions under money illusion},
  author={Li, Wenyuan and Wei, Pengyu},
  journal={Insurance: Mathematics and Economics},
  pages={103141},
  year={2025},
  publisher={Elsevier}
}

@article{bick2013solving,
	title={Solving constrained consumption--investment problems by simulation of artificial market strategies},
	author={Bick, Bj{\"o}rn and Kraft, Holger and Munk, Claus},
	journal={Management Science},
	volume={59},
	number={2},
	pages={485--503},
	year={2013},
	publisher={INFORMS}
}

@article{boyle2022annuity,
	title={Annuity and insurance choice under habit formation},
	author={Boyle, Phelim and Tan, Ken Seng and Wei, Pengyu and Zhuang, Sheng Chao},
	journal={Insurance: Mathematics and Economics},
	volume={105},
	pages={211--237},
	year={2022},
	publisher={Elsevier}
}

@article{chabakauri2013dynamic,
	title={Dynamic equilibrium with two stocks, heterogeneous investors, and portfolio constraints},
	author={Chabakauri, Georgy},
	journal={The Review of Financial Studies},
	volume={26},
	number={12},
	pages={3104--3141},
	year={2013},
	publisher={Oxford University Press}
}

@article{cuoco1997optimal,
	title={Optimal consumption and equilibrium prices with portfolio constraints and stochastic income},
	author={Cuoco, Domenico},
	journal={Journal of Economic Theory},
	volume={72},
	number={1},
	pages={33--73},
	year={1997},
	publisher={Elsevier}
}

@article{karatzas1991martingale,
	title={Martingale and duality methods for utility maximization in an incomplete market},
	author={Karatzas, Ioannis and Lehoczky, John P and Shreve, Steven E and Xu, Gan-Lin},
	journal={SIAM Journal on Control and optimization},
	volume={29},
	number={3},
	pages={702--730},
	year={1991},
	publisher={SIAM}
}

@article{cvitanic1992convex,
	title={Convex duality in constrained portfolio optimization},
	author={Cvitani{\'c}, Jak{\v{s}}a and Karatzas, Ioannis},
	journal={The Annals of Applied Probability},
	pages={767--818},
	year={1992},
	publisher={JSTOR}
}

@article{cvitanic1993hedging,
	title={Hedging contingent claims with constrained portfolios},
	author={Cvitani{\'c}, Jak{\v{s}}a and Karatzas, Ioannis},
	journal={The Annals of Applied Probability},
	pages={652--681},
	year={1993},
	publisher={JSTOR}
}

@book{dellacherie2011probabilities,
	title={Probabilities and Potential, C: Potential Theory for Discrete and Continuous Semigroups},
	author={Dellacherie, Claude and Meyer, P-A},
	year={2011},
	publisher={Elsevier}
}

@article{harrison1979martingales,
	title={Martingales and arbitrage in multiperiod securities markets},
	author={Harrison, J Michael and Kreps, David M},
	journal={Journal of Economic Theory},
	volume={20},
	number={3},
	pages={381--408},
	year={1979},
	publisher={Citeseer}
}

@article{haugh2006evaluating,
	title={Evaluating portfolio policies: A duality approach},
	author={Haugh, Martin B and Kogan, Leonid and Wang, Jiang},
	journal={Operations Research},
	volume={54},
	number={3},
	pages={405--418},
	year={2006},
	publisher={INFORMS}
}

@article{hambel2022solving,
	title={Solving life-cycle problems with biometric risk by artificial insurance markets},
	author={Hambel, Christoph and Kraft, Holger and Munk, Claus},
	journal={Scandinavian Actuarial Journal},
	volume={2022},
	number={4},
	pages={307--327},
	year={2022},
	publisher={Taylor \& Francis}
}

@article{he1993labor,
	title={Labor income, borrowing constraints, and equilibrium asset prices},
	author={He, Hua and Pages, Henri F},
	journal={Economic Theory},
	volume={3},
	number={4},
	pages={663--696},
	year={1993},
	publisher={Springer}
}

@article{huang2008portfolio,
	title={Portfolio choice and life insurance: The {CRRA} case},
	author={Huang, Huaxiong and Milevsky, Moshe A and Wang, Jin},
	journal={Journal of Risk and Insurance},
	volume={75},
	number={4},
	pages={847--872},
	year={2008},
	publisher={Wiley Online Library}
}

@article{rockafellar1970convex,
	title={Convex Analysis Princeton University Press},
	author={Rockafellar, R Tyrrell},
	journal={Princeton, NJ},
	year={1970}
}

@article{zeng2016optimal,
	title={Optimal life insurance with no-borrowing constraints: duality approach and example},
	author={Zeng, Xudong and Carson, James M and Chen, Qihong and Wang, Yuling},
	journal={Scandinavian Actuarial Journal},
	volume={2016},
	number={9},
	pages={793--816},
	year={2016},
	publisher={Taylor \& Francis}
}

\vspace{0.3in}
\noindent Wenyuan Li (Corresponding Author)\\
Department of Statistics and Actuarial Science\\
The University of Hong Kong\\
Pokfulam Road, Hong Kong SAR, China\\
Email: wylsaas@hku.hk\\

\noindent Pengyu Wei (In Memoriam)\\
Division of Banking and Finance\\
Nanyang Business School\\
Nanyang Technological University, Singapore\\
E-mail: pengyu.wei@ntu.edu.sg

\clearpage

\pagenumbering{arabic}
\setcounter{page}{1}
\appendix
\begin{flushleft}
\Large Online Appendix--Constrained portfolio optimization in a life-cycle model: A deep pricing kernel approach
\end{flushleft}

\section{Proof of Theorem \ref{theorem_1}}\label{appendix_1}
\setcounter{equation}{0}
\renewcommand{\theequation}{A.\arabic{equation}}
\begin{proof}
	``Only if'' part: ``$\Rightarrow$''\\
	By Ito's formula and equation \eqref{dynamic_budget_constraint}, we have
	\begin{equation*}
		d({_tp_x}\beta_{v,t} W_t)={_tp_x}\beta_{v,t} (-v_{0,t} \alpha_t dt - \theta^{\top}_t v_{-,t} dt +\theta^{\top}_t \sigma_t dZ_{v,t} - c_t dt -\lambda_{x+t}M_t dt + Y_t dt - dC_t).
	\end{equation*}
	Integrating on both hands sides, we obtain the inequality
	\begin{eqnarray}
		&& {_tp_x} \beta_{v,t}  W_t - w_0 + \int_0^t {_sp_x} \lambda_{x+s} \beta_{v,s}  M_s ds + \int_0^t {_sp_x}\beta_{v,s}(c_s-Y_s) ds \notag\\
		&&\leq \int_0^t {_sp_x}\beta_{v,s} \left[ -(\alpha_s, \theta^{\top}_s)\begin{pmatrix}
			v_{0,s}\\
			v_{-,s}
		\end{pmatrix}
		\right] ds + \int_0^t {_sp_x}\beta_{v,s}  \theta^{\top}_s \sigma_s dZ_{v,s}.\label{appendix_1_ineq_1}
	\end{eqnarray}
	Moreover, by the definition of supporting function \eqref{supporting_function}, together with the inequality \eqref{appendix_1_ineq_1}, we arrive at the following inequality
	\begin{eqnarray}
		&&{_tp_x}\beta_{v,t} W_t - w_0 + \int_0^t {_sp_x} \lambda_{x+s} \beta_{v,s}  M_s ds + \int_0^t {_sp_x}\beta_{v,s}(c_s-Y_s) ds \notag\\
		&&\leq \int_0^t {_sp_x}\beta_{v,s}\delta(v_s) ds + \int_0^t {_sp_x}\beta_{v,s} \theta^{\top}_s \sigma_s dZ_{v,s}.\label{appendix_1_ineq_2}	
	\end{eqnarray}
	
	Define the stopping time $\tau_n = T\wedge \inf \{t\in [0,T]: \int_0^t |\theta^{\top}_s \sigma_s|^2 ds \geq n\}$ for $n\in \mathbb{N_+}$ and $\inf(\emptyset)=\infty$. Since the stochastic integral in \eqref{appendix_1_ineq_2} is a $Q_v$ martingale in $[0,\tau_n]$, we have
	\begin{eqnarray}
		&&E^{Q_v}\left[{_{\tau_n}p_x}\beta_{v,\tau_n} W_{\tau_n} + \int_0^{\tau_n} {_tp_x}\lambda_{x+t} \beta_{v,t} M_t dt + \int_0^{\tau_n}{_tp_x}\beta_{v,t}(c_t-Y_t)dt \right] \notag\\
		&&\leq w_0 + E^{Q_v}\left[ \int_0^{\tau_n} {_tp_x}\beta_{v,t} \delta(v_t) dt\right].\label{appendix_1_ineq_3}
	\end{eqnarray}
	By the definition of the admissible strategy \eqref{admissible}, we have $\tau_n \nearrow T$ when $n \rightarrow \infty$. Because of $v_0\geq 0$ in Assumption \ref{assumption_3} and \eqref{income_constraint}, we have the boundedness of the income process
	\begin{equation*}
		E^{Q_v}\left[\int_0^T {_tp_x} \beta_{v,t} Y_t dt\right] \leq E^{Q_v}\left[\int_0^T {_tp_x}\beta_{0,t}Y_t dt\right]\leq K_y.
	\end{equation*}
	Therefore, the following equality holds by the monotone convergence theorem
	\begin{equation*}
		\lim \limits_{n\rightarrow \infty} E^{Q_v}\left[\int_0^{\tau_n} {_tp_x}\beta_{v,t}(c_t - Y_t) dt\right] = E^{Q_v}\left[\int_0^T {_tp_x}\beta_{v,t} (c_t-Y_t)dt\right].
	\end{equation*}
	According to Assumption \ref{assumption_3}, $\delta(v)$ is bounded above. Subsequently, by the monotone convergence theorem, we have
	\begin{equation*}
		\lim \limits_{n\rightarrow \infty} E^{Q_v}\left[\int_0^{\tau_n}{_tp_x}\beta_{v,t}\delta(v_t)dt\right] =  E^{Q_v}\left[\int_0^{T}{_tp_x}\beta_{v,t}\delta(v_t)dt\right].
	\end{equation*}	
	As for the wealth term in \eqref{appendix_1_ineq_3}, we derive from \eqref{wealth_lower_bound} and Assumption \ref{assumption_1}
	\begin{equation*}
		({_{\tau_n}p_x}\beta_{v,\tau_n}W_{\tau_n})^-\leq ({_{\tau_n}p_x}\beta_{0,\tau_n}W_{\tau_n})^-\leq K \exp\left(\int_0^T r^-_t dt\right) <\infty, ~\text{P-a.s.}
	\end{equation*}
	for all $n$. Next, by Assumption \ref{assumption_1} , we can use Fatou's lemma to show
	\begin{equation*}
		\liminf \limits_{n\rightarrow \infty} E^{Q_v}[{_{\tau_n}p_x} \beta_{v,\tau_n} W_{\tau_n}] \geq E^{Q_v}[{_Tp_x}\beta_{v,T}W_T]\geq 0.
	\end{equation*}
	Finally, we derive 
	\begin{eqnarray*}
		&&E^{Q_v}\left[ {_Tp_x}\beta_{v,T} W_T + \int_0^T {_tp_x}\lambda_{x+t} \beta_{v,t}  M_t dt + \int_0^T {_tp_x}\beta_{v,t} (c_t-Y_t)dt   \right]\\
		&\leq& \liminf \limits_{n\rightarrow \infty} E^{Q_v} \left[{_{\tau_n}p_x} \beta_{v,\tau_n}  W_{\tau_n} + \int_0^{\tau_n} {_tp_x}\lambda_{x+t} \beta_{v,t} M_t dt  + \int_0^{\tau_n} {_tp_x}\beta_{v,t} (c_t-Y_t)dt  \right]\\
		&\leq& w_0 + \liminf \limits_{n\rightarrow \infty} E^{Q_v}\left[\int_0^{\tau_n} {_tp_x}\beta_{v,t} \delta(v_t) dt\right]\\
		&=& w_0 + E^{Q_v}\left[\int_0^T {_tp_x}\beta_{v,t} \delta(v_t) dt\right],
	\end{eqnarray*}
	where the second inequality comes from inequality \eqref{appendix_1_ineq_3}. This completes the proof of ``only if'' part.
	
	Next, we prove the ``if'' part: ``$\Leftarrow$''\\
	To show the inverse, we use $\mathcal{T}$ to denote the set of stopping times $\tau$ with $\tau \leq T$, and for $\forall \tau \in \mathcal{T}$, define
	\begin{eqnarray}\label{appendix_1_wealth_tau}
		&&\widehat{W}_\tau = \sup \limits_{v\in \mathcal{N}^*} E^{Q_v} \left[ \int_{\tau}^T {_{t-\tau}p_{x+\tau}}\frac{\beta_{v,t}}{\beta_{v,\tau}} [c_t-Y_t + \lambda_{x+t}M_t - \delta(v_t)]dt  + {_{T-\tau}p_{x+\tau}}\frac{\beta_{v,T}}{\beta_{v,\tau}}  W_T\bigg|\mathcal{F}_{\tau}\right].
	\end{eqnarray}
	Since $(c,M,W_T) \in G^*_+$, Assumption \ref{assumption_3}, and Assumption \ref{assumption_4}, we have
	\begin{eqnarray}\label{appendix_1_wealth_tau_lower_bound}
		\widehat{W}_\tau &\geq& - \sup \limits_{v\in \mathcal{N}^*} E^{Q_v} \left[ \int_{\tau}^T {_{t-\tau}p_{x+\tau}}\frac{\beta_{v,t}}{\beta_{v,\tau}}  Y_t dt \bigg|\mathcal{F}_{\tau}\right]\geq -K_y,
	\end{eqnarray}
	which satisfies lower boundedness condition \eqref{wealth_lower_bound} of wealth process. Follow the same discussion in \cite{cvitanic1993hedging}, we have $\widehat{W}_t$ satisfies the dynamic programming principle
	\begin{equation}
		\widehat{W}_{\tau_1} =\sup \limits_{v\in \mathcal{N}^*} E^{Q_v}\left[ \int_{\tau_1}^{\tau_2} {_{t-\tau}p_{x+\tau_1}}\frac{\beta_{v,t}}{\beta_{v,\tau_1}} [c_t-Y_t + \lambda_{x+t}M_t -\delta(v_t)]dt +{_{\tau_2-\tau_1}p_{x+\tau_1}}\frac{\beta_{v,\tau_2}}{\beta_{v,\tau_1}} \widehat{W}_{\tau_2}\bigg|\mathcal{F}_{\tau_1}\right], \label{appendix_1_wealth_DPP}
	\end{equation}	
	for all $\tau_1\leq \tau_2$, $\tau_1, \tau_2 \in \mathcal{T}$. Setting $\tau_1 = t$, $\tau_2 = T$, and cancel out the supreme operator in \eqref{appendix_1_wealth_DPP}, we derive 
	\begin{equation}
		H_{v,t} = {_tp_x}\beta_{v,t}\widehat{W}_t + \int_0^t {_sp_x}\beta_{v,s} [c_s-Y_s+\lambda_{x+s}M_s -\delta(v_s)] ds \label{appendix_1_H_vt}
	\end{equation}
	is a $Q_v$-supermartingale for all $v\in \mathcal{N}^*$. By the Doob decomposition (see Theorem VII.12 in \cite{dellacherie2011probabilities}) and the martingale representation theorem, for each $v\in \mathcal{N}^*$ there exists an increasing real valued process $A_v$ and a $\mathbb{R}^n$-valued process $\psi_v$ with $\int_0^T |\psi_{v,t}|^2 dt<\infty$ such that
	\begin{equation}
		H_{v,t} = \widehat{W}_0 + \int_0^t \psi^{\top}_{v,s} dZ_{v,s} - A_{v,t}. \label{appendix_1_doob_decomp}
	\end{equation}
	By the definition of $H_{v,t}$ \eqref{appendix_1_H_vt}, we have
	\begin{eqnarray*}
		&&({_tp_x})^{-1}\beta_{v,t}^{-1}\left(H_{v,t} - \int_0^t {_sp_x}\beta_{v,s}[c_s-Y_s +\lambda_{x+s}M_s -\delta(v_s)]ds\right)\\
		&&=\widehat{W}_t = ({_tp_x})^{-1}\beta_{0,t}^{-1}\left(H_{0,t} - \int_0^t  {_sp_x}\beta_{0,s} [c_s-Y_s +\lambda_{x+s}M_s]ds\right).
	\end{eqnarray*}
	Then using Ito's formula and change of measure \eqref{change of measure}, we drive
	\begin{eqnarray}
		&&d\widehat{W}_t =d\left[({_tp_x})^{-1}\beta_{v,t}^{-1}\left(H_{v,t} - \int_0^t {_sp_x}\beta_{v,s} [c_s-Y_s +\lambda_{x+s}M_s -\delta(v_s)]ds\right)\right] \notag\\
		&&=(r_t+v_{0,t}+\lambda_{x+t})\widehat{W}_t dt + ({_tp_x})^{-1}\beta^{-1}_{v,t} \psi^{\top}_{v,t}[dZ_t + \sigma^{-1}_t( \mu_t + v_{-,t} - (r_t + v_{0,t})\bar{1}_n )dt] \notag\\
		&&-({_tp_x})^{-1}\beta^{-1}_{v,t} dA_{v,t} - [c_t-Y_t+\lambda_{x+t}M_t -\delta(v_t)] dt, \label{appendix_1_tildeW_1}
	\end{eqnarray}
	and
	\begin{eqnarray}
		&&d\widehat{W}_t =d\left[({_tp_x})^{-1}\beta_{0,t}^{-1}\left(H_{0,t} - \int_0^t {_sp_x}\beta_{0,s} [c_s-Y_s +\lambda_{x+s}M_s]ds\right)\right] \notag\\
		&&=(r_t+\lambda_{x+t})\widehat{W}_t dt + ({_tp_x})^{-1}\beta^{-1}_{0,t} \psi^{\top}_{0,t}[dZ_t + \sigma^{-1}_t( \mu_t - r_t \bar{1}_n )dt] \notag\\
		&&-({_tp_x})^{-1}\beta^{-1}_{0,t} dA_{0,t} - [c_t-Y_t+\lambda_{x+t}M_t]dt.\label{appendix_1_tildeW_2}
	\end{eqnarray}
	Compare \eqref{appendix_1_tildeW_1} and \eqref{appendix_1_tildeW_2}, we have 
	\begin{eqnarray}
		&&\beta^{-1}_{v,t}\psi^{\top}_{v,t} = \beta^{-1}_{0,t}\psi^{\top}_{0,t}, \label{appendix_1_eq1_v0}\\
		&&\int_0^t \{v_{0,s} \widehat{W}_s + ({_sp_x})^{-1}\beta^{-1}_{v,s} \psi^{\top}_{v,s}\sigma^{-1}_s[v_{-,s}-v_{0,s}\bar{1}_n] + \delta(v_s) \}ds \notag\\
		&&- \int_0^t ({_sp_x})^{-1}\beta^{-1}_{v,s} dA_{v,s} = -\int_0^t ({_sp_x})^{-1}\beta^{-1}_{0,s} dA_{0,s}, \label{appendix_1_eq2_v0}
	\end{eqnarray}
	for all $v\in \mathcal{N}^*$ and all $t \in [0,T]$. Let 
	\begin{equation}
		\theta^{\top}_t=({_tp_x})^{-1}\beta^{-1}_{0,t} \psi^{\top}_{0,t}\sigma^{-1}_t, ~\alpha_t = \widehat{W}_t - \theta^{\top}_t \bar{1}_n \label{appendix_1_strategy_1}
	\end{equation}
	Substitute them into \eqref{appendix_1_tildeW_2} and integrate, we derive
	\begin{eqnarray*}
		\widehat{W}_t &=& w_0 + \int_0^t (\alpha_s r_s +  \theta^{\top}_s\mu_s) ds + \int_0^t \theta^{\top}_s \sigma_s dZ_s - \int_0^t (c_s+I_s-Y_s)ds \\
		&&- \left(w_0 -\widehat{W}_0 + \int_0^t({_sp_x})^{-1}\beta^{-1}_{0,s}  dA_{0,s}\right)\\
		&:=& w_0 + \int_0^t (\alpha_s r_s +  \theta^{\top}_s\mu_s) ds  + \int_0^t \theta^{\top}_s \sigma_s dZ_s - \int_0^t (c_s+I_s-Y_s)ds - C_t,
	\end{eqnarray*}
	which is the same as the dynamic budget constraint \eqref{dynamic_budget_constraint} and $C_t$ is the free disposal equals
	\begin{equation*}
		C_t = w_0 -\widehat{W}_0 + \int_0^t ({_sp_x})^{-1}\beta^{-1}_{0,s} dA_{0,s}.
	\end{equation*}
	Finally, we only need to prove $(\alpha,\theta)\in A$ for the trading strategy \eqref{appendix_1_strategy_1}. Substituting $\widehat{W}_t = \alpha_t + \theta^{\top}_t \bar{1}_n$ into \eqref{appendix_1_eq2_v0}, we can derive 
	\begin{equation*}
		\int_0^t \alpha_s v_{0,s} + \theta^{\top}_s v_{-,s} + \delta(v_s) ds + \int_0^t({_sp_x})^{-1}\beta^{-1}_{0,s} dA_{0,s} = \int_0^t ({_sp_x})^{-1}\beta^{-1}_{v,s}  dA_{v,s} \geq 0.
	\end{equation*} 
	Since $v\in \mathcal{N}^*$ is arbitrage, $\widetilde{A}$ is a convex cone, and $\delta$ is positive homogeneous, if there exists some $(v_0,v_-)$ such that $\alpha_s v_{0,s} + \theta^{\top}_s v_{-,s} + \delta(v_s)<0$, then $\alpha_s b v_{0,s} + \theta^{\top}_s b v_{-,s} + \delta(b v_s)$ can be any negative number for $b>0$, which contradicts
	\begin{equation*}
		\int_0^t \alpha_s v_{0,s} + \theta^{\top}_s v_{-,s} + \delta(v_s) ds + \int_0^t ({_sp_x})^{-1}\beta^{-1}_{0,t} dA_{0,s} \geq 0.
	\end{equation*}
	
	Therefore, there exists a set $E$ having full $(\bar{\lambda} \times P)$ measure (where $(\bar{\lambda} \times P)$ is product measure on $[0,T] \times \Omega$) such that
	\begin{equation*}
		\delta(v) + \alpha(t,\omega) v_0 + \theta(t,\omega)^{\top} v_- \geq 0, \forall (t,\omega) \in E, v\in \widetilde{A}.
	\end{equation*}
	(see Step 3 of Theorem 9.1 in \cite{cvitanic1992convex}). By Theorem 13.1 in \cite{rockafellar1970convex}, we derive $(\alpha, \theta)\in A$, $(\bar{\lambda} \times P)\text{-a.s.}$
\end{proof}

\section{Proof of Lemma \ref{dual_utility_property}}\label{appendix_2}
\setcounter{equation}{0}
\renewcommand{\theequation}{B.\arabic{equation}}
\begin{proof}
	By the definition of $\widetilde{U}_1$, we have
	\begin{equation*}
		\widetilde{U}_1(z,t) = \sup \limits_{c\geq 0} \{U_1(c,t)-zc\} = U_1(c^*,t) - zc^*, ~~z>0
	\end{equation*}	
	where $c^*$ is the optimal consumption satisfying
	\begin{equation}\label{appendix_3_opt_c}
		U'_1(c^*,t) - z =0, z>0.
	\end{equation}
    Then, we have $c^*>0$ because $U_1$ satisfies Inada condition \eqref{Inada_condition}, $U_1$ is strictly concave with the first variable by Definition \ref{def_utility_function}, and $z>0$. 
	Moreover, by \eqref{appendix_3_opt_c}, we have the optimal $c^*$ is a function of $z$. Next, by the law of implicit differentiation, we can derive the first-order and second-order partial derivatives of $\widetilde{U}_1$ with respect to $z$
	\begin{eqnarray}
		\frac{\partial \widetilde{U}_1(z,t)}{\partial z} &=& U'_1(c^*,t)\frac{\partial c^*}{\partial z} - c^* - z\frac{\partial c^*}{\partial z} = - c^*<0,\label{appendix_3_dtildeUdz}\\
		\frac{\partial^2 \widetilde{U}_1(z,t)}{\partial z^2} &=& -\frac{\partial c^*}{\partial z} = -\frac{\partial U'^{-1}_1(z,t)}{\partial z} = -\frac{1}{U^{''}_1(U'^{-1}_1(z,t),t)} = -\frac{1}{U^{''}_1(c^*,t)}>0.\label{appendix_3_d2tildeUdz2}
	\end{eqnarray} 
	Therefore, $\widetilde{U}_1(z,t)$ is strictly decreasing and strictly convex in its first variable. The same arguments can be applied to $\widetilde{U}_2$ and $\widetilde{U}_3$. 
	
	The representation \eqref{tildeU1_derivation2} is a direct result by substituting $c^*$ in \eqref{opt_f} into \eqref{tildeU1_derivation}. The same arguments are for $\widetilde{U}_2$ and $\widetilde{U}_3$. 
	
	For $i=1,2,3$, by the Inada condition \eqref{Inada_condition}
	\begin{equation*}
		U'_i(0+,t) = \infty, U'_i(\infty,t) = 0+, ~\text{for}~ \forall t\in[0,T],
	\end{equation*}
    we have 
    \begin{equation*}
    	U'^{-1}_i(0+,t) = \infty, U'^{-1}_i(\infty,t) = 0+, ~\text{for}~ \forall t\in[0,T].
    \end{equation*}
i.e.
    \begin{equation*}
	f_i(0+,t) = \infty, f_i(\infty,t) = 0+, ~\text{for}~ \forall t\in[0,T].
    \end{equation*}
    When $z$ goes to infinity, we have
    \begin{eqnarray*}
    	\widetilde{U}_i(\infty,t) &\leq& U_i(f_i(\infty,t),t) = U_i(0+,t)\\
    	\widetilde{U}_i(\infty,t) &\geq& \lim \limits_{z \rightarrow \infty}\left[U\left(\frac{\epsilon}{z},t\right)-\epsilon\right] = U_i(0+,t) - \epsilon, \forall \epsilon>0.
    \end{eqnarray*}
    Therefore, $\widetilde{U}_i(\infty,t)=U_i(0+,t)$.\\
    The inverse transform from the dual utility to the primal utility is
    \begin{eqnarray*}
    	U_i(x,t) = \inf \limits_{y>0} [\widetilde{U}_i(y,t)+xy] = \widetilde{U}_i(U'_i(x,t),t) + xU'_i(x,t).
    \end{eqnarray*}
    Next, we can derive
    \begin{eqnarray*}
    	U_i(\infty,t) &\geq& \widetilde{U}_i(U'_i(\infty,t),t) = \widetilde{U}_i(0+,t)\\
    	U_i(\infty,t) &\leq& \lim \limits_{x\rightarrow \infty}\left[\widetilde{U}_i\left(\frac{\epsilon}{x},t\right)+\epsilon\right] = \widetilde{U}_i(0+,t)+\epsilon, ~\forall \epsilon>0.
    \end{eqnarray*}
    Thus, $\widetilde{U}_i(0+,t)=U_i(\infty,t)$, which completes the proof.
\end{proof}

\section{Proof of Proposition \ref{find_opt_psi_prop}}\label{appendix_3}
\setcounter{equation}{0}
\renewcommand{\theequation}{C.\arabic{equation}}
\begin{proof}
	From the dual problem (\ref{dual_problem}), we can obtain the following first-order partial derivative
	\begin{align*}
		&\frac{\partial \widetilde{J}(\psi,v)}{\partial \psi} =E\left[\int_0^T {_tp_x} \widetilde{U}'_1(\psi \pi_{v,t},t) \pi_{v,t} dt + \int_0^T {_tp_x}\lambda_{x+t} \widetilde{U}’_2(\psi \pi_{v,t},t) \pi_{v,t} dt\right.\notag\\
		&\left.+{_Tp_x} \widetilde{U}'_3(\psi \pi_{v,T},T)\pi_{v,T}\right]+ w_0 + E\left[\int_0^T {_tp_x}\pi_{v,t}[Y_t+\delta(v_t)]dt\right],
	\end{align*}
	where $\widetilde{U}'_i(z,t)$, $i=1, 2, 3,$ are the first-order partial derivatives of dual utilities in its first variables. 
	
	For dual utility $\widetilde{U}_1(z,t)$, based on \eqref{appendix_3_opt_c} and \eqref{appendix_3_dtildeUdz}, we derive
	\begin{equation} \label{appendix_3_opt_c_2}
		\frac{\partial \widetilde{U}_1(z,t)}{\partial z} = - c^* = - U'^{-1}_1(z,t).	
	\end{equation}
	Together with the Inada condition \eqref{Inada_condition}, we obtain
	\begin{equation*}
		\widetilde{U}'_1(0+,t) = -\infty,~\widetilde{U}'_1(\infty,t) = 0, ~\text{for}~ \forall t \in [0,T].
	\end{equation*}
	Moreover, by \eqref{appendix_3_d2tildeUdz2}, we have $\widetilde{U}'_1(z,t)$ increase from $-\infty$ to $0$ when $z$ moves from $0+$ to $\infty$. The same arguments can be applied to $\widetilde{U}'_2(z,t)$ and $\widetilde{U}'_3(z,t)$. In addition, since $\pi_{v,t}>0$ and $w_0 + E\left[\int_0^T {_tp_x}\pi_{v,t}[Y_t+\delta(v_t)]dt\right]>0$, we can always find a unique $\psi^v>0$ such that
	\begin{equation}\label{Jtilde_zero_equality}
		\frac{\partial \widetilde{J}(\psi^v,v)}{\partial \psi} = 0.
	\end{equation}
	Finally, because $\widetilde{J}(\psi,v)$ is convex in $\psi$, the zero point $\psi^v$ of $\frac{\partial \widetilde{J}(\psi,v)}{\partial \psi}$ minimizes $\widetilde{J}(\psi,v)$ under a given $v$.
	Lastly, by \eqref{opt_f} and \eqref{Jtilde_zero_equality}, we find the optimal strategy under $(\psi^v,v)$ satisfies the following static budget constraint 
	\begin{align*}
		&E\left[\int_0^T {_tp_x}\pi_{v,t}  [f_1(\psi_v \pi_{v,t}) + \lambda_{x+t}f_2(\psi_v \pi_{v,t})] dt   +{_Tp_x}\pi_{v,T}  f_3 (\psi_v \pi_{v,T})\right]\\
            &=w_0 +E\left[\int_0^T {_tp_x}\pi_{v,t} [Y_t + \delta(v_t)] dt  \right]. 
	\end{align*}
	From this static budget constraint, we can define the optimal wealth following \eqref{optimal_wealth_v}, which is a martingale. Therefore, the optimal free disposal equals zero.
\end{proof}

\section{Proof of Lemma \ref{explicit_solution_0}}\label{appendix_4}
\setcounter{equation}{0}
\renewcommand{\theequation}{D.\arabic{equation}}
\begin{proof}
By the definitions \eqref{wealth_process_after_death} and \eqref{VB_definiton}, we can apply dynamic programming principle to derive the following Hamilton–Jacobi–Bellman(HJB) equation
\begin{equation}\label{VB_HJB}
	0 = -\widetilde{\delta}V_B(t,W_t) + \frac{\partial V_B}{\partial t} + \frac{\partial V_B}{\partial W}r(t)W_{t} - \frac{1}{2}\kappa^2_{0,t}\left(\frac{\partial V_B}{\partial W}\right)^2/\frac{\partial^2 V_B}{\partial W^2} + \frac{\gamma}{1-\gamma}\left(\frac{\partial V_B}{\partial W}\right)^{-\frac{1-\gamma}{\gamma}}.  
\end{equation}
From \eqref{VB_explicit}, we obtain the following derivatives
\begin{align*}
	&\frac{\partial V_B}{\partial t} = -\frac{\gamma}{1-\gamma} W_t^{1-\gamma} F_{B}(t)^{\gamma-1}+ \frac{\gamma}{1-\gamma} W_t^{1-\gamma}F_{B}(t)^{\gamma}\left\{\frac{\widetilde{\delta}}{\gamma}+\frac{\gamma-1}{\gamma}r(t) + \frac{1}{2}\frac{\gamma-1}{\gamma^2}\kappa^2_{0,t}\right\},\\
	&\frac{\partial V_B}{\partial W} = W^{-\gamma}_t F_B(t)^{\gamma}, ~\frac{\partial^2 V_B}{\partial W^2} = -\gamma W^{-\gamma-1}_t F_B(t)^{\gamma}.
\end{align*}
Substitute these derivatives into \eqref{VB_HJB}, the equality holds. Therefore, \eqref{VB_explicit} is the explicit solution to \eqref{VB_HJB}.
\end{proof}

\section{Proof of Proposition \ref{explicit_solution_1} and Proposition \ref{explicit_solution_2}}\label{appendix_5}

\setcounter{equation}{0}
\renewcommand{\theequation}{E.\arabic{equation}}

\begin{proof}
	We first prove Proposition \ref{explicit_solution_1}. To begin with, we denote $(\alpha_v, \theta_v, c_v, I_v)$ as the general strategy and $((\alpha_v)^*, (\theta_v)^*, (c_v)^*, (I_v)^*)$ as the optimal strategy under artificial market $\mathcal{M}_v$. Then, we can obtain the optimal wealth $W_{v,t}$ under $\mathcal{M}_v$ by \eqref{optimal_wealth_v}
	\begin{eqnarray*}
		&& W_{v,t} = E^{Q_{v}} \left[  \int_t^T {_{s-t}p_{x+t}}e^{-\int_t^s r(u) + v_{0}(u) du}[c_{v,s} -Y_s + \lambda_{x+s}M_{v,s} - \delta(v(s))] ds \right. \notag\\
		&&\left.+ {_{T-t}p_{x+t}}e^{-\int_t^T r(s) + v_{0}(s) ds}W_{v,T}|\mathcal{F}_{t} \right].\label{appendix_8_static_budget_constraint} 
	\end{eqnarray*}
	Therefore,
	\begin{equation}
		H_{v,t} = {_{t}p_{x}} \beta_{v,t}  W_{v,t} + \int_0^t {_{s}p_{x}}\beta_{v,s}  [c_{v,s}-Y_s+\lambda_{x+s}M_{v,s} -\delta(v(s))] ds \label{appendix_8_eq1}
	\end{equation}
	is a $Q_v$-martingale for $v \in \mathcal{N}^*$. Next, by martingale presentation theorem, there exists a $\mathbb{R}$-valued process $\psi_v$ with $\int_0^T |\psi_{v,t}|^2 dt<\infty$, such that
	\begin{equation}
		H_{v,t} = W_{v,0} + \int_0^t \psi_{v,s} dZ_{v,s}.\label{appendix_8_eq2}
	\end{equation}
	Substitute \eqref{appendix_8_eq2} into \eqref{appendix_8_eq1}, we derive
	\begin{align*}
		&W_{v,t} = ({_{t}p_{x}})^{-1}\beta^{-1}_{v,t} \left\{ H_{v,t} - \int_0^t {_{s}p_{x}} \beta_{v,s} [c_{v,s}-Y_s+\lambda_{x+s}M_{v,s} - \delta(v(s))] ds \right \}\\
		&=({_{t}p_{x}})^{-1}\beta^{-1}_{v,t} \left\{  W_{v,0} + \int_0^t \psi_{v,s} dZ_{v,s} - \int_0^t {_{s}p_{x}} \beta_{v,s} [c_{v,s}-Y_s+\lambda_{x+s}M_{v,s} - \delta(v(s))] ds \right \}.
	\end{align*}
	By Ito's formula and change of measure \eqref{change of measure}, we obtain
	\begin{align}
		&dW_{v,t} = (r(t)+v_{0}(t) +\lambda_{x+t}) W_{v,t} dt \notag\\
		&+ ({_{t}p_{x}})^{-1}\beta^{-1}_{v,t}\psi_{v,t}[dZ_t + \sigma^{-1}(t) (\mu(t)+v_{-}(t)-(r(t)+v_{0}(t))) dt] \notag\\
		& - [c_{v,t}-Y_t + \lambda_{x+t}M_{v,t} -\delta(v(t))] dt.\label{appendix_8_eq3}
	\end{align}
	If we choose $\psi_{v,t} = {_{t}p_{x}}\beta_{v,t}\sigma(t)\theta_{v,t}$ and rewrite $M_{v,t} = W_{v,t} + I_{v,t}/\lambda_{x+t}$, then \eqref{appendix_8_eq3} can be simplified to
	\begin{eqnarray}\label{appendix_8_wealth_SDE_2}
		dW_{v,t} &=& [r(t)\alpha_{v,t}+\theta_{v,t} \mu(t)]dt + [\alpha_{v,t} v_{0}(t) + \theta_{v,t} v_{-}(t)+\delta(v(t))] dt\\
		&&+ \sigma(t)\theta_{v,t}dZ_t - (c_{v,t}+I_{v,t}-Y_t) dt,\notag\\
		W_{v,0} &=& w_0, (\alpha_v,\theta_v) \in \mathbb{R}^2.\notag
	\end{eqnarray} 
	which has no free disposal. Here, we enlarge the domain of $(\alpha_v,\theta_v)$ to $\mathbb{R}^2$ to find the upper bound for the primal problem. Specifically, by the definition \eqref{supporting_function}, we have $v_{0}(t)\alpha_{v,t} + v_{-}(t)\theta_{v,t}  + \delta(v(t))\geq 0$ for $(\alpha_v,\theta_v) \in A$. Therefore, the wealth process \eqref{appendix_8_wealth_SDE_2} has a higher drift term and is bigger and equal to the wealth process \eqref{numerical_wealth} almost surely for $(\alpha_v,\theta_v) \in A$. Moreover, since $A \subset \mathbb{R}^2$, maximizing the objective function $J(c_v,M_v,W_{v,T})$ under the wealth process \eqref{appendix_8_wealth_SDE_2} with a large admissible set $(\alpha_v,\theta_v) \in \mathbb{R}^2$ provides an upper bound for the optimal objective function $J(c_v,M_v,W_{v,T})$ under the wealth process \eqref{numerical_wealth} with small admissible set $(\alpha_v,\theta_v) \in A$. In other words, the expected utility of an individual who invests freely following \eqref{appendix_8_wealth_SDE_2} under artificial market $\mathcal{M}_v$ provides an upper bound for the primal problem. For $t\in [T_R,T]$, SDE \eqref{appendix_8_wealth_SDE_2} equals
	\begin{eqnarray*}
		dW_{v,t} &=& \{\alpha_{v,t}[r(t)+v_{0}(t)] + \theta_{v,t}[\mu(t)+v_{-}(t)]\} dt + \sigma(t) \theta_{v,t} dZ_t - (c_{v,t}+I_{v,t} -\delta(v(t))) dt\\
		&=&\{[r(t)+\lambda_{x+t}+v_{0}(t)]W_{v,t} + \theta_{v,t}[\mu(t)+v_{-}(t)-(r(t)+v_{0}(t))]\}dt + \theta_{v,t} \sigma(t) dZ_t \\
		&& - [c_{v,t} + \lambda_{x+t} M_{v,t} - \delta(v(t))] dt.
	\end{eqnarray*}
	Define the value function $\widetilde{J}_R(t,W_{v,t};v)$ as
	\begin{align*}
		&\widetilde{J}_R(t,W_{v,t};v) = \sup \limits_{\theta_v,c_v,M_v} E_t\left[\int_t^T {_{s-t}p_{x+t}}e^{- \widetilde{\delta}(s-t)} \frac{(c_{v,s})^{1-\gamma}}{1-\gamma} ds \right.\\
		&\left. + \int_t^T {_{s-t}p_{x+t}}\lambda_{x+s} e^{ - \widetilde{\delta}(s-t)} \frac{(M_{v,s})^{1-\gamma}}{1-\gamma}\varphi g(s)^{\gamma} ds + {_{T-t}p_{x+t}}e^{ - \widetilde{\delta}(T-t)}  \frac{(W_{v,T})^{1-\gamma}}{1-\gamma}\right].
	\end{align*}
	By the dynamic programming principal, we derive the HJB equation
	\begin{align}
		&0 = -(\lambda_{x+t}+\widetilde{\delta})\widetilde{J}_R(t,W_{v,t};v) + \frac{\partial \widetilde{J}_R}{\partial t} + \frac{\partial \widetilde{J}_R}{\partial W_v}[(r(t)+\lambda_{x+t}+v_{0}(t))W_{v,t} + \delta(v(t))] \notag\\
		&-\frac{1}{2 [\sigma(t)]^2 \frac{\partial^2 \widetilde{J}_R}{\partial (W_v)^2}}\left( \frac{\partial \widetilde{J}_R}{\partial W_v}  \right)^2 [\mu(t) + v_{-}(t) - (r(t)+v_{0}(t))]^2 + \frac{\gamma}{1-\gamma}[1+\lambda_{x+t}\varphi^{\frac{1}{\gamma}}g(t)]\left(\frac{\partial \widetilde{J}_R}{\partial W_v} \right)^{\frac{\gamma-1}{\gamma}}, \label{appendix_8_HJB_equation}
	\end{align}
	together with the optimal strategies
	\begin{eqnarray}
		&&(\theta_{v,t})^* = \text{Proj}_{[0,W_{v,t}]}\left\{\frac{\kappa_{v,t}}{\sigma(t)\frac{\partial^2 \widetilde{J}_R}{\partial (W_v)^2}}\frac{\partial \widetilde{J}_R}{\partial W_v}\right\}, \label{appendix_8_theta}\\
		&&(c_{v,t})^* =\left( \frac{\partial \widetilde{J}_R}{\partial W_v}  \right)^{-\frac{1}{\gamma}},~~ (M_{v,t})^* = \left( \frac{\partial \widetilde{J}_R}{\partial W_v} \right)^{-\frac{1}{\gamma}}\varphi^{\frac{1}{\gamma}}g(t).\label{appendix_8_c_M}
	\end{eqnarray}
	From \eqref{upper_bound_JR}, we can derive the formulas for derivatives $\frac{\partial \widetilde{J}_R}{\partial t}$, $\frac{\partial \widetilde{J}_R}{\partial W_v}$, and $\frac{\partial^2 \widetilde{J}_R}{\partial (W_v)^2}$.
	Plug these derivatives into the equation \eqref{appendix_8_HJB_equation}, the equality holds. Therefore, the value function $\widetilde{J}_R(t,W_{v,t};v)$ is the solution to \eqref{appendix_8_HJB_equation}. Moreover, substitute \eqref{upper_bound_JR} into the optimal strategies \eqref{appendix_8_theta} and \eqref{appendix_8_c_M}, we obtain \eqref{JR_strategies1} and \eqref{JR_strategies2}, respectively.
    
    The proof of Proposition \ref{explicit_solution_2} is same to that of Proposition \ref{explicit_solution_1}, except the wealth process has a stochastic income $Y_t$
	\begin{eqnarray*}
		dW_{v,t}
		&=&\{(r(t)+\lambda_{x+t}+v_{0}(t))W_{v,t} + \theta_{v,t}[\mu(t)+v_{-}(t)-(r(t)+v_{0}(t))]\}dt + \theta_{v,t} \sigma(t) dZ_t\\
		&& - [c_{v,t} + \lambda_{x+t} M_{v,t} -Y_t - \delta(v(t))] dt,
	\end{eqnarray*}
    and the value function follows
    \begin{align*}
		&\widetilde{J}(t,W_{v,t},Y_t;v) = \sup \limits_{\theta_v,c_v,M_v} E_t\left[\int_t^{T_R} {_{s-t}p_{x+t}}e^{ - \widetilde{\delta}(s-t)} \frac{(c_{v,s})^{1-\gamma}}{1-\gamma} ds\right.\\
		&\left. + \int_t^{T_R} {_{s-t}p_{x+t}}\lambda_{x+s} e^{ - \widetilde{\delta}(s-t)} \frac{(M_{v,s})^{1-\gamma}}{1-\gamma}\varphi g^{\gamma}(s) ds  + {_{T_R-t}p_{x+t}}e^{- \widetilde{\delta}(T_R-t)} J_R(T_R,W_{v,T_R};v)\right].
	\end{align*}

\end{proof}	

\section{Proofs of Proposition \ref{D_explicit_solution_1} and Proposition \ref{D_explicit_solution_2}}\label{appendix_6}
\setcounter{equation}{0}
\renewcommand{\theequation}{F.\arabic{equation}}

\begin{proof}
We first prove Proposition \ref{D_explicit_solution_1}. From \eqref{lag_upperbound}, we know that maximizing the Lagrangian $\mathcal{L}$ with $(c, M, W_T)$ and minimizing $(\psi, v)$ provides an upper bound for the primal problem.  We construct the following Lagrangian, a linear combination of the objective and static budget constraint under a fixed $v$, to obtain an upper bound at each time $t$.
\begin{align*}
    &\mathcal{L}(t, c, M, W_T, \psi, v) = E_t\left[\int_t^T e^{-\widetilde{\delta}(s-t)}{_{s-t}p_{x+t}} \frac{c_{s}^{1-\gamma}}{1-\gamma}ds + \int_t^T  {_{s-t}p_{x+t}}\lambda_{x+s} e^{-\widetilde{\delta}(s-t)}   \frac{M_{s}^{1-\gamma}}{1-\gamma} \varphi g^{\gamma}(s)ds \right.\\
    &\left.+ {_{T-t}p_{x+t}}e^{-\widetilde{\delta}(T-t)}\frac{W_{T}^{1-\gamma}}{1-\gamma} \right] + \psi \left\{W_{v,t} - E_t\left[\int_t^T {_{s-t}p_{x+t}}\frac{\pi_{v,s}}{\pi_{v,t}}[c_{s}-Y_s+\lambda_{x+s}M_{s}-\delta(v_s)]ds \right.\right.\\
    &\left.\left.+{_{T-t}p_{x+t}}\frac{\pi_{v,T}}{\pi_{v,t}}W_{T}\right]\right\},
\end{align*}
where $\psi>0$ is the Lagrangian multiplier, and $\delta(v)=0$ for the portfolio mixed constraint.

By the first-order condition, we can obtain the optimal candidate strategies
\begin{eqnarray*}
	\left\{
	\begin{array}{l}
		c^*_{v,s}=\psi^{-\frac{1}{\gamma}}e^{-\frac{\widetilde{\delta}}{\gamma}(s-t)}\left(\frac{\pi_{v,s}}{\pi_{v,t}}\right)^{-\frac{1}{\gamma}},\\
		M^*_{v,s}=\psi^{-\frac{1}{\gamma}}e^{-\frac{\widetilde{\delta}}{\gamma}(s-t)}\left(\frac{\pi_{v,s}}{\pi_{v,t}}\right)^{-\frac{1}{\gamma}}\varphi^{\frac{1}{\gamma}} g(s),\\
		W^*_{v,T}=\psi^{-\frac{1}{\gamma}}e^{-\frac{\widetilde{\delta}}{\gamma}(s-t)}\left(\frac{\pi_{v,T}}{\pi_{v,t}}\right)^{-\frac{1}{\gamma}}.
	\end{array}
	\right.
\end{eqnarray*}
Plug these strategies into the static budget constraint, we obtain the optimal Lagrangian multiplier
\begin{eqnarray*}
    \psi^* &=& \left\{ \int_t^T {_{s-t}p_{x+t}} e^{-\frac{\widetilde{\delta}}{\gamma}(s-t)} (1+\lambda_{x+s}\varphi^{\frac{1}{\gamma}} g(s)) E_t\left[\left(\frac{\pi_{v,s}}{\pi_{v,t}}\right)^{-\frac{1-\gamma}{\gamma}}\right]ds \right.\\
    &&\left.+ {_{T-t}p_{x+t}} e^{-\frac{\widetilde{\delta}}{\gamma}(T-t)} E_t\left[\left(\frac{\pi_{v,T}}{\pi_{v,t}}\right)^{-\frac{1-\gamma}{\gamma}}\right]\right\}^{\gamma} W^{-\gamma}_{v,t}.
\end{eqnarray*}
From \eqref{numerical_Yt_process}, we have 
\begin{eqnarray*}
		Y_s= Y_t = 0, ~~T_R \leq t\leq s\leq T,
\end{eqnarray*} 
Together with
\begin{eqnarray*}
    \left(\frac{\pi_{v,s}}{\pi_{v,t}}\right)^{-\frac{1-\gamma}{\gamma}}=\exp\left\{\frac{1-\gamma}{\gamma}\int_t^s r+ v_{0,u} du + \frac{1}{2}\frac{1-\gamma}{\gamma}\int_t^s \kappa^2_{v,u}du -\frac{1-\gamma}{\gamma}\int_t^s\kappa_{v,u}dZ_u\right\},
\end{eqnarray*}
we obtain that the conditional expectation $E_t\left[\left(\pi_{v,s}/\pi_{v,t}\right)^{-\frac{1-\gamma}{\gamma}}\right]$ is a function of $(t,s)$.
Finally, plug the optimal $(c^*_{v,s}, M^*_{v,s}, W^*_{v,T})$ and $\psi^*$ into the objective, we can obtain the optimal value function \eqref{Method3_value_function}.
By the dynamic programming principle, we can derive the HJB equation under the artificial market
\begin{eqnarray*}
    0&=&\sup \limits_{c_{v,t}}\left\{\frac{c^{1-\gamma}_{v,t}}{1-\gamma} - c_{v,t}\frac{\partial \widetilde{J}_R}{\partial W_v}\right\} + \sup \limits_{M_{v,t}}\left\{\lambda_{x+t}\frac{M^{1-\gamma}_{v,t}}{1-\gamma}\varphi g^{\gamma}(t)-\lambda_{x+t}M_{v,t}\frac{\partial \widetilde{J}_R}{\partial W_v}\right\}\\
    &&+\sup \limits_{\theta_{v,t}}\left\{\frac{\partial \widetilde{J}_R}{\partial W_v}\theta_{v,t} [\mu(t)+v_{-,t}-(r(t)+v_{0,t})] + \frac{1}{2}\frac{\partial^2 \widetilde{J}}{\partial W^2_v} \theta^2_{v,t} \sigma^2(t)\right\}\\
    &&-(\lambda_{x+t}+\widetilde{\delta})\widetilde{J}_R(t,W_{v,t}) + \frac{\partial \widetilde{J}_R}{\partial t} + \frac{\partial \widetilde{J}_R}{\partial W_v}\{[r(t)+\lambda_{x+t}+v_{0,t}]W_{v,t} + \delta(v_t)\}.
\end{eqnarray*}

Solving the HJB equation for the optimal strategies, we obtain

\begin{eqnarray*}
	\left\{
	\begin{array}{l}
		c^*_{v,t}=\left(\frac{\partial \widetilde{J}_R}{\partial W_v}\right)^{-\frac{1}{\gamma}}, M^*_{v,t}=\left(\frac{\partial \widetilde{J}_R}{\partial W_v}\right)^{-\frac{1}{\gamma}}\varphi^{\frac{1}{\gamma}} g(t),\\
		\theta^*_{v,t}=\text{Proj}_{[0,W_{v,t}]}\left\{ \frac{\frac{\partial \widetilde{J}_R}{\partial W_v}\kappa_{v,t}}{\frac{\partial^2 \widetilde{J}}{\partial W_v^2}\sigma(t)}   \right\},
	\end{array}
	\right.
\end{eqnarray*}
where $\text{Proj}_{[0,W_{v,t}]}\{\cdot\}$ is a truncation projection to the interval $[0,W_{v,t}]$. Here, we truncate $\theta^*_{v,t}$ because the trading constraint only binds under $v^*$, not for general $v$. The truncated $\theta^*_{v,t}$ is a good candidate strategy for simulating the lower bound. Finally, we can obtain \eqref{D_J_strategy_theta} and \eqref{D_J_strategy_c_M} by substituting the semi-explicit value function \eqref{Method3_value_function} into the optimal strategies above.

The proof of Proposition 5.4 is similar to that of Proposition 5.3, except for the income process 
\begin{eqnarray*}
	\left\{
	\begin{array}{ll}
		Y_s&= Y_t \exp\{(\mu_Y-\frac{1}{2}\sigma^2_Y)(s-t)+\sigma_Y (Z_{s}-Z_{t})\}, ~~ 0\leq t \leq s < T_R, \label{Yt_process_exp}\\
		Y_s&= Y_t= 0, ~~T_R\leq s\leq T,
	\end{array}
	\right.
\end{eqnarray*}
and the form of the value function 
\begin{align}
    \widetilde{J}^D(t,W_{v,t},Y_t;v)&=\frac{1}{1-\gamma}\left\{W_{v,t}+ \int_t^T {_{s-t}p_{x+t}} Y_t E_t\left[\frac{\pi_{v,s}Y_s}{\pi_{v,t}Y_t}\right]ds\right\}^{1-\gamma}\notag\\
    &\left\{ \int_t^T {_{s-t}p_{x+t}} e^{-\frac{\widetilde{\delta}}{\gamma}(s-t)} (1+\lambda_{x+s}\varphi^{\frac{1}{\gamma}}g(s)) E_t\left[\left(\frac{\pi_{v,s}}{\pi_{v,t}}\right)^{-\frac{1-\gamma}{\gamma}}\right]ds\right.\notag\\
    &\left.+ {_{T-t}p_{x+t}} e^{-\frac{\widetilde{\delta}}{\gamma}(T-t)} E_t\left[\left(\frac{\pi_{v,T}}{\pi_{v,t}}\right)^{-\frac{1-\gamma}{\gamma}}\right]\right\}^{\gamma}. \notag
\end{align}
\end{proof}

\end{document}